\documentclass[12pt]{report}
\usepackage[totalwidth=480pt,totalheight=680pt]{geometry}
\usepackage{amssymb}
\newcommand{\be}{\begin{equation}}
\newcommand{\ee}{\end{equation}}
\newcommand{\bea}{\begin{eqnarray}}
\newcommand{\eea}{\end{eqnarray}}

\def\order{O}
\usepackage{epsfig}
\def\calM{\mathcal{M}}
\def\Re{\mathop{\rm Re}}
\def\Im{\mathop{\rm Im}}

\def\sgn{\mathop{\rm sgn}}

\def\Tren{T^{\text{ren}}}
\def\Vmax{V_{\text{max}}}
\usepackage{amsmath}
\usepackage{amsfonts}
\usepackage{latexsym}
\def\bml{\begin{subequations}}
\def\eml{\end{subequations}}
\def\blea{\bml\bea}
\def\elea{\eea\eml}

\def\Rmax{R_{\text{max}}}
\def\bl{\boldsymbol{\ell}}
\def\bx{\mathbf{x}}

\def\bx{\mathbf{x}}

\def\bzeta{\mathbf{\zeta}}

\def\Tsplit{T^{\text{split}}}
\def\Rmax{R_{\text{max}}}

\def\Texotic{T_{\text{exotic}}}
\def\Gexotic{G_{\text{exotic}}}
\def\lpl{l_{\text{Planck}}}
\def\stwo{\sqrt{2}}
\def\tu{\tilde{u}}
\def\tv{\tilde{v}}

\def\Fermi{\mathop{\rm Fermi}\nolimits}

\begin{document}
\thispagestyle{plain}
\begin{center}
    \Large
    \textbf{Averaged null energy condition and quantum inequalities in curved spacetime}
    
    \vspace{0.4cm}
    \large
    Based on a dissertation submitted by \\
\textbf{Eleni-Alexandra Kontou} \\
in partial fulfillment of the requirements \\
for the degree of Doctor of Philosophy in Physics \\
TUFTS UNIVERSITY
    
    \vspace{0.4cm}
    
Advisor: \textbf{Ken D. Olum}
    
    \vspace{0.9cm}
    \textbf{Abstract}
\end{center}
The Averaged Null Energy Condition (ANEC) states that the integral along a complete null geodesic of the projection of the stress-energy tensor onto the tangent vector to the geodesic cannot be negative.  ANEC can be used to rule out spacetimes with exotic phenomena, such as closed timelike curves, superluminal travel and wormholes. We prove that ANEC is obeyed by a
minimally-coupled, free quantum scalar field on any achronal null
geodesic (not two points can be connected with a timelike curve) surrounded by a tubular neighborhood whose curvature
is produced by a classical source. To prove ANEC we use a null-projected quantum inequality, which provides constraints on how negative the weighted average of the renormalized stress-energy tensor of a quantum field can be. Starting with a general result of Fewster and Smith, we first
derive a timelike projected quantum inequality for a minimally-coupled scalar field on flat spacetime with a background potential. Using that result we proceed to find the bound of a quantum inequality on a geodesic in a spacetime with small curvature, working to first
order in the Ricci tensor and its derivatives.  The last step is to derive a bound for the null-projected quantum inequality on a general timelike path. Finally we use that result to prove achronal ANEC in spacetimes with small curvature.

\tableofcontents

\listoffigures

\chapter*{Acknowlegements}

First and foremost I would like to thank my advisor Professor Ken Olum without whom this work would not be possible. Ken was my collaborator in all the work presented in this thesis, and moreover acted as my academic mentor as I worked toward my PhD. Next, I would like to thank Professor Larry Ford for his many valuable comments on this work and for offering his knowledge through thoughtful research discussions. Many thanks go to Professor Christopher Fewster for several discussions, comments and corrections on much of this work. I would also like to acknowledge Professor Thomas Roman and Dr.\ Douglas Urban for their insights. Thanks to Professor Jose-Juan Blanco-Pillado and Dr.\ Ben Shlaer for helpful discussions regarding multi-step Fermi coordinates. Finally I would like to thank Professor Alex Vilenkin, Professor Krzysztof Sliwa and Professor Mark Hertzberg for comments and corrections on the final manusrcipt of this thesis.

For their financial support I am grateful to the Foundational Questions Institute, the Burlingame fellowship and especially the Onassis Foundation which allowed me to pursue this research. 

\chapter{Introduction}

In the context of General Relativity, it is always possible to invent a spacetime with exotic features, such
as wormholes, superluminal travel, or the construction of time
machines, and then determine what stress-energy tensor is necessary to
support the given spacetime according to Einstein's Equations (units: $\hbar=c=1$)
\be
G_{ab}=8\pi T_{ab} \,.
\ee
However, in quantum field theory, there are restrictions
on $T_{ab}$ that could rule out exotic spacetimes.  Two examples of these are quantum inequalities and energy conditions.

Quantum inequalities (also called Quantum Energy Inequalities) are bounds on the weighted time averages of the stress-energy tensor. They were first introduced by Ford \cite{Ford:1978qya}
to prevent the violation of the second law of thermodynamics.  The general form
of a quantum inequality is
\be\label{QI}
\int_{-\infty}^\infty d\tau\,f(t) T_{ab}(w(t)) V^a V^b > - B\,,
\ee
where $w(t)$ is a timelike path parameterized by proper time $t$
with tangent vector $V$, and $f$ is a sampling function.  The quantity
$B$ is a bound, depending on the function $f$, the path $w$ and the quantum field of
interest.

Since then, they have been derived for a wide range of spacetimes, fields,
and weighting functions. We concentrate here on quantum inequalities
for minimally coupled scalar fields in curved spacetime. However, quantum inequalities for interacting fields have been derived in 1+1 dimensions \cite{math-ph/0412028,arXiv:1304.7682}.  For systems
with boundaries, there are difference quantum inequalities
\cite{Ford:1994bj,Fewster:1999gj}, in which $T_{ab}$ in Eq.~(\ref{QI})
is replaced by the difference between $T_{ab}$ in some state of
interest and $T_{ab}$ in a reference state.  The bound $B$ may also
then depend on the reference state. However, such difference
inequalities cannot be used to rule out exotic spacetimes, at least in
the case where the exotic matter that supports the spacetime comes
from the vacuum state in the presence of the boundaries.

Pointwise energy conditions bound the stress-energy tensor at individual spacetime points. One example is the Null Energy Condition (NEC) which requires that the null contracted stress-energy tensor cannot be negative,
\be \label{eqn:nec}
T_{ab}\ell^a \ell^b \geq 0 \,,
\ee
for $\ell^a$ a null vector.
Classically, pointwise energy conditions seem reasonable, but in the quantum context they are violated. Quantum field theory allows arbitrary negative energy densities at individual points, a well known example being the Casimir effect. Even in the simple case of a minimally coupled free scalar field, all known pointwise energy conditions fail and even local averages must admit negative expectation values \cite{Epstein:1965zza}.

On the other hand, averaged energy conditions bound the stress-energy tensor integrated along a
complete geodesic; they are weaker and have been proven to hold in a
variety of spacetimes. One example is the Averaged Null Energy Condition (ANEC) which bounds the null-projected stress-energy tensor integrated along a null geodesic $\gamma$
\be \label{eqn:anec}
\int_\gamma T_{ab} \ell^a \ell^b d\lambda \geq 0 \,.
\ee

To rule out exotic spacetimes such as those with wormholes and closed timelike curves, we
would like to prove energy conditions that restrict the stress-energy
tensor that might arise from quantum fields and show that the
stress-energy necessary to support these spacetimes is impossible.
We need a condition which is strong enough to rule out exotic cases,
while simultaneously weak enough to be proven correct.

The best possibility for such a condition seems to be the achronal
ANEC  \cite{Graham:2007va}, which requires
that $\gamma$ of Eq.~(\ref{eqn:anec}) is a complete achronal null geodesic i.e., no two points of $\gamma$ can be connected by a timelike curve.  
That is to say, we require that the projection of the stress-energy
tensor along a null geodesic integrate to a non-negative value, but only
for geodesics that are achronal.  As far as we know, there is no
example of achronal ANEC violation in spacetimes satisfying Einstein's
equations with classical matter or free quantum fields as
sources.\footnote{Except for the case of non-minimally coupled quantum
  scalar fields.}  Achronal ANEC is sufficient
to rule out many exotic spacetimes \cite{Graham:2007va}.

It has been proven that ANEC holds in Minkowski space and Ref.~\cite{Fewster:2006uf} showed that it also holds for geodesics traveling through empty, flat
space, even if elsewhere in the spacetime there are boundaries or
spacetime curvature, providing that these stay some minimum distance
from the geodesic and do not affect the causal structure of the
spacetime near the geodesic. This proof made use of quantum
inequalities for null contractions of the stress tensor averaged over
timelike geodesics \cite{Fewster:2002ne}.

This work however, does not really addresses the possibility of exotic
spacetimes.  The quantum inequalities on which it depends apply only
in flat spacetime, so they cannot be used to rule out spacetimes with
exotic curvature.  For that, we need limits on the stress-energy
tensor in curved spacetimes, the work presented here. 

This thesis presents a complete proof of achronal ANEC for minimally coupled scalar fields in spacetimes with curvature in a classical background that obeys NEC, using a null projected quantum inequality. In the first chapter we present a general quantum inequality derived by Fewster and Smith \cite{Fewster:2007rh} that we use in later chapters to derive a bound. In the second chapter we derive the bound for flat spacetime with a background potential \cite{Kontou:2014eka}, a case similar to the curved spacetime one. In chapter three we present the timelike projected quantum inequality in curved spacetime \cite{Kontou:2014tha} and discuss the importance of this result. In chapter four we use that result to derive a null projected quantum inequality, which we proceed to use to prove achronal ANEC \cite{Kontou:2012ve}. Finally, in the Appendix we present a new class of coordinates, called multi-step Fermi coordinates and use them to write the connection and the metric in terms of the curvature \cite{Kontou:2012kx}, results we use throughout this work.

We use the sign convention $(-,-,-)$ in the classification of Misner,
Thorne and Wheeler \cite{MTW}. Indices $a,b,c, \dots$ denote all
spacetime coordinates while $i,j,k \dots$ denote only spatial coordinates. We denote normal derivatives with comma and covariant derivatives with semicolon.

\chapter{Absolute quantum energy inequality}
\label{ch:QI}

In this chapter we present a general quantum inequality derived by Fewster and Smith \cite{Fewster:2007rh}, result which we are going to use throughout the thesis.
First we consider a minimally-coupled scalar field with the usual
classical stress-energy tensor,
\be
T_{ab}=\nabla_a \Phi \nabla_b \Phi-\frac{1}{2} g_{ab} g^{cd} \nabla_c
\Phi \nabla_d \Phi+\frac{1}{2}g_{ab}\mu^2 \,.
\ee
where $\mu$ is the mass. Following Ref.~\cite{Fewster:2007rh}, we define the renormalized stress-energy tensor
\be\label{eqn:Tren}
\langle \Tren_{ab} \rangle \equiv \lim_{x\to x'}  \Tsplit_{ab'} \left( \langle \phi(x)\phi(x') \rangle-H(x,x') \right)-Qg_{ab}+C_{ab}\,.
\ee
The quantities appear in Eq.~(\ref{eqn:Tren}) will be defined below.  $\Tsplit_{ab'}$ is  the point-split energy density operator,
\be\label{eqn:tsplit}
\Tsplit_{ab'}=\nabla_a \otimes \nabla_{b'}-g_{ab'} g^{cd'} \nabla_c \otimes \nabla_{d'}+\frac{1}{2}g_{ab'}\mu^2 \,.
\ee
The point-split energy density operator acts on the difference between the two-point function and the Hadamard series,
\bea \label{hadamard}
H(x,x')&=&\frac{1}{4\pi^2} \bigg[ \frac{1}{\sigma_+(x,x')}+\sum_{j=0}^{\infty}v_j(x,x') \sigma_+^j (x,x') \ln{\left(\frac{\sigma_+(x,x')}{l^2}\right)} \nonumber\\
&& \qquad +\sum_{j=0}^{\infty}w_j (x,x')\sigma^j (x,x') \bigg]\,,
\eea
where $\sigma$ is the
squared invariant length of the geodesic between $x$ and $x'$,
negative for timelike separation. In flat space
\be
\sigma(x,x')=-\eta_{ab} (x-x')^a (x-x')^b \,.
\ee
By $F(\sigma_+)$, for some function $F$, we mean the distributional
limit
\be 
F(\sigma_+)=\lim_{\epsilon \to 0^+} F(\sigma_{\epsilon}) \,,
\ee
where
\be
\sigma_{\epsilon}(x,x')=\sigma(x,x')+2i \epsilon(t(x)-t(x'))+\epsilon^2 \,.
\ee
In some parts of the calculation it is possible to assume that the points have only timelike separation, so we define
\be
\tau=t-t'
\ee
and write
\be
F(\sigma_+)=F(\tau_-)=\lim_{\epsilon \to 0} F(\tau_\epsilon) \,,
\ee
where
\be
\tau_\epsilon=\tau-i\epsilon \,.
\ee
We have introduced a length $l$ so that the argument of the
logarithm in Eq.~(\ref{hadamard}) is dimensionless.  The possibility
of changing this scale creates an ambiguity in the definition of $H$,
but this ambiguity for curved spacetime can be absorbed into the ambiguity involving local
curvature terms discussed below. The ambiguity in the case of a field in flat spacetime with background potential is discussed in Ch.~\ref{ch:potential}. For simplicity of notation, we will
assume we are working in units where $l = 1$.

The function $\Delta$ is the Van Vleck-Morette determinant bi-scalar, given by
\be \label{delta}
\Delta(x,x')=-\frac{\det (\nabla_a \otimes \nabla_{b'} \sigma(x,x')/2)}{\sqrt{-g(x)}\sqrt{-g(x')}} \,.
\ee
The Hadamard coefficients are given by the Hadamard recursion
relations, which are the solutions to
\be
(\Box+\mu^2) H(x,x')=0\,.
\ee
The recursion relations for the minimally coupled field in a curved background are \cite{Fewster:2007rh}
\be \label{recursion1}
(\Box+\mu^2) \Delta^{1/2}+2 v_{0,a} \sigma^{,a}+4 v_0+v_0 \Box \sigma=0 \,,
\ee
\be \label{recursion2}
(\Box+\mu^2) v_j +2(j+1) v_{j+1,a}\sigma^{,a}-4j(j+1)v_{j+1}+(j+1)v_{j+1}\Box \sigma=0 \,,
\ee
\be
2 w_{1,a} \sigma^{,a}+w_1 \Box \sigma+2 v_{1,a}\sigma^{,a}-4v_1+v_1 \Box \sigma=0 \,,
\ee
\bea
&&(\Box+\mu^2)w_j+2(j+1)w_{j+1,a}\sigma^{,a}-4j(j+1)w_{j+1}+(j+1)w_{j+1}\Box \sigma \nonumber\\
&&+2v_{j+1,a}\sigma^{,a}-4(2j+1)v_{j+1}+v_{j+1} \Box \sigma=0 \,.
\eea
All the $v_j$, and the $w_j$ for $j\ge1$ are determined by the
differential equations discussed above, but $w_0$ is undetermined.
Here we will follow Wald \cite{Wald:qft} and choose $w_0 = 0$.

From Ref.~\cite{Fewster:2007rh} we have the definition
\be \label{tilde}
\tilde{H}(x,x')=\frac{1}{2} \left[ H(x,x')+H(x',x)+iE(x,x') \right] \,,
\ee
where $iE$ is the antisymmetric part of the two-point function.
We can write $H_j(x,x')$, $j=-1,0,1,\ldots$, to denote the term in $H$ involving $\sigma^{j}$
(with or without $\ln{(\sigma_+)}$), and $H_{(j)}$ to denote the sum of all
terms up through $H_j$ .  We will split up $E(x,x')$ into terms
labeled $E_j$ that are proportional to $\sigma^{j}$, define a
``remainder term''
\be
R_j = E - \sum_{k=-1}^j E_k\,,
\ee
and let
\blea
\tilde
H_{j}(x,x') &=& \frac12\left[H_j(x,x') + H_j(x',x) + iE_j(x,x'))\right]\\
\tilde H_{(j)}(x,x') &=& \frac12\left[H_{(j)}(x,x') + H_{(j)}(x',x) + iE(x,x'))\right]\,.
\elea

The term $Q$ in Eq.~(\ref{eqn:Tren}) is the one introduced by Wald to preserve the
conservation of the stress-energy tensor. Wald \cite{Wald:1978pj}
calculated this term in the coincidence limit,
\be \label{Q}
Q=\frac{1}{12\pi^2}w_1(x,x) \,.
\ee

The term $C_{ab}$ handles the ambiguities in the definition of the
stress-energy tensor $T$ in curved spacetime.  We will adopt the
axiomatic definition given by Wald \cite{Wald:qft}, but there remains
the ambiguity of adding local curvature terms with arbitrary
coefficients.  From Ref.~\cite{Birrellbook} we find that these terms
include
\bml\label{eqn:12H}
\bea 
^{(1)}H_{ab} &\equiv& \frac{\delta}{\sqrt{-g}\delta g^{ab}} \int \sqrt{-g} R^2 d^4 x= 2R_{;ab} -2g_{ab}\Box R - \frac{g_{ab}R^2}{2} + 2RR_{ab} \label{1H}\\
^{(2)}H_{ab} &\equiv&  \frac{\delta}{\sqrt{-g}\delta g^{ab}} \int \sqrt{-g} R^{ab} R_{ab}=R_{;ab} -\Box R_{ab} -\frac{g_{ab}\Box R}{2} \nonumber\\
&&\qquad \qquad \qquad \qquad \qquad - \frac{g_{ab}R^{cd} R_{cd}}{2}  + 2R^{cd}R_{acbd} \label{2H}\,.
\elea
Thus in Eq.~(\ref{qinequality}) we must include a term given by a
linear combination of Eqs.~(\ref{1H}) and (\ref{2H}) to first order in $R$,
\be \label{localc}
C_{ab}=a\, ^{(1)}\!H_{ab}+b\, ^{(2)}\!H_{ab}
\ee
where $a$ and $b$ are undetermined constants.\footnote{There are also
  ambiguities corresponding to adding multiples of the metric and the
  Einstein tensor to the stress tensor.  The first can be considered
  renormalization of the cosmological constant and the second
  renormalization of Newton's constant.  We will assume that these
  renormalization have been performed, and that the cosmological
  constant is considered part of the gravitational sector, so neither
  of these affects $T_{ab}$.}

A spacetime is globally hyperbolic when it contains a Cauchy surface, a subset of spacetime which is intersected by every causal curve exactly once. Global hyperbolicity requires the existence of unique advanced and retarded Green functions. We define $w(t)$, a timelike path contained in a globally hyperbolic convex \footnote{A convex normal neighborhood $N$ is one such that any point $q \in N$ can be connected to any other point $p \in N$ with a unique geodesic totally contained in $N$. For more detailed discussion of normal neighborhoods and their properties see Ref.~\cite{HawkingEllis}} normal neighborhood $N$ and for this path we can state the quantum inequality of Ref.~\cite{Fewster:2007rh}.
\bea \label{qinequality}
\int_{-\infty}^\infty dt \, g(t)^2 \langle \Tren_{ab}V^a V^b \rangle_{\omega} w(t) &\geq& -\int_0^\infty\frac{d\xi}{\pi} \left[  g \otimes g (\theta^* \Tsplit_{ab'} V^a V^{b'} \tilde{H}_{(5)}) \right]^{\wedge} (-\xi,\xi) \nonumber\\
&&+\int_{-\infty}^\infty dt \, g^2(t) (-Q g_{ab}+C_{ab})V^a V^b \,.
\eea
The Fourier transform convention we use is
\be \label{Fourier}
\hat{f}(k) \text{ or } f^{\wedge}[k]=\int_{-\infty}^\infty dxf(x) e^{ixk} \,.
\ee
In the inequality we take the transform with respect to both arguments and
evaluate at $\xi$ and $-\xi$.
The operator $\theta^*$ denotes the pullback of the function to
the geodesic,
\be
(\theta^*\Tsplit_{ab'}\tilde{H}_{(5)})(t,t') \equiv (\Tsplit_{ab'} \tilde{H}_{(5)})(w(t),w(t'))\,,
\ee
and the subscript $(5)$ means that we include
only terms through $j = 5$ in the sums of
Eq.~(\ref{hadamard}). However, we will prove that terms of order $j >1$ make no contribution
to Eq.~(\ref{qinequality}).

We will use the general inequality of Eq.~(\ref{qinequality}) to provide a bound for the integral of the renormalized stress-energy tensor in three different cases. In Ch.~\ref{ch:potential} for the timelike projected $T_{\mu \nu}$ in flat spacetime with a background potential, in Ch.~\ref{ch:curvature} for the energy density in curved spacetime and in Ch.~\ref{ch:ANEC} for the null projected stress-energy tensor in spacetimes with curvature.

\chapter{Quantum Inequality for a scalar field with a background potential}
\label{ch:potential}

As a first step toward deriving a bound for the quantum inequality in a spacetime with bounded curvature we first derive a quantum
inequality in a flat spacetime with a background potential, i.e., a
field with a mass depending on spacetime position. This is a simpler
system that has many of the important features of quantum fields in
curved spacetime. For a scalar field $\Phi$ in a background potential,
the Lagrangian is
\be
L=\frac{1}{2} \left[\partial_\mu \Phi \partial^\mu \Phi-V(x) \Phi^2 \right]\,,
\ee
the equation of motion is
\be
(\Box+V(x)) \Phi=0\,,
\ee
and the classical energy density is
\be\label{classicalT00}
T_{tt}=\frac12\left[(\partial_t \Phi)^2+(\nabla \Phi)^2+V(x)\Phi^2\right]\,.
\ee
So the $\Tsplit_{ab'}$ operator of Eq.~(\ref{eqn:tsplit}) contracted with timelike vectors, for a scalar field with background potential $V$ becomes
\be\label{eqn:tsplitV}
\Tsplit_{tt'}=\frac{1}{2} \left[ \sum_{a=0}^3 \partial_a \partial_{a'}
 +\frac{V(x)+V(x')}2 \right]\,,
\ee
where the potential is analogous to the mass of Eq.~(\ref{eqn:tsplit}). Since we will take the limit where $x$ and $x'$ coincide the
location of evaluation of $V$ does not matter, but the form above will
be convenient later. The renormalized stress-energy tensor in this case is
\be \label{TVren}
\langle \Tren_{tt}(x') \rangle \equiv \lim_{x\to x'}  \Tsplit_{tt'} \left( \langle \phi(x)\phi(x') \rangle-H(x,x') \right)-Q(x')\,,
\ee
As discussed in Chapter \ref{ch:QI} there is an ambiguity in the above procedure.  In order
to take logarithms, we must divide $\sigma$ by the square of some
length scale $l$.  Changing the scale to some other scale $l'$
decreases $H$ by $\delta H =2 (v_0+v_1\sigma+\cdots)\ln(l'/l)$.  This
results in increasing $T_{ab}$ by $\lim_{x\to x'}
(\partial_a\partial_b- (1/2) \eta_{ab}\partial^c\partial_c)\delta H$.
Using the values for $v_0$ and $v_1$ computed below, this becomes
$(1/12) (V_{,ab} -\eta_{ab}\Box V)\ln(l'/l)$.  Thus we see that
the definition of $T_{ab}$ must include arbitrary multiple of $(V_{,ab}
-\eta_{ab}\Box V)$.  This ambiguity can also be understood as the
possibility of including in the Lagrangian density a term of the form
$R(x) V(x)$, where $R$ is the scalar curvature.  Varying the metric to
obtain $T_{ab}$ and then going to flat space yields the above term.
The situation is very much analogous to the possible addition of terms
of the form $R^2$ and $R_{ab} R^{ab}$ in the case of a field in curved
spacetime, which give rise to the local curvature terms in Eq.~(\ref{localc}).

Thus we rewrite Eq.~(\ref{TVren}) to include the ambiguous term,
\be
\langle \Tren_{tt}(x') \rangle \equiv \lim_{x\to x'}  \Tsplit \left(
\langle \phi(x)\phi(x') \rangle-H(x,x') \right)-Q(x')+CV_{,ii}\,,
\ee
where $C$ is some constant.  Whatever definition of $T_{tt}$ one is
trying to use, one can pick an arbitrary scale $l$ and adjust $C$
accordingly.

So the quantum inequality of Eq.~(\ref{qinequality}) becomes
\be\label{eqn:Vqinequality}
\int_{-\infty}^\infty d\tau\,g(t)^2 \langle \Tren_{00} \rangle (t,0) 
\geq -B\,,
\ee 
where
\be\label{eqn:BV}
B = \int_0^\infty\frac{d\xi}{\pi}\hat F(-\xi,\xi)
+\int_{-\infty}^\infty dt\,g^2(t) (Q-CV_{,ii})\,,
\ee
and
\be\label{FV}
F(t,t') = g(t)g(t')\Tsplit \tilde H_{(5)}((t,0),(t',0))\,,
\ee
$\hat F$ denotes the Fourier transform in both arguments according to
Eq.~(\ref{Fourier}).

We work only in first order in $V$ but don't otherwise assume that it is small. We can express the maximum values of the background potential and its derivatives as 
\bea\label{Vmax}
\begin{array}{cc}
|V| \leq \Vmax & |V_{,a}| \leq \Vmax' \\
|V_{,ab}| \leq \Vmax'' & |V_{,abc}| \leq \Vmax'''\,,
\end{array}
\eea
where $\Vmax$, $\Vmax'$, $\Vmax''$ and $\Vmax'''$ are positive
numbers, finite but not necessarily small.

First, we discuss the $\Tsplit_{tt'}$ operator, then we compute the Hadamard series and we apply the operator. After that we perform the Fourier transform, leading to
the final quantum inequality.

\section{General considerations}
\label{sec:general}

We will now compute the quantum inequality bound $B$ to first order in
the potential $V$ and its derivatives.  In the next subsection, we
will make some general remarks about terms in $\tilde H$ coming from
$H$, which are symmetrical under the exchange of $x$ and $x'$, and
show that we need keep terms only through first order, not fifth order
as in Eq.~(\ref{qinequality}).  Then we will simplify the operator $\Tsplit_{tt'}$
defined in Eq.~(\ref{eqn:tsplitV}).  In the next section, we will compute,
order by order, the terms and $\tilde H$.  We will then take the
Fourier transform to find the quantum inequality bound.

\subsection{Smooth, symmetrical contributions}
\label{sec:symmetry}

Define $\bar{x}=(x-x')/2$, $\bar t = (t+t')/2$ and $\tau = t-t'$.  Let
\be\label{AF}
A(\tau) = \int_{-\infty}^\infty d\bar t\,F\left(\bar t+\frac{\tau}{2},\bar t-\frac{\tau}{2}\right)\,.
\ee
so that $\hat F(-\xi,\xi) = \hat A(-\xi)$.  The presence of the $g$
functions in Eq.~(\ref{FV}) makes $F$ have compact support in $t$
and $t'$, so $A$ has compact support in $\tau$.

Suppose $F$ contains some term $f$ that is symmetrical in $t$ and
$t'$.  Let $a$ be the corresponding term in $A$ according to
Eq.~(\ref{AF}).  Then $a$ will be even in $\tau$, so $\hat a$ will
be even also.  If $a\in C^1$, then $\hat a\in L^2$, and we can perform
the integral of this term separately, giving an inverse Fourier
transform,
\be
\int_0^\infty\frac{d\xi}{\pi}\hat f(-\xi,\xi)
= \int_{-\infty}^\infty\frac{d\xi}{2\pi} \hat a(\xi) = a(0)\,.
\ee
In particular, if
\be
\lim_{t' \to t} f(t,t') = f(t)\,,
\ee
then
\be
\int_0^\infty\frac{d\xi}{\pi}\hat f(-\xi,\xi)
= \int_{-\infty}^\infty dt\,g(t)^2 f(t)\,,
\ee 
and if $f(t)=0$ there is no contribution.  

Terms arising from $H$ appear symmetrically in $\tilde H$.  At orders
$j>1$ they have at least 4 powers of $\tau$, so they vanish in the
coincidence limit even when differentiated twice by the operators of
$\Tsplit_{tt'}$.  Thus such terms make no contribution to
Eq.~(\ref{B}).

\subsection{Simplification of $\Tsplit_{tt'}$}
\label{sec:simplification}

We would like to write the operator $\Tsplit_{tt'}$ of Eq.~(\ref{eqn:tsplitV}) in terms of separate
derivatives on the center point $\bar x$ and the difference between the
points.  First we separate the derivatives in $\Tsplit_{tt'}$ into time and
space,
\be \label{Vspti}
\sum_{a=0}^3 \partial_a \partial_{a'}=\partial_t
\partial_{t'}+\nabla_x \cdot \nabla_{x'}\,.
\ee
We can write the spatial derivative with respect to\footnote{When a
  derivative is with respect to $t$ or $t'$ (or $x$ or $x'$), we mean
  to keep the other of these fixed, while when the derivative is with
  respect to $\bar t$ or $\tau$, we mean to keep the other of these
  fixed.  When the derivative is with respect to $\bar x$ we mean to
  keep $x-x'$ fixed.}  $\bar x$ in terms of the derivatives at the endpoints,
\be\label{Vbarxd}
\nabla_{\bar{x}}^2=\nabla_x^2+2\nabla_x\cdot \nabla_{x'}+\nabla_{x'}^2\,.
\ee
Then Eqs.~(\ref{eqn:tsplitV}), (\ref{Vspti}),(\ref{Vbarxd}) give
\bea
\Tsplit_{tt'}&=&\frac{1}{2}\left[\partial_t \partial_{t'}+\frac{1}{2}
  \left(\nabla_{\bar{x}}^2-\nabla_x^2-\nabla_{x'}^2\right)
+\frac12\left(V(x)+V(x')\right)\right]=\nonumber\\
&=&\frac{1}{4}\left[\nabla_{\bar{x}}^2+\Box_x-\partial_t^2
+\Box_{x'}-\partial_{t'}^2+2\partial_t\partial_{t'}+V(x)+V(x')\right]\,,
\eea
where $\Box_x$ and $\Box_{x'}$ denote the D'Alembertian operator with
respect to $x$ and $x'$.  Then using
\be\label{Vbartd}
\partial_\tau^2=\frac14\left[\partial_t^2-2\partial_t \partial_{t'}+\partial_{t'}^2\right]\,,
\ee
we can write
\be\label{TsplitEOM}
\Tsplit_{tt'} \tilde{H}=\frac{1}{4}\left[\left( \Box_x+V(x) \right) \tilde{H}
+\left( \Box_{x'}+V(x') \right)\tilde{H}+\nabla_{\bar{x}}^2 \tilde{H}\right]-\partial_\tau^2 \tilde{H}\,.
\ee
Consider the first term.
The function $H(x,x')$ obeys the equation of motion in $x$, and so
does $E(x,x')$.  Thus all of $\tilde H(x,x')$ is annihilated by
$\Box_x+V(x)$, except for $H(x',x)$,
\be \label{Veqmd}
\left( \Box_x+V(x) \right) \tilde{H}=\frac{1}{2}(\Box_x+V) H(x',x)\,.
\ee
The quantities $v_j(x,x')$ are symmetric, so the
only sources of asymmetry in $H$ are the functions $w_j$, which are
real but may not be symmetric, and the fact that the imaginary part of
$\sigma_+(x,x')$ is antisymmetric.  Thus
\be \label{Hasym}
H(x',x) = H(x,x')^* + \frac{1}{4\pi^2}\sum_j(w_j(x',x)-w_j(x,x'))\sigma^j(x,x')\,,
\ee
where $*$ denotes complex conjugation.  Since $\Box_x$ is real, if we use Eq.~(\ref{Hasym}) in Eq.~(\ref{Veqmd}), we have
$(\Box_x+V)H(x,x')^*=0$, and we ignore $V w_j$, because it is second
order in $V$, leaving
\be\label{VEofMH1}
\left( \Box_x+V(x) \right) \tilde{H}
= \frac{1}{4\pi^2}\Box_x\sum_j(w_j(x',x)-w_j(x,x'))\sigma^j(x,x')\,.
\ee
By the same argument,
\be\label{VEofMH2}
\left( \Box_{x'}+V(x') \right) \tilde{H}
= \frac{1}{4\pi^2}\Box_{x'}\sum_j(w_j(x,x')-w_j(x',x))\sigma^j(x,x')\,.
\ee
The first two terms in the brackets in Eq.~(\ref{TsplitEOM}) are the
sum of Eqs.~(\ref{VEofMH1}), (\ref{VEofMH2}).  This sum is smooth,
symmetric in $x$ and $x'$, and vanishes in the coincidence limit.
Thus according to the analysis of Sec.~\ref{sec:symmetry}, it makes no
contribution in the Fourier transform of Eq.~(\ref{B}), and for our
purposes we can take
\be \label{eqn:TVsplitfinal}
\Tsplit_{tt'} \tilde{H}=\left[\frac{1}{4}\nabla_{\bar{x}}^2-\partial_\tau^2
\right] \tilde{H}\,.
\ee

\section{Computation of $\tilde H$}
\label{sec:VH}

Examining Eq.~(\ref{T}) we see that is sufficient to compute $\tilde
H$ for purely temporal separation as a function of $t$, $t'$, and
$\bx$, the common spatial position of the points.  The function
$H(t,t')$ is a series of terms with decreasing degree of singularity
at coincidence: $\tau^{-2}$, $\ln\tau$, $\tau^2\ln\tau$, etc.  For the
first term in Eq.~(\ref{T}), terms in $H$ that have any positive
powers of $\tau$ will not contribute by the analysis of
Sec.~\ref{sec:symmetry}.  For the second term we need to keep terms in
$H$ up to order $\tau^2$, because the derivatives will reduce the
order by 2.

The symmetrical combination $H(t,t') + H(t',t)$, will lead to
something whose Fourier transform does not decline rapidly for
positive $\xi$, so that if this alone were put into
Eq.~(\ref{B}) the integral over $\xi$ would not converge.
But each term in $H(t,t')+H(t',t)$ will combine with a term coming
from $iE(x,x')$ to give something whose Fourier transform does decline
rapidly.

We will work order by order in $\tau$.

\subsection{General computation of $E$}

We will need the Green's functions for the background potential,
including only first order in $V$, so we write
\be
G=G^{(0)}+G^{(1)}+ \cdots.
\ee
The equation of motion is
\be \label{eqm}
(\Box+V(x)) G(x,x')=\delta^{(4)}(x-x')\,.
\ee
Using $\Box G^{(0)}(x,x')=\delta^{(4)}(x,x')$ and keeping only first-order
terms we have
\be
\Box G^{(1)}(x,x')=-V(x)G^{(0)}(x,x')\,,
\ee
so
\be
G^{(1)}(x,x')=-\int d^4 x'' G^{(0)}(x,x'') V(x'') G^{(0)}(x'',x')\,.
\ee
For $t>t''>t'$ we have for the retarded Green's function,
\be
G_R^{(0)}(x'',x')=\frac{1}{2\pi} \delta((t''-t')^2-|\bx''-\bx'|^2)=\frac{1}{4\pi} \frac{\delta(t''-t'-|\bx''-\bx'|)}{|\bx''-\bx'|}\,.
\ee
So we can write
\be
G_R^{(1)}(x,x')=- \frac{1}{8\pi^2} \int d^3 \bx'' \int dt''\delta((t-t'')^2-|\bx-\bx''|^2) \frac{\delta(t''-t'-|\bx''-\bx'|)}{|\bx''-\bx'|} V(t'',\bx'')\,.
\ee
Integrating over the second delta function we find
$t''=t'+|\bx''-\bx'|$.  Again considering purely temporal separation and
defining $\bzeta''=\bx''-\bx'$ and $\zeta'' = |\bzeta''|$, we find
\be
G_R^{(1)}(t,t')=-\frac{1}{8\pi^2}\int d\Omega \int d\zeta'' \zeta''^2  \frac{\delta(\tau^2-2\tau \zeta'')}{\zeta''} V(t'+\zeta'',\bx'+\zeta'' \hat{\Omega})\,,
\ee 
where $\int d\Omega$ denotes integration over solid angle, and
$\hat{\Omega}$ varies over all unit vectors.  We can integrate over
$\zeta''$ to get $\zeta''=\tau/2$ and
\be
G_R^{(1)}(t,t')= - \frac{1}{32\pi^2}\int d\Omega\,V(\bar{t},\bx'+\frac{\tau}{2}\hat{\Omega})\,.
\ee
If we define a 4-vector $\Omega=(0,\hat{\Omega})$ we can write
\be \label{G1V}
G_R^{(1)}(t,t')=-\frac{1}{32 \pi^2} \int d\Omega\,V(\bar{x}+\frac{\tau}{2} \Omega)\,.
\ee
The advanced Green's functions are the same with $t$ and $t'$ reversed.
Since $E$ is the advanced minus the retarded function, we have
\be\label{Egeneral}
E^{(1)}(t,t') = \frac{1}{32 \pi^2} \int d\Omega\,V(\bar{x}+\frac{|\tau|}{2} \Omega)\sgn\tau\,.
\ee

\subsection{Terms of order $\tau^{-2}$}

We now compute the various $H_j$, $\tilde H_j$, and $E_j$, starting
with terms that go as $\sigma^{-1}$ or $\tau^{-2}$.  These terms are
exactly what one would have for flat space without potential.
Equation~(\ref{hadamard}) gives
\be
H_{-1}(x,x')=\frac{1}{4\pi^2\sigma_+(x,x')}=-\frac{1}{4\pi^2(\tau_-^2-\zeta^2)}
\,,
\ee
where
\be
\bzeta=\bx-\bx'
\ee
and
\be
\zeta = |\bzeta|\,.
\ee

Similarly, the advanced minus retarded Green's function to this order is
\be
E_{-1}(x,x') = G_A(x,x') - G_R(x,x')
= \frac{\delta(\tau-\zeta) - \delta(\tau+\zeta)}{4\pi \zeta}\,,
\ee
so
\be
\tilde H_{-1}(t,t') = \lim_{\zeta \to 0} \frac{1}{8\pi^2}\left[
-\frac{1}{\tau_+^2-\zeta^2}-\frac{1}{\tau_-^2-\zeta^2}+i \pi
\frac{\delta(\tau+\zeta)-\delta(\tau-\zeta)}{\zeta} \right]\,,
\ee
where 
\be
F(\tau_+)=\lim_{\epsilon \to 0} F(\tau+i\epsilon)\,.
\ee
Taking the $\epsilon\to0$ limit in $\tau_+$ and $\tau_-$ gives the
formula
\be
-\frac{1}{\tau_+^2-\zeta^2}+\frac{1}{\tau_-^2-\zeta^2}=-i \pi \frac{\delta(\tau+\zeta)-\delta(\tau-\zeta)}{\zeta} 
\ee
so
\be\label{VH-1}
\tilde H_{-1}(t,t') =-\frac{1}{4\pi^2\tau_-^2} = H_{-1}(t,t')
\ee
as discussed in Ref~\cite{Fewster:2007rh}.

\subsection{Terms with no powers of $\tau$}

We can find the Hadamard coefficients from the recursion relations,
Eq.~(\ref{recursion1}), (\ref{recursion2}).  To find the zeroth order of the Hadamard series we
need only $v_0$.  For flat space, $\sigma_{,a}=-2\eta_{ab}(x''-x')^b$
and $\Box{\sigma}=-8$.  Putting these in Eq.~(\ref{recursion1}) we have
\be \label{v01V}
(x''-x')^a v_{0,a} + v_0 = \frac{V(x'')}{4}\,,
\ee
Now let $x'' = x' + \lambda(x-x')$ to integrate along the geodesic going
from $x'$ to $x$.  We observe that
\be
\frac{dv_0(x'',x')}{d\lambda} = (x-x')^a v_{0,a}(x'',x')\,.
\ee
So Eq.~(\ref{v01V}) gives
\be
\lambda\frac{ dv_0(x'',x')}{d\lambda} + v_0(x'',x') = \frac{V(x'')}{4}\,,
\ee
or 
\be
\frac{d(\lambda v_0(x'',x'))}{ d \lambda} = \frac{V(x'')}{4}\,,
\ee
from which we immediately find
\be \label{v0V}
v_0(x,x') = \int_0^1 d\lambda \frac{V(x' + \lambda(x-x'))}{4}\,.
\ee
Now we consider purely temporal separation so the background potential
is evaluated at $(t' + \lambda \tau, \bx)$.  We expand $V$ in a
Taylor series in $\tau$ around $0$ with $\bar t$ fixed,
\be\label{VTaylor}
V(t' + \lambda \tau)=V(\bar{t})+\tau (\lambda-\frac{1}{2})V_{,t} (\bar{t})+\frac{\tau^2}{2}(\lambda-\frac{1}{2})^2V_{,tt}(\bar{t})+\cdots.
\ee
We are calculating the zeroth order so we keep only the first term of
Eq.~(\ref{VTaylor}), and Eq.~(\ref{v0V}) gives
\be
v_0(t,t')=\frac{1}{4}V(\bar{t})+O(\tau^2)
\ee
and thus
\be
H_0(x,x')=\frac{1}{16\pi^2} V(\bar{x}) \ln{(-\tau_-^2)}\,,
\ee
and
\be\label{H02}
H_0(x,x')+H_0(x',x)=  \frac{1}{4\pi^2}V(\bar{x}) \ln{|\tau|}\,.
\ee

We can expand $V$ around $\bar{x}$,
\be 
V(\bar{x}+\frac{\tau}{2}\Omega)=V(\bar{x})+V^{(1)} (\bar{x}+\frac{\tau}{2} \Omega)\,,
\ee
where $V^{(1)}$ is the remainder of the Taylor series
\be\label{V(1)}
V^{(1)}(\bar{x}+\frac{\tau}{2} \Omega)=V(\bar{x}+\frac{\tau}{2} \Omega) - V(\bar{x})
= \int_0^{\tau/2} dr\,V_{,i}(\bar{x}+r\Omega) \Omega^i\,.
\ee
Then from Eq.~(\ref{Egeneral}),
\blea
E_0(x,x')&=&\frac{1}{8\pi} V(\bar{x})\sgn\tau\label{E0V}\\
R_0(x,x')&=&\frac{1}{32\pi^2}\int d\Omega\,V^{(1)} (\bar{x}+\frac{|\tau|}{2} \Omega)\sgn{\tau}\label{R0V}\,.
\elea
Using
\be\label{lnsgn}
2\ln{|\tau|}+\pi i \sgn{\tau}= \ln{(-\tau_-^2)}\,,
\ee
we combine Eqs.~(\ref{H02}), (\ref{E0V}) to find
\be\label{Ht0V}
\tilde H_0(t,t')=\frac{1}{16\pi^2} V(\bar{x})\ln{(-\tau_-^2)}\,.
\ee
Combining all terms through order $0$ gives
\be\label{Ht-10V}
\tilde H_{(0)}(t,t')=\tilde H_{-1}(t,t')+\tilde H_0(t,t')+\frac12i
R_0(t,t')\,.
\ee

\subsection{Terms of order $\tau^2$}

Now we compute the terms of order $\tau^2$ in $H$ and $E$.  First we
need $v_0$ at this order, so we use Eq.~(\ref{VTaylor}) in
Eq.~(\ref{v0V}).  The $V_{,t}$ term in Eq.~(\ref{VTaylor}) does not
contribute, because it is odd in $\lambda -1/2$, and the others give
\be\label{v02}
v_0(x,x')= \frac{1}{4}V(\bar{x})+\tau^2 \frac{1}{96} V_{,tt} (\bar{x})+\cdots\,.
\ee

Next we need to know $v_1$, but since $v_1$ is multiplied by $\tau^2$
in $H$, we need only the $\tau$-independent term $v_1(x,x)$.  From
Eq.~(\ref{recursion2}),
\be  \label{recv1}
(\Box+V(x))v_0(x,x')+2\eta^{ab}v_{1,a}(x,x')
\sigma_{,b}(x,x')+v_1(x,x')\Box_{x}\sigma(x,x') =0\,.
\ee
We neglect the $V(x)v_0$ term because it is second order in $V$.  At
$x=x'$, $\sigma_{,b} = 0$, so
\be\label{v1v0}
v_1(x,x)=\frac{1}{8}\lim_{x'\to x}\Box_x v_0(x,x')\,.
\ee
Using Eq.~(\ref{v0}) we find
\be
\Box_x v_0(x,x')=\frac{1}{4}\int_0^1 d\lambda \Box_x V(x' + \lambda(x-x'))=\frac{1}{4}\int_0^1 d\lambda\,\lambda^2 (\Box V)(x' + \lambda(x-x'))\,,
\ee
and Eq.~(\ref{v1v0}) gives
\be\label{v1V}
v_1(x,x) =  \frac{1}{96}\Box V(\bar{x})\,.
\ee

We also need to know $w_1$, but again only at coincidence.
Reference~\cite{Wald:1978pj} gives
\be\label{w1V}
w_1(x,x)=-\frac{3}{2} v_1(x,x)= -\frac{1}{64}\Box V(x)\,.
\ee
Combining the second term of Eq.~(\ref{v02}) with
Eqs.~(\ref{v1V}), (\ref{w1V}) gives
\be
H_1(t,t')=\frac{\tau^2 }{128\pi^2}
\left[\frac{1}{3} V_{,ii}(\bar{x})\ln{(-\tau_-^2)}
+\frac{1}{2}\Box V(\bar{x}) \right]\,.
\ee
Then $H_1(x',x)$ is given by symmetry, so
\be\label{H12}
H_1(x,x')+H_1(x',x)= \frac{\tau^2 }{64\pi^2} \left[\frac{2}{3} V_{,ii}(\bar{x}) \ln{|\tau|}+ \frac{1}{2}\Box V(\bar{x})\right]\,.
\ee

The calculation of $E_1$ is similar to that of $E_0$, but
now we have to include more terms in the Taylor expansion of $V$
around $\bar{x}$. So we expand
\be 
V(\bar{x}+\frac{\tau}{2}\Omega)=V(\bar{x})+\frac{1}{2}V_{,i}(\bar{x})\Omega^i\tau+\frac{1}{8}V_{,ij}(\bar{x})\Omega^i \Omega^j \tau^2+V^{(3)} (\bar{x}+\frac{\tau}{2} \Omega)\,,
\ee
where the remainder of the Taylor series $V^{(3)}$ is
\be\label{V3}
V^{(3)}(\bar{x}+\frac{\tau}{2} \Omega)=\frac12\int_0^{\tau/2} dr\,V_{,ijk}(\bar{x}+r\Omega)\left(\frac{\tau}{2}-r \right)^2\Omega^i \Omega^j \Omega^k dr\,.
\ee
Since $\int d\Omega\,\Omega^i = 0$ and
$\int d\Omega\,\Omega^i \Omega^j =(4\pi/3) \delta^{ij}$,
Eq.~(\ref{Egeneral}) gives
\blea
E_1(x,x')&=& \frac{1}{192 \pi} V_{,ii}(\bar{x})\tau^2\sgn\tau\label{E1V}\,,\\
R_1(x,x')&=& \frac{1}{32\pi^2} \int d\Omega\,V^{(3)}
(\bar{x}+\frac{|\tau|}{2} \Omega)\sgn{\tau}\label{R1V}\,.
\elea
Again using Eq.~(\ref{lnsgn}), we combine Eqs.~(\ref{H12}),(\ref{E1V}) to get
\be\label{Ht1V}
\tilde H_1(x,x')=\frac{\tau^2}{128\pi^2} \left[\frac{1}{3} \ln{(-\tau_-^2)} V_{,ii}+\frac{1}{2}\Box V(\bar{x}) \right]\,.
\ee
Combining all terms through order 1 gives
\be\label{Ht-11V}
\tilde H_{(1)}(t,t')=\tilde H_{-1}(t,t')+\tilde H_0(t,t')+\tilde H_1(t,t')+\frac12
iR_1(t,t')\,.
\ee

\section{The $\Tsplit_{tt'} \tilde H$}
\label{sec:TH}

Using Eqs.~(\ref{FV}), (\ref{T}), we need to compute
\be
\int_0^\infty \frac{d\xi}{\pi}\hat F(-\xi,\xi')\,,
\ee
where
\be
F(t,t') = g(t) g(t')\left[\frac{1}{4}\nabla_{\bar{x}}^2\tilde H_{(0)}(t,t')
-\partial_\tau^2  \tilde H_{(1)}(t,t')\right]\,.
\ee
Using
Eqs.~(\ref{VH-1}), (\ref{R0V}), (\ref{Ht0V}), (\ref{Ht-10V}), (\ref{R1V}, (\ref{Ht1V}), (\ref{Ht-11V})
we can write this
\be
F(t,t') = g(t) g(t')\sum_{i=1}^6 f_i(t,t')\,,
\ee
with
\blea
f_1&=&\frac{3}{2\pi^2\tau_-^4}\\\
f_2&=&\frac{1}{8\pi^2\tau_-^2}V(\bar{x})\\
f_3&=&\frac{1}{96\pi^2} V_{,ii}(\bar{x}) \ln{(-\tau_-^2)}\\
f_4&=&-\frac{1}{128\pi^2}\left[V_{,tt}(\bar{x})+V_{,ii}(\bar{x})\right]\\
f_5&=&\frac{1}{256\pi^2} \int d\Omega\,\nabla_{\bar{x}}^2 \left[
  V^{(1)} (\bar{x}+\frac{|\tau|}{2} \Omega)\right]  i\sgn{\tau}\label{f5V}  \\
f_6&=& -\frac{1}{64\pi^2} \int d\Omega \, \partial_{\tau}^2\left[ V^{(3)} (\bar{x}+\frac{|\tau|}{2} \Omega)  i\sgn{\tau} \right]\,.\label{f6V}
\elea

\section{The Fourier transform}
\label{sec:Fourier}

We want to calculate the quantum inequality bound $B$, given
by Eq.~(\ref{eqn:BV}).  We can write it
\be
B=\sum_{i=1}^8 B_i\,,
\ee
where
\blea
B_i&=&\int_0^\infty  \frac{d\xi}{\pi} \int_{-\infty}^\infty dt
\int_{-\infty}^\infty dt' g(t) g(t') f_i(t,t') e^{i\xi(t'-t)}\nonumber\\
&=&\int_0^\infty \frac{d\xi}{\pi} \int_{-\infty}^\infty d\tau
\int_{-\infty}^\infty d\bar{t}\,
g(\bar{t}-\frac{\tau}{2})g(\bar{t}+\frac{\tau}{2})f_i(\bar t,\tau)e^{-i\xi \tau}, i = 1\ldots6\\
\label{B7V}
B_7 &=& \int_{-\infty}^\infty dt\,g^2(t) Q(t)
= -\frac{1}{768\pi^2}\int_{-\infty}^\infty dt\,g^2(t)\Box V(t)\\
\label{B8V}B_8 &=& -\int_{-\infty}^\infty dt\,g^2(t)CV_{,ii} (t)\,,
\elea
using Eqs.~(\ref{Q}), (\ref{B}) and (\ref{w1V}).

\subsection{The singular terms}

For $i=1,2,3$, $f_i$ consists of a singular function of $\tau$ times a
function of $\bar t$ (or a constant).  So we will separate the
singular part by writing
\be
f_i(\bar t,\tau) = g_i(\bar t) s_i(\tau)\,.
\ee
Then we define
\be
G_i(\tau)=\int_{-\infty}^\infty d\bar{t}\, g_i(\bar{t}) g(\bar{t}-\frac{\tau}{2})g(\bar{t}+\frac{\tau}{2})\,,
\ee
so
\be\label{Bi1}
B_i=\int_0^\infty \frac{d\xi}{\pi} \int_{-\infty}^\infty d\tau\,G_i(\tau) s_i(\tau) e^{-i\xi\tau}\,.
\ee
This is a Fourier transform of a product, so we can write it as a
convolution.  The $G_i$ are all real, even functions, and thus
their Fourier transforms are also, and we have
\be
B_i=\frac{1}{2\pi^2}\int_0^{\infty} d \xi\int_{-\infty}^{\infty} d\zeta\,\hat{G_i}(\xi+\zeta)\hat{s_i}(\zeta)\,.
\ee
Now if we change the order of integrals we can perform another change of variables $\eta=\xi+\zeta$, so we have
\be \label{general}
B_i=\frac{1}{2\pi^2}\int_{-\infty}^{\infty} d\zeta\int_\zeta^{-\infty} d\eta\,\hat{G_i}(\eta)\hat{s_i}(\zeta)
=\frac{1}{2\pi^2}\int_{-\infty}^{\infty}  d\eta\hat{G_i}(\eta) h_i(\eta)\,,
\ee
where
\be\label{hi}
h_i(\eta)=\int_{-\infty}^{\eta}d\zeta\,\hat{s_i}(\zeta)\,.
\ee
The arguments of Ref.~\cite{Fewster:2007rh} show that the integrals
over $\xi$ in Eq.~(\ref{Bi1}) and $\eta$ in Eq.~(\ref{hi}) converge.

We now calculate the Fourier transforms in turn, starting with $B_1$.  We have
\blea
g_1(\bar{t})&=&\frac{3}{2\pi^2} \\
s_1(\tau)&=&\frac{1}{\tau_-^4}\,.
\elea
The Fourier transform of $s_1$ is \cite{Gelfand:functions}
\be
\hat{s_1}(\zeta)=\frac{\pi}{3} \zeta^3 \Theta(\zeta)\,,
\ee
so
\be
h_1(\eta)=\int_0^{\eta} d\zeta\,\frac{\pi}{3}\zeta^3\Theta(\eta) =\frac{\pi}{12} \eta^4\Theta(\eta)\,.
\ee
From Eq.~(\ref{general}) we have
\be
B_1=\frac{1}{24\pi} \int_{0}^{\infty} d\eta\,\hat{G_1}(\eta) \eta^4\,.
\ee
Using $\widehat{f'}(\xi)=i \xi \hat{f}(\xi)$, we get
\be
B_1=\frac{1}{24\pi} \int_0^{\infty} d\eta\,\widehat{G_1''''}(\eta)\,.
\ee
The function $G_1$ is even, so its Fourier transform is also
even and we can extend the integral
\be
B_1=\frac{1}{48\pi} \int_{-\infty}^{\infty} d\eta\,\widehat{G_1''''}(\eta)=\frac{1}{24} G_1''''(0)\,.
\ee
For $G_1$ we have
\be
G_1(\tau)=\frac{3}{2\pi^2} \int d\bar{t}\, g(\bar{t}-\frac{\tau}{2})
g(\bar{t}+\frac{\tau}{2})\,,
\ee
and taking the derivatives and integrating by parts gives
\be \label{B1}
B_1=\frac{1}{16\pi^2} \int_{-\infty}^{\infty} d\bar{t}\, g''(\bar{t})^2\,,
\ee
reproducing a result of Ref.~\cite{Fewster:2007rh}.

For $B_2$ we have
\blea
g_2(\bar{t})&=&\frac{1}{8\pi^2}V(\bar{t}) \\
s_2(\tau)&=&\frac{1}{\tau_-^2}\,.
\elea
This calculation is the same as before except the Fourier transform of $s_2$ is
\cite{Gelfand:functions}
\be
\hat{s_2}(\zeta)=-2\pi \zeta \Theta(\zeta)\,.
\ee
So we have
\be
B_2=\frac{1}{2} G_2''(0)\,,
\ee
where
\be
G_2(\tau)=\frac{1}{8\pi^2} \int_{-\infty}^\infty d\bar{t}\, V(\bar{t}) g(\bar{t}-\frac{\tau}{2}) g(\bar{t}+\frac{\tau}{2})\,.
\ee
After taking the derivatives
\be \label{B2}
B_2=\frac{1}{32\pi^2} \int_{-\infty}^\infty d\bar{t}\, V(\bar{t})
[g(\bar{t}) g''(\bar{t}) - g'(\bar{t})^2]\,.
\ee

For $B_3$ we have
\bea
s_3(\tau)&=&\ln(-\tau_-^2)\,.
\eea
In the appendix, we find the Fourier transform of $s_3$ as a distribution,
\be\label{s3hat}
\hat s_3[f] = 4\pi\int_0^\infty dk\,f'(k) \ln|k| -4\pi \gamma f(0)\,.
\ee
From Eq.~(\ref{hi}), we can write
\be
h_3(\eta)=\int_{-\infty}^\infty d\zeta\,\hat{s_3}(\zeta)\Theta(\eta-\zeta)\,,
\ee
which is given by Eq.~(\ref{s3hat}) with $f(\zeta)
=\Theta(\eta-\zeta)$, so
\be
h_3(\eta)=-4\pi\int_0^\infty d\zeta\,\delta(\eta-\zeta) \ln|\zeta| - 4\pi\gamma \Theta(\eta)=
-4\pi\Theta(\eta) ( \ln\eta + \gamma )\,.
\ee
Then Eq.~(\ref{general}) gives
\be
B_3= -\frac{2}{\pi}\int_0^\infty d\eta\,\hat{G_3}(\eta)
 \left( \ln\eta + \gamma \right)
= -\frac{1}{\pi}\int_{-\infty}^\infty d\eta\,\hat{G_3}(\eta)
( \ln|\eta| + \gamma )\,,
\ee
since $G_3$ is even.  The integral is just the distribution $w$ of
Eq.~(\ref{distw}) applied to $\hat G_3$, which is by definition $\hat
w[G_3]$, so Eq.~(\ref{distwhat}) gives
\be
B_3=-\int_{-\infty}^{\infty} d\tau\,G_3'(\tau)\ln{|\tau|}\sgn{\tau}\,,
\ee
with 
\be
G_3(\tau)=\frac{1}{96\pi^2}\int_{-\infty}^\infty dt
V_{,ii}(\bar{t}) g(\bar{t}-\frac{\tau}{2})g(\bar{t}+\frac{\tau}{2})\,,
\ee
so
\be\label{B3}
B_3 = -\frac{1}{96\pi^2}\int_{-\infty}^{\infty} d\tau\,\ln{|\tau|}\sgn{\tau}
\int_{-\infty}^{\infty} d\bar t\,V_{,ii}(\bar{t})
g(\bar{t}-\frac{\tau}{2})g'(\bar{t}+\frac{\tau}{2})\,.
\ee

\subsection{The non-singular terms}

For $i=4,5,6$, $f_i$ is not singular at $\tau=0$.  We include everything in
\be\label{Fi}
F_i(\tau)=\int_{-\infty}^\infty d\bar{t}\, f_i(\tau,\bar{t}) g(\bar{t}-\frac{\tau}{2})g(\bar{t}+\frac{\tau}{2})\,,
\ee  
so
\be
B_i=\int_0^\infty \frac{d\xi}{\pi}\int_{-\infty}^\infty d\tau
F_i(\tau) e^{-i\xi\tau}= \int_0^\infty \frac{d\xi}{\pi}
\hat{F_i}(-\xi)=\frac{1}{\pi}
\int_{-\infty}^\infty d\xi\,\Theta(\xi) \hat{F_i}(-\xi)\,.
\ee
The integral is the distribution $\Theta$ applied to $\hat
F_i(-\xi))$, which is the Fourier transform of $\Theta$ applied to
$F_i(-\tau)$.  The Fourier transform of the $\Theta$ function acts on
a function $f$ as \cite{Gelfand:functions}
\be \label{FTheta}
\Theta[f] = i P \int_{-\infty}^\infty d\tau  
\left(\frac{1}{\tau} f(\tau) \right) + \pi f(0)\,,
\ee
where $P$ denotes principal value, so
\be\label{B46}
B_i=-\frac{i}{\pi} P \int_{-\infty}^\infty d\tau  \left(
\frac{1}{\tau} F_i(\tau) \right) + F_i(0)\,.
\ee

The first of the non-singular terms is a constant: $f_4$ does not
depend on $\tau$.  Thus $F_4$ is even in $\tau$, and only the second term
of Eq.~(\ref{B46}) contributes, giving
\be\label{B4}
B_4=F_4(0)=-\frac{1}{128\pi^2} \int_{-\infty}^{\infty}d\bar{t}\,
g(\bar{t})^2 \left[V_{,tt}(\bar{t}) + V_{,ii}(\bar{t}) \right]\,.
\ee
The functions $f_5$ and $f_6$ are odd in $\tau$ and vanish as $\tau\to
0$, so in these cases only the first term in Eq.~(\ref{B46})
contributes and the principal value symbol is not needed.
Equations~(\ref{f5V},\ref{Fi},\ref{B46}) give
\be\label{B5}
B_5=\frac{1}{256\pi^3}  \int_{-\infty}^\infty d\tau \frac{1}{\tau}
\int_{-\infty}^\infty d\bar{t}\,g(\bar{t}-\frac{\tau}{2})g(\bar{t}+\frac{\tau}{2}) \int d\Omega\,\nabla_{\bar{x}}^2 V^{(1)} (\bar{x}+\frac{|\tau|}{2} \Omega) \sgn{\tau}
\ee
and Eqs.~(\ref{f6V}), (\ref{Fi}) and (\ref{B46}) give
\be
B_6=-\frac{1}{64\pi^3}  \int_{-\infty}^\infty d\tau \frac{1}{\tau} \int_{-\infty}^\infty d\bar{t}\,g(\bar{t}-\frac{\tau}{2})g(\bar{t}+\frac{\tau}{2}) \int d\Omega\,\partial_{\tau}^2\left[ V^{(3)}(\bar{x}+\frac{|\tau|}{2} \Omega) \sgn{\tau} \right]\,.
\ee
Here we can integrate by parts twice, giving
\be \label{B6}
B_6=-\frac{1}{64\pi^3}  \int_{-\infty}^\infty d\tau
\int_{-\infty}^\infty d\bar t\,
\partial_\tau^2 \left[\frac{1}{\tau} g(\bar{t}-\frac{\tau}{2})g(\bar{t}+\frac{\tau}{2})  \right]
\int d\Omega\,V^{(3)}(\bar{x}+\frac{|\tau|}{2} \Omega) \sgn{\tau}\,.
\ee
From Eqs.~(\ref{V(1)}), (\ref{V3}), $V^{(1)}$ goes as $\tau$ and $V^{(3)}$
as $\tau^3$ for small $\tau$, so the $\tau$ integrals converge.

\section{The Quantum Inequality}
\label{sec:QIV}

Now we can collect all the terms of $B$ from
Eqs.~(\ref{B7V}), (\ref{B8V}), (\ref{B1}), (\ref{B2}), (\ref{B3}), (\ref{B4}), (\ref{B5}) and (\ref{B6}).
Since $B_7$ is made of the same quantities as $B_4$, we
merge these together.  We find
\be\label{BforV}
B = \frac{1}{16\pi^2}\left[ I_1
+\frac{1}{2} I^V_2
-\frac{1}{6} I^V_3
-\frac{1}{8} I^V_4
+\frac{1}{16\pi} I^V_5
-\frac{1}{4\pi} I^V_6\right]
-I^V_7\,,
\ee
where
\bml\label{IV}\bea
I_1&=&\int_{-\infty}^{\infty} dt\,g''(t)^2 \\
I^V_2&=&
\int_{-\infty}^\infty d\bar{t}\, V(\bar{t})
[g(\bar{t}) g''(\bar{t}) - g'(\bar{t})^2]\\
\label{I3V}I^V_3  &=&\int_{-\infty}^{\infty} d\tau \ln{|\tau|}\sgn{\tau}
\int_{-\infty}^{\infty} d\bar t\,V_{,ii}(\bar{t})
g(\bar{t}-\frac{\tau}{2})g'(\bar{t}+\frac{\tau}{2})\\
\label{I4V}I^V_4 &=& \int_{-\infty}^{\infty}d\bar{t}\,g(\bar{t})^2 
\left[\frac{7}{6}V_{,tt} (\bar{t})+\frac{5}{6}
 V_{,ii}(\bar{t}) \right]
\\
I^V_5 &=&\int_{-\infty}^{\infty} d\tau \frac{1}{\tau}\int_{-\infty}^\infty d\bar{t}\,
g(\bar{t}-\frac{\tau}{2})g(\bar{t}+\frac{\tau}{2})
\int d\Omega \nabla_{\bar{x}}^2 V^{(1)} (\bar{x}+\frac{|\tau|}{2} \Omega)
\sgn\tau \qquad
\\
\label{I6V}I^V_6&=&\int_{-\infty}^{\infty} d\tau\int_{-\infty}^\infty d\bar t\,
\partial_\tau^2
\left[\frac{1}{\tau}g(\bar{t}-\frac{\tau}{2})g(\bar{t}+\frac{\tau}{2}) \right]
\int d\Omega\,V^{(3)}(\bar{x}+\frac{|\tau|}{2} \Omega)
\sgn{\tau}\qquad \\
\label{I7V}I^V_7 &=&C\int_{-\infty}^{\infty}d\bar{t}\, g(\bar{t})^2 V_{,ii}(\bar{t}) \,.
\elea
In the case with no potential, only $I_1$ remains, reproducing a
result of Fewster and Eveson \cite{Fewster:1998pu}.

In Eq.~(\ref{I3V}), $\ln|\tau|$ really means $\ln(|\tau|/l)$, where $l$ is
the arbitrary length discussed in Ch.~\ref{ch:QI}.  The choice of
a different length changes Eqs.~(\ref{I3V}) and (\ref{I4V}) in compensating
ways so that $B$ is unchanged.

Equations~(\ref{QI},\ref{BforV},\ref{IV}) give a quantum inequality
useful when the potential $V$ is known and so the integrals in
Eqs.~(\ref{IV}) can be done.  If we only know that $V$ and its
derivatives are restricted by the bounds of Eq.~(\ref{Vmax}), then we
can restrict the magnitude of each term of Eq.~(\ref{BforV}) and add
those magnitudes.  We start with
\bea
|I^V_2| \le \int_{-\infty}^\infty d\bar{t} |V(\bar{t})|
|g(\bar{t}) g''(\bar{t}) - g'(\bar{t})^2 |
\le \Vmax \int_{-\infty}^\infty d\bar{t}\left[g(\bar{t})|g''(\bar{t})|
+  g'(\bar{t})^2\right]  \,.
\eea
The cases of $I^V_3$, $I^V_4$, and $I^V_7$ are similar.  For $I^V_5$ and $I^V_6$, it is
useful to take explicit forms for the Taylor series remainders.  From
Eq.~(\ref{V(1)}), we see that
\be
\left|\int d\Omega\,\nabla_{\bar{x}}^2 V^{(1)}
(\bar{x}+\frac{|\tau|}{2} \Omega)\right|
\le \frac{|\tau|}{2}\int d\Omega|\nabla^2 V_{,i}||\Omega^i|
\le \frac{3|\tau|}{2} \Vmax'''\sum_i\int d\Omega|\Omega^i|
= 9\pi|\tau| \Vmax'''\,.
\ee
Similarly from Eq.~(\ref{V3}) we have
\bea
\left|\int d\Omega\,V^{(3)}(\bar{x}+\frac{|\tau|}{2} \Omega)\right|
&\le&\frac{|\tau|^3}{48}\int d\Omega |V_{,ijk}||\Omega^i \Omega^j \Omega^k|\\
&\le& \frac{|\tau|^3}{48}\Vmax'''\sum_{ijk}
\int d\Omega|\Omega^i \Omega^j \Omega^k|\nonumber
=\frac{2\pi+1}{8}|\tau|^3\Vmax'''\,,
\eea
We can then perform the derivatives in Eq.~(\ref{I6V}) and take the
absolute value of each resulting term separately.

We define
\bml\label{J17V}\bea
J_2&=&\int_{-\infty}^\infty dt\left[g(t)|g''(t)|+g'(t)^2\right]\\
J_3&=&\int_{-\infty}^\infty dt \int_{-\infty}^\infty dt' |g'(t')|g(t) |\!\ln{|t'-t|}| \\ J_4&=&\int_{-\infty}^\infty dt\,g(t)^2\\
J_5&=&\int_{-\infty}^\infty dt \int_{-\infty}^\infty dt' g(t)g(t') \\
J_6&=&\int_{-\infty}^\infty dt \int_{-\infty}^\infty dt' |g'(t')|g(t)|t'-t|\\
J_7&=&\int_{-\infty}^\infty dt \int_{-\infty}^\infty dt' 
\left [g(t)|g''(t')| +g'(t)g'(t')\right] (t'-t)^2
\elea
and find
\blea
|I^V_2| &\le & \Vmax J_2\\
|I^V_3| &\le &3\Vmax'' J_3\\
|I^V_4| &\le & \frac{11}{3}\Vmax''J_4\\
|I^V_5|&\le &9\pi\Vmax''' J_5\\
|I^V_6|&\le &\frac{2\pi+1}{16}\Vmax'''\left(4J_5+4J_6+J_7\right)\\
|I^V_7|&\le & 3|C|\Vmax''J_4\,.
\elea

Thus we have
\bea
\label{finalV}
\int_{\mathbb{R}} dt \,g(t)^2\langle T^{ren}_{tt}\rangle_{\omega}
(t,0) \geq- \frac{1}{16\pi^2} &&\bigg\{I_1+\frac12\Vmax J_2
+\Vmax''\left[\frac{1}{2} J_3+\left(\frac{11}{24}+48\pi^2|C|\right) J_4\right]
\nonumber\\
&&+\Vmax''' \left[\frac{11\pi+1}{16\pi}J_5
+\frac{2\pi+1}{64\pi}(4J_6+J_7)\right] \bigg\}\qquad \,.
\eea

\subsection{An example for a specific sampling function}
\label{sec:sampling}

An example of the quantum inequality with a specific sampling function
$g$ is the following. Consider a Gaussian sampling function
\be
g(t)=e^{-t^2/t_0^2}\,,
\ee
where $t_0$ is a positive number with the dimensions of $t$.
Then the integrals of Eqs.~(\ref{J17V}), calculated numerically, become
\bea
J_1= 3.75 t_0^{-3}
&& J_2= 3.15 t_0^{-1} \nonumber\\
J_3= 2.70 t_0
&& J_4=1.25 t_0 \\
J_5= 3.14 t_0^2
&& J_6= 3.57 t_0^2\nonumber\\
J_7= 3.58 t_0^2\,,\nonumber
\eea
so the right hand side of Eq.~(\ref{finalV}) becomes
\be
-\frac{1}{16 \pi^2 t_0^3} \left\{ 3.75+3.15\Vmax t_0^2+(1.63+591.25 |C|) \Vmax'' t_0^4+ 2.86 \Vmax''' t_0^5 \right\}\,.
\ee

\section{Discussion of the result}

In this chapter we have demonstrated a quantum inequality for a flat
spacetime with a background potential, considered as a first-order
correction, using a general inequality derived by Fewster and Smith, which we presented in Ch.~\ref{ch:QI} . We calculated the necessary terms from the
Hadamard series and the antisymmetric part of the two-point function
to get $\tilde{H}$. Next we Fourier transformed the terms, which are,
as expected, free of divergences, to derive a bound for a given
background potential.  We then calculated the maximum values of these
terms to give a bound that applies to any potential whose value and
first three derivatives are bounded.

To show the meaning of this result, in the last section we presented
an example for a specific sampling function. By studying the result we
can see the meaning of the right hand side of our quantum
inequality. The first term of the bound goes as $t_0^{-3}$, where
$t_0$ is the sampling time, and agrees with the quantum inequality
with no potential \cite{Fewster:1998pu}. The rest of the terms show the
effects of the potential to first order.  These corrections will
be small, provided that
\blea
\Vmax t_0^2 &\ll& 1\label{Vsmall}\\
\Vmax'' t_0^4&\ll& 1\label{V2small}\\\
\Vmax''' t_0^5&\ll& 1\label{V3small}\,.
\elea

Equation~(\ref{Vsmall}) says that the potential is small when its
effect over the distance $t_0$ is considered.  Given
Eq.~(\ref{Vsmall}), Eqs.~(\ref{V2small}) and (\ref{V3small}) say,
essentially, that the distance over which $V$ varies is large compared
to $t_0$, so that each additional derivative introduces a factor less
than $t_0^{-1}$. Unlike the flat spacetime case the bound does not go to zero when the sampling time $t_0 \to \infty$ so we cannot obtain the Averaged Weak Energy Condition (AWEC).

Finally, it is interesting to note the relation of this result to
the case of a spacetime with bounded curvature. Since the Hadamard
coefficients in that case are components of the Riemann tensor and its
derivatives, we expect that the bound will be the flat space term plus
correction terms that depend on the maximum values of the curvature
and its derivatives, just as in our case they depend on the the
potential and its derivatives. We will demonstrate that this hypothesis is true in the next chapter.

\chapter{Quantum Inequality in spacetimes with small curvature}
\label{ch:curvature}

In this chapter we present a derivation of a timelike-projected quantum inequality in spacetimes with curveture. First we consider a massless, minimally-coupled scalar field with the usual
classical stress-energy tensor,
\be
T_{ab}=\nabla_a \Phi \nabla_b \Phi-\frac{1}{2} g_{ab} g^{cd} \nabla_c
\Phi \nabla_d \Phi \,.
\ee
Let $\gamma$ be any timelike geodesic parametrized by proper time $t$,
and let $g(t)$ be any any smooth, positive, compactly-supported
sampling function. 

Let's construct Fermi normal coordinates \cite{Manasse:1963zz} in
the usual way: We let the vector $e_0(t)$ be the unit tangent to the
geodesic $\gamma$, and construct a tetrad by choosing arbitrary
normalized vectors $e_i(0), i=1,2,3$, orthogonal to $e_0(0)$ and to
each other, and define $\{e_i(t)\}$ by parallel transport along
$\gamma$.  The point with coordinates $(x^0,x^1,x^2,x^3)$ is found by
traveling unit distance along the geodesic given by $x^i e_i(0)$
from the point $\gamma(0)$.

We work only in first order in the curvature and its derivatives, but
don't otherwise assume that it is small.  We assume that the
components of the Ricci tensor in any Fermi coordinate system, and
their derivatives, are bounded,
\bml\label{Rmax}
\be 
|R_{ab}| \leq \Rmax \qquad |R_{ab,cd}| \leq \Rmax'' \qquad |R_{ab,cde}| \leq \Rmax'''\,.
\ee
These lead to bounds on the Ricci scalar and its derivatives,
\be \label{Rsmax}
|R| \leq 4\Rmax \qquad |R_{,cd}| \leq 4\Rmax'' \qquad |R_{,cde}| \leq
4\Rmax'''\,,
\ee
\eml
since we are working in four dimensions.

Eqs.~(\ref{Rmax}) are intended as universal bounds which hold without
regard to the specific choice of Fermi coordinate system above.  We
will not need a bound on the first derivative.  The reason that we
bound the Ricci tensor and not the Riemann tensor is that, as we will
prove, the additional terms of the quantum inequality do not depend on
any other components of the Riemann tensor.

Thus we can write  Eq.~(\ref{qinequality}) in our case as
\be\label{qinequality2}
\int_{-\infty}^\infty d\tau\,g(t)^2 \langle \Tren_{tt} \rangle (t,0)
\geq -B\,,
\ee 
where
\be\label{B}
B = \int_0^\infty\frac{d\xi}{\pi}\hat F(-\xi,\xi)
+\int_{-\infty}^\infty dt\,g^2(t) \left(Q-2aR_{,ii}-\frac{b}{2}(R_{tt,tt}+R_{ii,tt}-3R_{tt,ii}+R_{ii,jj})\right)\,,
\ee
\be\label{F}
F(t,t') = g(t)g(t')\Tsplit \tilde H_{(1)}((t,0),(t',0))\,,
\ee
and $\hat F$ denotes the Fourier transform in both arguments according to
Eq.~(\ref{Fourier}).

\section{Simplification of $\Tsplit$}
\label{sec:tsplit}

The $\Tsplit_{tt'}$ operator of Eq.~(\ref{eqn:tsplit}) for a massless field can be written
\be \label{Rspti}
\Tsplit_{tt'}=\frac12\left[\partial_t \partial_{t'}+\sum_{i=1}^3\partial_i \partial_{i'}\right]\,.
\ee
To simplify it, we will define the following operator,
\be\label{Rbarxd}
\nabla_{\bar{x}}^2=\nabla_x^2+2\sum_{i=1}^3\partial_i\partial_{i'}+\nabla_{x'}^2\,,
\ee
which in flat space would be the derivative with respect to the center point.
Then Eqs.~(\ref{Rspti}) and (\ref{Rbarxd}) give
\bea
\Tsplit_{tt'}&=&\frac{1}{2}\left[\partial_t \partial_{t'}+\frac{1}{2}
  \left(\nabla_{\bar{x}}^2-\nabla_x^2-\nabla_{x'}^2\right)
\right]\nonumber\\
&=&\frac{1}{4}\left[\nabla_{\bar{x}}^2+\Box_x-\partial_t^2
+\Box_{x'}-\partial_{t'}^2+2\partial_t\partial_{t'}\right]\,,
\eea
where $\Box_x$ and $\Box_{x'}$ denote the D'Alembertian operator with
respect to $x$ and $x'$.  Because we are using Fermi coordinates and
are on the generating geodesic, the D'Alembertian and Laplacian
operators have the same form with respect to Fermi coordinates as they
do in flat space.  Then using
\be\label{Rbartd}
\partial_\tau^2=\frac14\left[\partial_t^2-2\partial_t \partial_{t'}+\partial_{t'}^2\right]\,,
\ee
we can write
\be\label{tsplit1}
\Tsplit_{tt'} \tilde{H}=\frac{1}{4}\left[ \Box_x \tilde{H}
+ \Box_{x'} \tilde{H}+\nabla_{\bar{x}}^2 \tilde{H} \right]-\partial_\tau^2 \tilde{H}\,.
\ee
Consider the first term.  The function $H(x,x')$ obeys the equation of
motion\footnote{In general the sums in Eq.~(\ref{hadamard}) do not
  converge and we should work only to some finite order in $\sigma$.
  In that case $H(x,x')$ obeys the equation of
  motion to that order.}
in $x$ and so does $E(x,x')$.  Thus
\be \label{Reqmd}
\Box_x \tilde{H}=\frac{1}{2}\Box_x H(x',x)\,.
\ee
As we discussed in Sec.~\ref{sec:simplification} we have
\be
H(x',x) = H(x,x')^* + \frac{1}{4\pi^2}\sum_j(w_j(x',x)-w_j(x,x'))\sigma^j(x,x')\,,
\ee
 Since $\Box_x$ is real,
$\Box_xH(x,x')^*= 0$, and we have
\be\label{EofMH1}
\Box_x  \tilde{H}
= \frac{1}{8\pi^2}\Box_x\sum_j(w_j(x',x)-w_j(x,x'))\sigma^j(x,x')\,.
\ee
In the coincidence limit Eq.~(\ref{EofMH1}) vanishes.  There is no
$j=0$ term because $w_0 = 0$.  In the $j = 1$ term, we have $\sigma$,
which vanishes at coincidence unless both derivatives of the $\Box$
are applied to it, in which case $w_1$ cancel each other, and for $j >
1$, even $\Box_x\sigma^j$ vanishes.

The second term in brackets in Eq.~(\ref{tsplit1}) gives
\be\label{EofMH2}
\Box_{x'} \tilde{H}
= \frac{1}{8\pi^2}\Box_{x'}\sum_j(w_j(x,x')-w_j(x',x))\sigma^j(x,x')\,.
\ee
Adding together Eqs.~(\ref{EofMH1}) and (\ref{EofMH2}), we get
something which is smooth, symmetric in $x$ and $x'$, and vanishes in
the coincidence limit.  Following the analysis of Sec.~\ref{sec:simplification}, such a term makes no contribution to
Eq.~(\ref{B}) and for our purposes we can take
\be \label{T}
\Tsplit \tilde{H}=\left[\frac{1}{4}\nabla_{\bar{x}}^2-\partial_\tau^2
\right] \tilde{H}\,.
\ee

\section{General computation of $E$}
\label{sec:iE}

The function $E$ is the advanced minus the retarded Green's function,
\be \label{theE}
E(x,x')=G_A(x,x')-G_R(x,x')\,,
\ee
and $iE$ is the imaginary, antisymmetric part of the two-point
function.  The Green's functions satisfy
\be \label{green}
\Box G(x,x')=\frac{\delta^{(4)}(x-x')}{\sqrt{-g}} \,.
\ee
Following Poisson, et al. \cite{Poisson:2011nh} and adjusting for different
sign and normalization conventions,
\be
G(x,x')=\frac{1}{4\pi} \left(2U(x,x')\delta(\sigma)+V(x,x')\Theta(-\sigma)\right)\,,
\ee
where $U(x,x') = \Delta^{1/2}(x,x')$ and $V(x,x')$ are smooth
biscalars.

For points $y$ null separated from $x'$, $V$ is called $\check V$
\cite{Poisson:2011nh} and satisfies
\be \label{vcheck}
\check V_{,a}\sigma^{,a}+\left[\frac{1}{2}\Box\sigma+2\right]\check V=-\Box U \,,
\ee
with all derivatives with respect to $y$.  Now $\check V$ is first
order in the curvature, so we will do the rest of the calculation as
though we were in flat space.  Under this approximation, we will
neglect coefficients which depend on the curvature, and also evaluate
curvature components at locations that would be relevant if we were in
flat space.  The distance between these locations and the proper
locations is first order in the curvature, so the overall inaccuracy
will always be second order in the curvature and its derivatives.

Thus we use $\sigma^{,a}=-2(y-x')^a$ and
$\Box\sigma=-8$ in Eq.~(\ref{vcheck}) to get
\be \label{vcheck2}
(y-x')^a \check V_{,a}(y)+\check V(y)=\frac{1}{2} \Box U(y) \,.
\ee
Now suppose we want to compute $\check V$ at some point $x''$.  We
need to integrate along the geodesic going from $x'$ to $x''$.  So let
$y=x'+\lambda(x''-x')$ and observe that
\be
\frac{d(\lambda \check V(y))}{d \lambda}
= \lambda\frac{ d \check V(y)}{d\lambda} + \check V(y)
= \lambda(x''-x')^a \check V_{,a} + \check V(y)
= (y-x')^a \check V_{,a} + \check V(y)
= \frac{1}{2} \Box U(y)\,,
\ee
so
\be\label{vcheckfinal}
\check V(x'',x') = \frac{1}{2} \int_0^1 d\lambda \Box U(y)\,.
\ee

The function $V$ obeys \cite{Poisson:2011nh}
\be
\Box_x V(x,x')=0 \,.
\ee
Let us consider points $x$ and $x'$ on the geodesic $\gamma$, which in
the flat-space approximation means they are separated only in time,
let $\bar x = (x+x')/2$, and establish a spherical coordinate
system $(r,\theta,\phi)$ with origin at the common spatial position of
$x$ and $x'$.
Now $V(x,x')$ can be found in terms of $V$ and its derivatives
evaluated at the time $\bar t$ (the time component of $\bar x$) using
Kirchhoff's formula,
\be \label{Vu1}
V(x,x')=\frac{1}{4\pi} \int d\Omega \left[\check
  V(x'',x')+\frac{\tau}2\frac{\partial}{\partial r}\check
  V(x'',x')
+\frac{\tau}2\frac{\partial}{\partial t}\check V(x'',x') \right] \,,
\ee
where the derivatives act on the first argument of $\check V$,
$\int d\Omega$ means to integrate over all spatial unit vectors
$\Hat\Omega$, and we now set
\be
x'' = \bar{x}+(\tau/2)\Omega
\ee
with the 4-vector $\Omega$ given by $\hat\Omega$ with zero time
component.

Now define null coordinates $u=t+r$ and $v=t-r$.  Then $x''$ has
$u=\tau$, $v = 0$.  The derivative $\partial/\partial u$ can be
written $(\partial/\partial t + \partial/\partial r)/2$ and so
\be \label{Vu}
V(x,x')= \frac{1}{4\pi} \int d\Omega \, \frac{d}{du}\left[
u \check V((u/2)\Omega_1,x')\right]_{u=\tau} \,,
\ee
where $\Omega_1$ is $\hat\Omega$ with unit time component.
From Eq.~(\ref{vcheckfinal}), 
\be
u \check V\big(\frac{u}2\Omega,x'\big)
= \frac{1}{2} \int_0^u du'(\Box U)((u'/2)\Omega_1,x')\,,
\ee
with the D'Alembertian applied to the first argument, and so
\be\label{Vint}
V(x,x')=\frac{1}{8\pi} \int d\Omega\Box_{x''} U(x'',x') \,.
\ee

We are only interested in the first order of curvature, so we can
expand U, which is just the square root of the Van Vleck determinant,
to first order.  From Ref.~\cite{Visser:1992pz},
\be \label{deltaexp}
\Delta^{1/2}(x, x')=1-\frac{1}{2}
\int_0^1 ds (1-s)s R_{ab}(sx+(1-s)x')(x-x')^a (x-x')^b+O(R^2) \,,
\ee
so in the case at hand we can use
\be
U(x'',x') = \Delta^{1/2}(x'',x')=1-\frac{1}{2} \int_0^1 ds (1-s)s
R_{ab}(y) X^a X^b \,
\ee 
where $y=sx'' = (su'',sv'',\theta'', \phi'')$ is a point between 0 and
$x''$, and the tangent vector $X=dy/ds$.  We are interested in
$\Box_{x''} U(x'',0)$.  To bring the $\Box$ inside the integral, we
define $Y=sX = (su'',sv'',0,0)$, and
\be
\Box U(x'', 0) = -\frac{1}{2} \int_0^1 ds (1-s)s \Box_{x''}[R_{ab}(y) X^a X^b]
 = -\frac{1}{2} \int_0^1 ds (1-s)s \Box_{y}[R_{ab}(y) Y^a Y^b] \,.
\ee
For the rest of this section, all occurrences of $u$, $v$, $\theta$,
$\phi$, and derivatives with respect to these variables will refer to
these components of $y$ or $Y$.

Now we expand the D'Alembertian in Eq.~(\ref{Vint}), in terms of an
angular part,
\be
\nabla^2_\Omega = \frac{4}{(v-u)^2 \sin{\theta}}
\frac{\partial}{\partial \theta} \left(\sin{\theta}
\frac{\partial}{\partial \theta} \right)+\frac{4}{(v-u)^2
  \sin{\theta}^2}\frac{\partial^2}{\partial \phi^2} \,.
\ee
and a radial and temporal part, which we can write in terms of
derivatives in $u$ and $v$,
\be
4\frac{\partial^2}{\partial v \partial u}-\frac{4}{u-v} \left(\frac{\partial}{\partial u}-\frac{\partial}{\partial v} \right) \,.
\ee
The angular part vanishes on integration, leaving
\be \label{Vs}
V(x,x')=-\frac{1}{4\pi} \int d\Omega \int_0^1 ds s(1-s) \left[\partial_u \partial_v-\frac{1}{u-v} \left(\partial_u-\partial_v \right) \right](R_{ab}(y) Y^a Y^b) \,.
\ee
Outside the derivatives, we can take $v=0$ and change variables to $u=
s\tau$, giving
\bea
V(x,x')&=&-\frac{1}{4\pi\tau^3} \int d\Omega \int_0^\tau du (\tau-u)\left[
u\partial_u \partial_v- \partial_u+\partial_v \right](R_{ab}(y) Y^a Y^b)\\
&=&-\frac{1}{4\pi\tau^3} \int d\Omega \int_0^\tau du (\tau-u)
\partial_u[(u \partial_v- 1)(R_{ab}(y) Y^a Y^b)] \,.
\eea
We can integrate by parts with no surface contribution, giving
\bea
V(x,x')&=&\frac{1}{4\pi\tau^3} \int d\Omega \int_0^\tau du 
(1-u \partial_v)(R_{ab}(y) Y^a Y^b)\\
&=&\frac{1}{4\pi\tau^3} \int d\Omega \int_0^\tau du 
u^2 \left[ -u R_{uu,v}(y)-2 R_{uv}(y)+R_{uu}(y) \right]\nonumber \,.
\eea

Now
\be \label{RG}
R_{ab} = G_{ab}- (1/2)g_{ab}G \,,
\ee
where $G_{ab}$ is the Einstein tensor and $G$ its trace. Thus
\be\label{VG}
V(x,x')=\frac{1}{4\pi\tau^3} \int d\Omega \int_0^\tau du\,
u^2 \left[ -u G_{uu,v}(y)-2 G_{uv}(y)+(1/2)G(y)+G_{uu}(y) \right] \,.
\ee

Now define a vector field $Q_a(y) = G_{ab}(y)Y^b$.  Then
\be
Q_{a;c} =G_{ab;c}(y)Y^b+G_{ab}(y){Y^b}_{;c}\,.
\ee
We write the covariant derivative only because we are working in
null-spherical coordinates, rather than because of spacetime
curvature, which we are ignoring because we already have first order
quantities.

Since the covariant divergence of $G$ vanishes,
\be
g^{ac}Q_{a;c} =g^{ac} G_{ab}(y){Y^b}_{;c} \,.
\ee
In Cartesian coordinates, $Y^b = y^b$, and ${y^b}_{;c} = \delta^b_c$,
which means that (in any coordinate system).
\be
g^{ac}Q_{a;c} = G \,.
\ee
Explicit expansion gives
\be
g^{ac}Q_{a;c} = 2 (Q_{v,u} + Q_{u,v}) 
-\frac{4}{u-v} (Q_u-Q_v)
- \frac{4}{(v-u)^2}\left[ \frac1{\sin\theta}
\frac{\partial}{\partial \theta} (\sin{\theta} Q_\theta)
+\frac{1}{\sin\theta^2}Q_{\phi,\phi}\right] \,,
\ee
but the angular terms vanish on integration.  Now we expand
the derivatives in $u$ and $v$ and set $v=0$, giving
\blea
Q_{v,u}& = & uG_{uv,u}+ G_{uv}\\
Q_{u,v}& = & uG_{uu,v}+ G_{uv} \,,
\elea
so
\be\label{Qfinal}
\int d\Omega \,\left(2uG_{uv,u} + 2uG_{uu,v}+8G_{uv}- 4G_{uu}\right) = 
\int d\Omega \, G \,.
\ee
Substituting Eq.~(\ref{Qfinal}) into Eq.~(\ref{VG}), we find
\be
V(x,x')=\frac{1}{4\pi\tau^3} \int d\Omega \int_0^\tau du \,
u^2 \left[u G_{uv,u}(y)+2 G_{uv}(y)-G_{uu}(y) \right]
\ee
and integration by parts yields
\be
V(x,x')=\frac{1}{4\pi} \int d\Omega \left[G_{uv}(x'') -
\frac{1}{\tau^3} \int_0^\tau du \, u^2 \left(G_{uv}(y)+G_{uu}(y) \right)\right] \,.
\ee
Now
\bea
\int d\Omega \int_0^\tau du \, u^2 \left(G_{uv}(y)+G_{uu}(y) \right)
&=&\frac12\int d\Omega\int_0^\tau du \, u^2 \left(G_{tt}(y)+G_{tr}(y) \right)\nonumber\\
&=& \frac12\int d\Omega\int_0^\tau du \, u^2 \left(G^{tt}(y)-G^{tr}(y) \right) \qquad
\eea
which is 4 times the total flux of $G^{ta}$ crossing inward through
the light cone.  Since this quantity is conserved, ${G^{ta}}_{;a}=0$,
we can integrate instead over a ball at constant time $\bar t$, giving
\be
4\int d\Omega\int_0^{\tau/2} dr\, r^2 G^{tt}(\bar x + r\Omega )
= \frac{\tau^3}2\int d\Omega\int_0^1 ds\, s^2 G^{tt}(\bar x + s(\tau/2)\Omega)
\ee
so
\be
V(x,x')=\frac{1}{8\pi} \int d\Omega \left[\frac{1}{2} \left[ G_{tt}(x'')-G_{rr}(x'') \right]
-\int_0^1 ds\, s^2 G_{tt}(x''_s)\right] \,,
\ee
where $x''_s=\bar{x}+s(\tau/2)\Omega$, and
\bea
G_R(x,x')&=&\Delta^{1/2}(x,x') \frac{\delta(\sigma)}{2\pi}+\frac{1}{32 \pi^2} \int d\Omega\bigg\{\frac{1}{2} \left[ G_{tt}(x'')-G_{rr}(x'') \right] \nonumber\\
&&-\int_0^1 ds \, s^2 G_{tt}(x''_s) \bigg\} \qquad
\eea
\bea \label{finaliE}
E(x,x')&=&\Delta^{1/2}(x,x') \frac{\delta(\tau-|\bx-\bx'|)-\delta(\tau+|\bx-\bx'|)}{4\pi|\bx-\bx'|} \\
&&+\frac{1}{32 \pi^2} \int d\Omega\bigg\{ \frac{1}{2} \left[ G_{tt}(x'')-G_{rr}(x'') \right]-\int_0^1 ds \,  s^2  G_{tt}(x''_s) \bigg\} \sgn{\tau} \nonumber \,.
\eea

\section{Computation of $\tilde{H}$}
\label{sec:H}

We now need to compute $\tilde H(x,x')$ and apply $\Tsplit_{tt'}$.  First we
consider the term in $\tilde H(x,x')$ that has no dependence on the
curvature.  It has the same form as it would in flat space as shown in Ref~\cite{Fewster:2007rh} and Ch.~\ref{ch:potential}
\be\label{H-1}
\tilde H_{-1}(x,x')=H_{-1}(x,x')=\frac{1}{4\pi^2\sigma_+(x,x')}\,.
\ee
In Sec.~\ref{sec:tsplitH}, we will apply the fully general $\Tsplit_{tt'}$
from Eq.~(\ref{T}) with $\nabla_{\bar{x}}$ defined in
Eq.~(\ref{Rbarxd}) to $\tilde H_{-1}(x,x')$.

All the remaining terms that we need are first order in the curvature,
so for these it is sufficient to take $\nabla_{\bar{x}}$ as the
flat-space Laplacian with respect to the center point, $\bar{x}$.
For this we only need to compute $\tilde H$ at positions given by time
coordinates $t$ and $t'$ but the same spatial position.

As we discussed in Sec~\ref{sec:symmetry}, we only need to keep terms in $\tilde H$ with powers
of $\tau$ up to $\tau^2$, but we need $E$ exactly.  The terms from $H$
alone give a function whose Fourier transform does not decline fast
enough for positive $\xi$ for the integral in Eq.~(\ref{B}) to
converge.  Thus we extract the leading order terms from $iE$ and
combine these with the terms from $H$.  This combination gives a
result that has the appropriate behavior after the Fourier transform.

\subsection{Terms with no powers of $\tau$}

First we want to calculate the zeroth order of the Hadamard
series. The Hadamard coefficients are given by Eqs.~(\ref{recursion1},\ref{recursion2}) for a massless field. To find the zeroth order of the Hadamard series we need only
$v_0(x,x')$, which we find by integrating Eq.~(\ref{recursion1}) along
the geodesic from $x'$ to $x$.  Since we are computing a first-order
quantity, we can work in flat space by letting $y'=x'+\lambda(x-x')$
and using the first-order formulas $\Box \sigma=-8$ and
$\sigma^{,a}=-2 (y'-x')^a$.  From Eq.~(\ref{recursion1}), we have
\be
(y'-x')^a v_{0,a}+v_0=\frac{1}{4} \Box \Delta^{1/2} (y',x') \,,
\ee
and thus
\be\label{v01}
v_0(x,x')=\frac{1}{4} \int_0^1 d\lambda (\Box \Delta^{1/2}) (x'+\lambda(x-x'),x') \,.
\ee
by the same analysis as Eq.~(\ref{vcheckfinal}).

Using the expansion for $\Delta^{1/2}$ from Eq.~(\ref{deltaexp}) gives
\bea
&&v_0(x,x')=-\frac{1}{8}\int_0^1 d\lambda \int_0^1 ds (1-s)s
\Box_{y'}[R_ {ab}(sy'+ (1 -s)x') (y' -x') ^a (y' -x') ^b]\nonumber\\
&&=-\frac{1}{8} \int_0^1 d\lambda \int_0^1 ds(1-s)s \bigg[ (\lambda s)^2 (\Box R_{ab})(x'+s\lambda(x-x'))(x-x')^a (x-x')^b\nonumber\\
&&+2\lambda s R_{,b} (x'+s\lambda(x-x')) (x-x')^b+2 R(x'+s\lambda (x-x')) \bigg] \,.
\eea
We can combine the $s$ and $\lambda$ integrals by defining a new variable $\sigma=s\lambda$
\bea \label{lambdasigma}
&&\int_0^1 d\lambda \int_0^1 ds (1-s)s f(\lambda s)=\int_0^1 d\lambda \int_0^\lambda d\sigma \left(\frac{\sigma}{\lambda^2}-\frac{\sigma^2}{\lambda^3} \right) f(\sigma) \\
&&=\int_0^1 d\sigma \, f(\sigma) \int_\sigma^1 d\lambda \left(\frac{\sigma}{\lambda^2}-\frac{\sigma^2}{\lambda^3} \right)=\int_0^1 d\sigma \, f(\sigma) \left[ -\frac{\sigma}{\lambda}+\frac{\sigma^2}{2\lambda^2} \right]_\sigma^1 \nonumber\\
&&=\frac{1}{2}\int_0^1 d\sigma \, f(\sigma) (1-\sigma)^2 \,.
\eea
Then, changing $\sigma$ to $s$, we find
\bea\label{v0s}
v_0(x,x')&=&-\frac{1}{16}\int_0^1 ds (1-s)^2  \bigg[s^2 (\Box R_{ab})(x'+s(x-x'))(x-x')^a (x-x')^b\nonumber\\
&&+2  s R_{,b} (x'+s(x-x')) (x-x')^b+2 R(x'+s (x-x')) \bigg] \,.
\eea
or when the two points are on the geodesic,
\be
v_0(t,t')=-\frac{1}{16} \int_0^1 ds (1-s)^2 \bigg[ s^2 (\Box R_{tt})(x'+s \tau) \tau^2+4s \eta^{cd}  R_{ct,d}(x'+s \tau)\tau+2 R(x'+s \tau) \bigg] \,.
\ee
In the second term we use the contracted Bianchi identity, $\eta^{cd}
R_{ct,d}= R_{,t}/2$, giving
\bea
&&2\int_0^1 ds (1-s)^2 s\tau R_{,t}(x'+s\tau)= 2\int_0^1 ds (1-s)^2 s \frac{d}{ds} R(x'+s\tau) \nonumber\\
&&=-2\int_0^1 ds (1-s)(1-3s)R(x'+s\tau)\,,
\eea
so the final expression for $v_0$ is
\be \label{v0}
v_0(t,t')=-\frac{1}{16} \int_0^1 ds  (1-s) \bigg[ s^2 (1-s) \Box R_{tt} (\bar{x}+(s-1/2)\tau) \tau^2+4 s R(\bar{x}+(s-1/2)\tau) \bigg] \,.
\ee 
To calculate $H_0$ we only need the zeroth order in $\tau$ from
$v_0$, so the first term does not contribute.  In the second term, we
make a Taylor series expansion,
\be\label{Rt1}
R(\bar{x}+(s-1/2)\tau)=R(\bar{x})+R_{,t}(\bar{x})\tau (s-1/2)+\frac{1}{2}R_{,tt}(\bar{x})\tau^2 (s-1/2)^2+O(\tau^3) \,,
\ee
but only the first term is relevant here.  Thus
\be\label{v0final}
v_0(t,t')=-\frac{1}{4}\int_0^1 ds (1-s)s R(\bar{x})=-\frac{1}{24} R(\bar{x}) \,.
\ee
We also need to expand the Van Vleck determinant appearing in the
Hadamard series.  From Eq.~(\ref{deltaexp}),
\be \label{vanvleck}
\Delta^{1/2}(t,t')=1-\frac{1}{12}R_{tt}(\bar{x})\tau^2-\frac{1}{480}R_{tt,tt}(\bar{x})\tau^4+O(\tau^6)\,.
\ee
Keeping the first order term from Eq.~(\ref{vanvleck}) and using
Eq.~(\ref{v0final}), we have
\be \label{had0}
H_0(x,x')=\frac{1}{48\pi^2} \left[R_{tt}(\bar{x})-\frac{1}{2}R(\bar{x}) \ln{(-\tau_-^2)} \right] \,.
\ee
Now we can add the $H_0(x',x)$ which is the same except that $t$ and $t'$ interchange
\be \label{H0}
H_0(x,x')+H_0(x',x)=\frac{1}{24\pi^2} \left[R_{tt}(\bar{x})-R(\bar{x}) \ln{|\tau|} \right] \,.
\ee

Next we must include $E$ from Eq.~(\ref{finaliE}).  We can expand the
components of the Einstein tensor around $\bar{x}$,
\be\label{G1}
G_{ab}(x'')=G_{ab}(\bar{x})+G_{ab}^{(1)}(x'') \,,
\ee
where $G_{ab}^{(1)}$ is the remainder of the Taylor series
\be \label{rem1}
G_{ab}^{(1)}(x'')=G_{ab}(x'')-G_{ab}(\bar{x})=\int_0^{\tau/2} dr \,G_{ab,i}(\bar{x}+r\Omega)\Omega^i \,.
\ee
To find $E_0$, we put the first term of Eq.~(\ref{G1}) into the second term of
Eq.~(\ref{finaliE}).  We use $G_{rr}=G_{ij}\Omega^i\Omega^j$ and 
$\int d\Omega\, \Omega^i \Omega^j=(4\pi/3) \delta^{ij}$ and find
\bea \label{E0}
E_0(x,x')&=&\frac{1}{8 \pi} \left\{ \frac{1}{2}G_{tt}(\bar{x})-\frac{1}{6}G_{ii}(\bar{x})-\int_0^1 ds \, s^2 G_{tt}(\bar{x}) \right\} \sgn{\tau}\nonumber\\
&=& \frac{1}{48 \pi}  G(\bar{x})\sgn{\tau} =-\frac{1}{48 \pi} R(\bar{x})\sgn{\tau} \,.
\eea
For the remainder term $R_0$, we put the second term of Eq.~(\ref{G1})
into the second term of Eq.~(\ref{finaliE}),
\bea \label{R0}
R_0(x,x')=\frac{1}{32\pi^2}\int d\Omega \left\{ \frac{1}{2}\left[G_{tt}^{(1)}(x'')-G_{rr}^{(1)}(x'')\right]-\int_0^1 ds\,s^2 G_{tt}^{(1)}(x''_s) \right\} \sgn{\tau} \,.
\eea
Using
\be \label{log}
2\ln{|\tau|}+\pi i\sgn{\tau}=\ln{(-\tau_-^2)} \,,
\ee
we combine Eqs.~(\ref{H0}) and (\ref{E0}) to find
\be \label{Ht0}
\tilde{H}_0(t,t')=\frac{1}{48\pi^2} \left[ R_{tt}(\bar{x})-\frac{1}{2}R(\bar{x}) \ln{(-\tau_-^2)} \right] \,.
\ee
Combining all terms through order 0 gives
\be \label{Ht-10}
\tilde{H}_{(0)}(t,t')=\tilde{H}_{-1}(t,t')+\tilde{H}_0(t,t')+\frac{1}{2}i R_0(t,t') \,.
\ee

\subsection{Terms of order $\tau^2$}

Now we compute the terms of order $\tau^2$ in $H$ and $E$. To find
$v_0$ at this order we take Eqs.~(\ref{v0}) and (\ref{Rt1}) and
include terms through second order in $\tau$.  The first-order term
vanishes, leaving
\bea \label{v0ab}
v_0(x,x')&=&-\frac{1}{24}R(\bar{x})-\frac{1}{16} \int_0^1 ds (1-s) [ s^2(1-s) \Box R_{tt}(\bar{x}) \nonumber\\
&&\qquad \qquad \qquad +2s(s-1/2)^2 R_{,tt}(\bar{x})]\tau^2+ \dots \\
&=&-\frac{1}{24}R(\bar{x})-\frac{1}{480} \left(\Box R_{tt}(\bar{x})+\frac{1}{2}R_{,tt}(\bar{x}) \right)\tau^2+\dots \nonumber\,.
\eea

Next we need $v_1$ but since it is multiplied by $\tau^2$ in $H$ we
need only the $\tau$ independent term. From Eq.~(\ref{recursion2})
\be
\Box v_0+2v_{1,a} \sigma^{,a}+v_1 \Box \sigma=0 \,,
\ee
At $x=x'$, $ \sigma^{,a}=0$ so
\be\label{v1lim}
v_1(x,x)=\frac{1}{8} \lim_{x \to x'} \Box_x v_0(x,x') \,.
\ee
Using Eq.~(\ref{v0s}) in  Eq.~(\ref{v1lim}), the only terms that
survive in the coincidence limit are those that have no powers of
$x-x'$ after differentiation, so
\be \label{v1}
v_1(x,x) = -\frac{1}{16} \int_0^1 ds (1-s)^2 s^2 \Box R(\bar{x})=-\frac{1}{480} \Box R(\bar{x}) \,.
\ee
Equations~(\ref{v0}), (\ref{v0ab}) and (\ref{v1}) agree with
Ref.~\cite{Decanini:2005gt} if we note that their expansions are
around $x$ instead of $\bar{x}$.

The $w_1$ at coincidence is given by Ref.~\cite{Wald:1978pj},
\be \label{w1}
w_1(x,x)=-\frac{3}{2} v_1(x,x)=\frac{1}{320} \Box R(\bar{x}) \,.
\ee

Combining Eqs.~(\ref{v0ab}), (\ref{v1}), and (\ref{w1}), and the
fourth order term from the Van Vleck determinant of
Eq.~(\ref{vanvleck}), and keeping in mind that $\sigma=-\tau^2$ when
both points are on the geodesic, we find
\be
H_1(x,x')=\frac{1}{640\pi^2} \bigg[\frac{1}{3}R_{tt,tt}(\bar{x})-\frac{1}{2} \Box R(\bar{x})
-\frac{1}{3}\left(\Box R_{ii}(\bar{x})+\frac{1}{2}R_{,tt}(\bar{x})\right)\ln{(-\tau_-^2)} \bigg]\tau^2  \,.
\ee
Then $H_1(x',x)$ is given by symmetry so
\be \label{H1}
H_1(x,x')+H_1(x',x)=\frac{1}{160\pi^2} \left[\frac{1}{6}R_{tt,tt}(\bar{x})-\frac{1}{4}\Box R(\bar{x})-\frac{1}{3}\left(\Box R_{ii}(\bar{x})+\frac{1}{2}R_{,tt}(\bar{x})\right)\ln{|\tau|}  \right]\tau^2 \,.
\ee

The calculation of $E_1$ is similar to $E_0$, but now we have to include more terms to the Taylor expansion,
\be\label{G3}
G_{ab}(x'')=G_{ab}(\bar{x})+\frac{\tau}{2}G_{ab,i}(\bar{x})\Omega^i+\frac{\tau^2}{8}G_{ab,ij}\Omega^i \Omega^j(\bar{x})+G_{ab}^{(3)}(x'') \,,
\ee
where the remainder of the Taylor series is
\be \label{rem3}
G_{ab}^{(3)}(x'')=\frac{1}{2}\int_0^{\tau/2} dr G_{ab,ijk}(\bar{x}+r\Omega) \left( \frac{\tau}{2}-r\right)^2 \Omega^i \Omega^j \Omega^k \,.
\ee
Now we put Eq.~(\ref{G3}) into Eq.~(\ref{finaliE}), and again use
$G_{rr}=G_{ij}\Omega^i\Omega^j$.  The first term of Eq.~(\ref{G3})
gives $E_0$, which we computed before, and the second term gives nothing,
because $\int d\Omega \, \Omega^i= \int d\Omega \, \Omega^i \Omega^j
\Omega^k=0$.  Using $\int d\Omega \, \Omega^i \Omega^j=4\pi/3
\delta^{ij}$ and $\int d\Omega \, \Omega^i \Omega^j \Omega^k
\Omega^l=(4\pi/15) (\delta^{ij} \delta^{kl}+\delta^{ik}\delta^{jl}
+\delta^{il}\delta^{jk})$, the third term gives
\bea 
E_1(x,x')&=&-\frac{1}{192 \pi} \bigg[ \frac{1}{10} G_{ii,jj}(\bar{x})+\frac{1}{5} G_{ij,ij}(\bar{x}) -\frac{1}{2} G_{tt,ii}(\bar{x}) \nonumber\\
&&\qquad \qquad+\int_0^1 ds \, s^4 G_{tt,ii}(\bar{x}) \bigg]\tau^2 \sgn{\tau} \\
&=&-\frac{1}{320\pi} \left[ \frac{1}{6} G_{ii,jj}(\bar{x})+\frac{1}{3} G_{ij,ij}(\bar{x})-\frac{1}{2} G_{tt,ii} (\bar{x}) \right] \tau^2 \sgn{\tau}\nonumber \,.
\eea
Using the conservation of the Einstein tensor, $0 = \eta^{ab} G_{ia,b} = G_{it,t} - G_{ij,j}$ and 
$0 = \eta^{ab} G_{ta,b} = G_{tt,t} - G_{it,i}$ we can write
\be
G_{ij,ij}(\bar{x})=G_{tt,tt}(\bar{x}) \,.
\ee
So
\be\label{E11}
E_1(x,x')=-\frac{1}{960\pi}  \left(\frac{1}{2}G_{ii,jj}(\bar{x})+
G_{tt,tt}(\bar{x})-\frac{3}{2} G_{tt,ii} (\bar{x})\right) \tau^2
\sgn{\tau}\,.
\ee
Now $G_{ab} = R_{ab} - (1/2)R$, so
\blea
G_{ii} &=& (3/2) R_{tt}-(1/2)R_{ii}\\
G_{tt} &=& (1/2) R_{tt}+(1/2) R_{ii}\,.\label{Gtt}
\elea
Putting these in Eq.~(\ref{E11}) gives
\bea\label{E1}
E_1(x,x')&=&-\frac{1}{960\pi}  \left(R_{ii, jj}(\bar{x})+\frac{1}{2}
R_{tt,tt}(\bar{x})+\frac{1}{2} R_{ii,tt}(\bar{x}) \right) \tau^2 \sgn{\tau}]\nonumber\\
&=&-\frac{1}{960\pi}  \left( \Box R_{ii}(\bar{x})+\frac{1}{2} R_{,tt}(\bar{x}) \right) \tau^2 \sgn{\tau}]\,.
\eea

The fourth term of Eq.~(\ref{G3}) gives the remainder
\be \label{R1}
R_1(x,x')=\frac{1}{32\pi^2} \int d\Omega \bigg\{ \frac{1}{2} \left[G_{tt}^{(3)}(x'')-G_{rr}^{(3)}(x'') \right]-\int_0^1 ds \, s^2 G_{tt}^{(3)}(x''_s)  \bigg\} \sgn{\tau} \,.
\ee
To calculate $\tilde{H}_1$, we combine Eqs.~(\ref{H1}) and (\ref{E1})
and use Eq.~(\ref{log}) to get
\be \label{Ht1}
\tilde{H}_1(x,x')=\frac{\tau^2}{640\pi^2} \left[ \frac{1}{3}R_{tt,tt}(\bar{x})-\frac{1}{2} \Box R(\bar{x})-\frac{1}{3} \left( \Box R_{ii}(\bar{x})+\frac{1}{2} R_{,tt}(\bar{x}) \right) \ln{(-\tau_-^2)} \right] \,.
\ee
All terms through order 1 are then given by
\be \label{Ht-11}
\tilde{H}_{(1)}(t,t')=\tilde{H}_{-1}(t,t')+\tilde{H}_0(t,t')+\tilde{H}_1(t,t')+\frac{1}{2}iR_1(t,t') \,.
\ee

\section{The $\Tsplit_{tt'} \tilde{H}$}
\label{sec:tsplitH}

We can easily take the derivatives of $\tilde{H}_0$ and $\tilde{H}_1$
using Eq.~(\ref{T}), because they are already first order in
$R$. However in the case of the term $\nabla_{\bar{x}}^2
\tilde{H}_{-1}$ we have to proceed more carefully. From
Eqs.~(\ref{Rbarxd}) and (\ref{H-1}) we have
\bea \label{TH-1}
\nabla_{\bar{x}}^2 \tilde{H}_{-1}&=&\frac{1}{4\pi^2} \sum_{i=1}^3 \left( \frac{\partial^2}{\partial (x^i)^2}+2\frac{\partial}{\partial x^i} \frac{\partial}{\partial x'^i}+\frac{\partial^2}{\partial (x'^i)^2} \right)\left(\frac{1}{\sigma_+}\right)\nonumber\\
&=&-\frac{1}{4\pi^2\sigma_+^2}  \sum_{i=1}^3 \left( \frac{\partial^2\sigma}{\partial (x^i)^2}+2\frac{\partial^2 \sigma}{\partial x^i\partial x'^i}+\frac{\partial^2 \sigma}{\partial (x'^i)^2} \right) \,,
\eea
where we used $\partial \sigma/ \partial x^i=\partial \sigma/
\partial x'^i=0$ when the two points are on the geodesic. From \cite{Decanini:2005gt}, after we shift the Taylor series so that the Riemann tensor is evaluated at $\bar{x}$, we have 
\bml\label{d2sigma}\bea
\frac{\partial^2 \sigma}{\partial (x^i)^2}&=&=-2\eta_{ii}-\frac{2}{3}R_{itit}(\bar{x})\tau^2-\frac{1}{2} R_{itit,t}(\bar{x})\tau^3-\frac{1}{5} R_{itit,tt}\tau^4+O(\tau^5) \label{sigma1}\\
\frac{\partial^2 \sigma}{\partial (x'^i)^2}&=&=-2\eta_{ii}-\frac{2}{3}R_{itit}(\bar{x})\tau^2+\frac{1}{2} R_{itit,t}(\bar{x})\tau^3-\frac{1}{5} R_{itit,tt}\tau^4+O(\tau^5) \label{sigma2}\\
\frac{\partial^2 \sigma}{\partial x^i\partial x'^i} &=&2\eta_{ii}-\frac{1}{3}R_{itit}(\bar{x})\tau^2-\frac{7}{40} R_{itit,tt}\tau^4+O(\tau^5)\label{sigma3} \,.
\elea
From Eqs.~(\ref{TH-1}) and (\ref{d2sigma}), and using $R_{itit}=-R_{tt}$ we have
\be \label{TH}
\nabla_{\bar{x}}^2 \tilde{H}_{-1}=-\frac{1}{4\pi^2} \left[\frac{2}{\tau_-^2}R_{tt}(\bar{x})+\frac{3}{4} R_{tt,tt}(\bar{x}) \right] \,.
\ee
From Eqs.~(\ref{F}) and (\ref{T}), we need to compute
\be
\int_0^\infty \frac{d\xi}{\pi}\hat F(-\xi,\xi')\,,
\ee
where
\be
F(t,t') = g(t) g(t')\left[\frac{1}{4}\nabla_{\bar{x}}^2 \tilde H_{(0)}(t,t')
-\partial_\tau^2  \tilde{H}_{(1)}(t,t') \right]\,.
\ee
In the first term in brackets it is sufficient to use $\tilde
H_{(0)}(t,t')$, because higher order terms in $H$ are smooth, even in
$\tau$, and vanish at coincidence, and so they do not contribute, as
discussed in Sec.~\ref{sec:symmetry}.  In the second term,
two powers of $\tau$ are removed by differentiation, so we need 
$\tilde{H}_{(1)}(t,t')$.

Using
Eqs.~(\ref{H-1}), (\ref{R0}), (\ref{Ht0}), (\ref{Ht-10}), (\ref{R1}), (\ref{Ht1}), (\ref{Ht-11}) and (\ref{TH})
we can combine all terms in $F$ to write
\be
F(t,t') = g(t) g(t')\sum_{i=1}^6 f_i(t,t')\,,
\ee
with
\blea
f_1&=&\frac{3}{2\pi^2\tau_-^4}\\\
f_2&=&\frac{1}{48\pi^2\tau_-^2}[R_{ii}(\bar{x})-7 R_{tt}(\bar{x})]\\
f_3&=&\frac{1}{384\pi^2} \left[ \frac{1}{5} R_{tt,tt}(\bar{x})+\frac{1}{5}R_{ii,tt}(\bar{x})-R_{tt,ii}(\bar{x})+\frac{3}{5}R_{ii,jj}(\bar{x}) \right] \ln{(-\tau_-^2)}\\
f_4&=&\frac{1}{320\pi^2}\bigg[-\frac{43}{3}R_{tt,tt}(\bar{x})+\frac{7}{6} R_{tt,ii}(\bar{x})-\frac{1}{2}R_{ii,jj}(\bar{x}) \bigg] \\
f_5&=&\frac{1}{256\pi^2} \int d\Omega\,\nabla_{\bar{x}}^2 \left\{ \frac{1}{2} \left[ G_{tt}^{(1)}(x'')-G_{rr}^{(1)}(x'') \right] -\int_0^1 dss^2 G_{tt}^{(1)}(x''_s) \right\} i\sgn{\tau} \label{f5} \nonumber\\
&&\\
f_6&=& -\frac{1}{64\pi^2} \int d\Omega\,\partial_{\tau}^2  \bigg\{ \frac{1}{2} \left[G_{tt}^{(3)}(x'')-G_{rr}^{(3)}(x'') \right]-\int_0^1 ds s^2 G_{tt}^{(3)}(x''_s)  \bigg\} i\sgn{\tau} \label{f6} \,. \nonumber\\
&&
\elea

\section{The quantum inequality}
\label{sec:QI}

We want to calculate the quantum inequality bound $B$, given
by Eq.~(\ref{B}).  We can write it
\be
B=\sum_{i=1}^8 B_i\,,
\ee
where
\blea
B_i&=&\int_0^\infty  \frac{d\xi}{\pi} \int_{-\infty}^\infty dt
\int_{-\infty}^\infty dt' g(t) g(t') f_i(t,t') e^{i\xi(t'-t)}\nonumber\\
&=&\int_0^\infty \frac{d\xi}{\pi} \int_{-\infty}^\infty d\tau
\int_{-\infty}^\infty d\bar{t}\,
g(\bar{t}-\frac{\tau}{2})g(\bar{t}+\frac{\tau}{2})f_i(\bar t,\tau)e^{-i\xi \tau}, i = 1\ldots6 \nonumber\\
&&\\
\label{B7}B_7 &=& \int_{-\infty}^\infty dt\,g^2(t) Q(t)
= \frac{1}{3840\pi^2}\int_{-\infty}^\infty dt\,g^2(t)\Box R(\bar{t})\\
\label{B8}
B_8 &=& -\int_{-\infty}^\infty dt\,g^2(t)\bigg[2aR_{,ii}(\bar{x})-\frac{b}{2}(R_{tt,tt}(\bar{x})+R_{ii,tt}(\bar{x})-3R_{tt,ii}(\bar{x}) \nonumber\\
&&\qquad \qquad \qquad+R_{ii,jj}(\bar{x}))\bigg] 
\elea
using Eqs.~(\ref{Q}), (\ref{localc}), (\ref{B}) and (\ref{w1}).  The first 6 terms
have the same $\tau$ dependence as the corresponding terms in
Ch.~\ref{ch:potential}. So the Fourier transform proceeds in the
same way, except that instead of dependence on the potential and its
derivatives, we have dependence on the Ricci tensor and its
derivatives.  After the Fourier transform, we see that $B_4$ and $B_7$
have exactly the same form so we merge them in one term. Thus
\be\label{BforR}
B = \frac{1}{16\pi^2}\left[ I_1
+\frac{1}{12} I_2^R
-\frac{1}{24} I_3^R
+\frac{1}{240} I_4^R
+\frac{1}{16\pi} I_5^R
-\frac{1}{4\pi} I_6^R \right]
-I_7^R \,,
\ee
where
\bml\label{I}\bea
I_1&=&\int_{-\infty}^{\infty} dt\,g''(t)^2 \\
I_2^R&=&
\int_{-\infty}^\infty d\bar{t}\,[7R_{tt}(\bar{x})-R_{ii}(\bar{x})]
(g(\bar{t}) g''(\bar{t}) - g'(\bar{t}) g'(\bar{t}))\\
\label{I3}I_3^R  &=&\int_{-\infty}^{\infty} d\tau\,\ln{|\tau|}\sgn{\tau}
\int_{-\infty}^{\infty} d\bar t\,\bigg[\frac{1}{5} R_{tt,tt}(\bar{x})+\frac{1}{5}R_{ii,tt}(\bar{x})-R_{tt,ii}(\bar{x})  \nonumber \\
&&+\frac{3}{5}R_{ii,jj}(\bar{x})\bigg] g(\bar{t}-\frac{\tau}{2})g'(\bar{t}+\frac{\tau}{2})\\
\label{I4}I_4^R &=& \int_{-\infty}^{\infty}d\bar{t}\,
g(\bar{t})^2\bigg[-171 R_{tt,tt}(\bar{x})-R_{ii,tt}(\bar{x})+13R_{tt,ii}(\bar{x})-5R_{ii,jj}(\bar{x})\bigg]
\\
\label{I5}I_5^R &=& \int_{-\infty}^\infty d\tau \, \frac{1}{\tau}
\int_{-\infty}^\infty d\bar{t}\,g(\bar{t}-\tau/2)g(\bar{t}+\tau/2) \int d\Omega\,\nabla_{\bar{x}}^2 \bigg\{ \frac{1}{2}\left[G_{tt}^{(1)}(x'')-G_{rr}^{(1)}(x'')\right] \nonumber\\
&&-\int_0^1 ds \, s^2 \left[G_{tt}^{(1)}(x''_s)\right] \bigg\} \sgn{\tau}
\\
\label{I6}I_6^R &=&\int_{-\infty}^\infty d\tau
\int_{-\infty}^\infty d\bar t\,
\partial_\tau^2 \left[\frac{1}{\tau} g(\bar{t}-\tau/2)g(\bar{t}+\tau/2)  \right]
\int d\Omega\,\bigg\{ \frac{1}{2} \left[G_{tt}^{(3)}(x'')-G_{rr}^{(3)}(x'') \right]\nonumber\\
&&-\int_0^1 ds \, s^2 G_{tt}^{(3)}(x''_s) \bigg\}\sgn{\tau}\\
\label{I7}I_7^R &=&\int_{-\infty}^\infty dt\,g^2(t)\bigg[2aR_{,ii}(\bar{x})-\frac{b}{2}(R_{tt,tt}(\bar{x})+R_{ii,tt}(\bar{x})-3R_{tt,ii}(\bar{x}) \nonumber\\
&& \qquad \qquad \qquad+R_{ii,jj}(\bar{x}))\bigg]\,.
\elea
If we only know that the Ricci tensor and its derivatives are bounded,
as in Eqs.~(\ref{Rmax}), we can restrict the magnitude of each term of
Eq.~(\ref{BforR}). We start with the second term
\bea
|I_2^R| &\leq& \int_{-\infty}^\infty d\bar{t}\, \left|7R_{tt}(\bar{x})-R_{ii}(\bar{x})\right|
|g(\bar{t}) g''(\bar{t}) - g'(\bar{t}) g'(\bar{t})|  \nonumber\\
&\leq& 10 \Rmax \int_{-\infty}^\infty d\bar{t} [g(\bar{t}) |g''(\bar{t})| + g'(\bar{t})^2] \,.
\eea
Terms $I_3^R$, $I_4^R$ and $I_7^R$ are similar. For $I_5^R$ and $I_6^R$, we need
bounds on the components of $G$.  From Eq.~(\ref{Gtt}),
$|G_{tt}|<2\Rmax$.  Since Eq.~(\ref{Rmax}) holds regardless of
rotation, we can bound $G_{rr}$ at any given point by taking the
$x$-axis to point in the radial direction.  Then $G_{rr} = G_{xx} =
(1/2)[R_{xx} - R_{yy} -R_{zz} +R_{tt}]$ and $|G_{rr}| < 2\Rmax$.
Taking derivatives of G just differentiates the corresponding
components of $R$, which are also bounded.  In particular, since there
are 3 terms in $\nabla_{\bar{x}}^2$, we have
$|\nabla_{\bar{x}}^2G_{tt}|,|\nabla_{\bar{x}}^2G_{rr}| < 6\Rmax''$.
Using these results and Eq.~(\ref{rem1}) for the remainder we have
\bea
&& \left| \int d\Omega\,\nabla_{\bar{x}}^2 \bigg\{ \frac{1}{2}\left[G_{rr}^{(1)}(x'')-G_{tt}^{(1)}(x'')\right]+\int_0^1 ds \, s^2 G_{tt}^{(1)}(x''_s) \bigg\} \right|  \nonumber\\
&& \leq \frac{|\tau|}{2} \int d\Omega \left\{ \frac{1}{2} \left[
  |\nabla^2 G_{rr,i}(\bar x)|+|\nabla^2 G_{tt,i}(\bar x)| \right]+\int_0^1 ds \, s^3 |\nabla^2 G_{tt,i}(\bar x)| \right\} |\Omega^i| \nonumber\\
&& \leq \Rmax''' \frac{15|\tau|}{4} \sum_i \int d\Omega |\Omega^i|=\frac{45\pi}{2} |\tau| \Rmax''' \,.
\eea
For $I_6^R$ we use Eq.~(\ref{rem3}) for the remainder
\bea
&&\left| \int d\Omega\,\bigg\{ \frac{1}{2} \left[G_{rr}^{(3)}(x'')-G_{tt}^{(3)}(x'') \right]+\int_0^1 ds \, s^2 G_{tt}^{(3)}(x''_s) \bigg\} \right|  \nonumber\\
&& \leq \frac{|\tau|^3}{48} \int d\Omega \left\{ \frac{1}{2} \left[ |G_{rr,ijk}(\bar x)|+|G_{tt,ijk}(\bar x)| \right]+\int_0^1 ds \,s^5 |G_{tt,ijk}(\bar x)| \right\} |\Omega^i||\Omega^j||\Omega^k| \nonumber\\
&& \leq \Rmax''' \frac{7 |\tau|^3}{144} \sum_{i,j,k}\int d\Omega |\Omega^i| |\Omega^j| |\Omega^k|=\frac{7( 2 \pi+1)}{24} |\tau|^3 \Rmax''' \,.
\eea
After we bound all the terms and calculate the derivatives in $I_6^R$ we can define
\bml\label{J17}\bea
J_2&=&\int_{-\infty}^\infty dt\left[g(t)|g''(t)|+g'(t)^2\right]\\
J_3&=&\int_{-\infty}^\infty dt \int_{-\infty}^\infty dt' |g'(t')|g(t) |\!\ln{|t'-t|}| \\ J_4&=&\int_{-\infty}^\infty dt\,g(t)^2\\
J_5&=&\int_{-\infty}^\infty dt \int_{-\infty}^\infty dt' g(t)g(t') \\
J_6&=&\int_{-\infty}^\infty dt \int_{-\infty}^\infty dt' |g'(t')|g(t)|t'-t|\\
J_7&=&\int_{-\infty}^\infty dt \int_{-\infty}^\infty dt' 
\left [g(t)|g''(t')| +g'(t)g'(t')\right] (t'-t)^2
\elea
and find
\blea
|I_2^R| &\le & 10 \Rmax J_2\\
|I_3^R| &\le &\frac{46}{5} \Rmax'' J_3\\
|I_4^R| &\le & 258 \Rmax''J_4\\
|I_5^R|&\le & \frac{45\pi}{2} \Rmax''' J_5\\
|I_6^R|&\le &\frac{7(2\pi+1)}{48}\Rmax'''\left(4J_5+4J_6+J_7\right)\\
|I_7^R|&\le & (24|a|+11|b|)\Rmax''J_4\,.
\elea
Thus the final form of the inequality is
\bea
\label{final}
\int_{\mathbb{R}} d\tau\,g(t)^2\langle T^{ren}_{tt}\rangle_{\omega}
(t,0) \geq- \frac{1}{16\pi^2} \bigg\{& &I_1+\frac{5}{6}\Rmax J_2\\
&+&\Rmax''\left[\frac{23}{60}J_3+\left(\frac{43}{40}+16\pi^2(24|a|+11|b|)\right) J_4\right]
\nonumber\\
&+&\Rmax''' \left[\frac{163\pi+14}{96\pi}J_5
+\frac{7(2\pi+1)}{192\pi}(4J_6+J_7)\right] \bigg\}\,.\nonumber
\eea

Once we have a specific sampling function $g$, we can compute the
integrals of Eqs.~(\ref{J17}) to get a specific bound.  In the case of
a Gaussian sampling function,
\be\label{Gaussiang}
g(t)=e^{-t^2/t_0^2}\,,
\ee
we computed these integrals numerically in Sec.~\ref{sec:sampling}.
Using those results, the right hand side of Eq.~(\ref{final}) becomes
\be\label{finalGaussian}
-\frac{1}{16 \pi^2 t_0^3} \left\{ 3.76+2.63 \Rmax t_0^2+[1.71+ 197.9(24|a|+11|b|)] \Rmax'' t_0^4+ 6.99 \Rmax''' t_0^5 \right\}\,.
\ee
The leading term is just the flat spacetime bound of
Ref.~\cite{Fewster:1998pu} for $g$ given by Eq.~(\ref{Gaussiang}).
The possibility of curvature weakens the bound by introducing
additional terms, which have the same dependence on $t_0$ as in
Ch.~\ref{ch:potential}, with the Ricci tensor bounds in place of
the bounds on the potential.

\section{Discussion of the result}

In this chapter, using the general quantum inequality of Fewster and Smith
we presented in Ch.~\ref{ch:QI}, we derived an inequality for a minimally-coupled
quantum scalar field on spacetimes with small curvature. We calculated
the necessary Hadamard series terms and the Green's function for this
problem to the first order in curvature. Combining these terms gives
$\tilde{H}$ and taking the Fourier transform gives a bound in terms of
the Ricci tensor and its derivatives.

If we know the spacetime explicitly, Eqs.~(\ref{qinequality2}),
(\ref{BforR}), and (\ref{I}) give an explicit bound on the weighted
average of the energy density along the geodesic.  This bound depends
on integrals of the Ricci tensor and its derivatives combined with the
weighting function $g$.

If we do not know the spacetime explicitly but know that the Ricci
tensor and its first 3 derivatives are bounded, Eqs.~(\ref{J17}) and
(\ref{final}) give a quantum inequality depending on the bounds and the
weighting function.  If we take a Gaussian weighting function,
Eq.~(\ref{finalGaussian}) gives a bound depending on the Ricci tensor
bounds and the width of the Gaussian, $t_0$.

As expected, the result shows that the corrections due to curvature
are small if the quantities $\Rmax t_0^2$, $\Rmax'' t_0^4$, and
$\Rmax''' t_0^5$ are all much less than 1.  That will be true if the
curvature is small when we consider its effect over a distance equal
to the characteristic sampling time $t_0$ (or equivalently if $t_0$ is
much smaller than any curvature radius), and if the scale of variation
of the curvature is also small compared to $t_0$.

In all bounds, there is unfortunately an ambiguity resulting from the
unknown coefficients of local curvature terms in the gravitational
Lagrangian.  This ambiguity is parametrized by the quantities $a$ and
$b$.

Ford and Roman \cite{Ford:1995wg} have argued that flat-space quantum
inequalities can be applied in curved spacetime, so long as the radius
of curvature is small as compared to the sampling time.  The present
chapter explicitly confirms this claim and calculates the magnitude of
the deviation.  The curvature must be small not only on the path where
the quantum inequality is to be applied but also at any point that is
in both the causal future of some point of this path and the causal
past of another.  All such points are included in the integrals in
Eq.~(\ref{I5}) and (\ref{I6}).

Is is interesting to consider vacuum spacetimes, i.e., those whose
Ricci tensor vanishes.  These include, for example, the Schwarzschild
and Kerr spacetimes, and those consisting only of gravitational waves.
In such spacetimes, the flat-space quantum inequality will hold to
first order without modification.  There are, of course, second-order
corrections.  For the Schwarzschild spacetime, for example, these were
calculated explicitly by Visser
\cite{Visser:1996iw,Visser:1996iv,Visser:1997sd}.

\chapter{Average Null Energy Condition in a classical curved background}
\label{ch:ANEC}

In this chapter we present the proof of the achronal ANEC in spacetimes with curvature using a null-projected quantum inequality. It is structured as follows.  First
we state our assumptions and present the ANEC theorem we will
prove. We begin the proof by constructing a parallelogram which can be
understood as a congruence of null geodesic segments or of timelike
paths.  Then we apply the general
inequality presented in Ch.~\ref{ch:QI} to the specific case needed here, using results from Ch.~\ref{ch:curvature}. Finally we present the proof
of the ANEC theorem using the quantum
inequality.

\section{Assumptions}
\subsection{Congruence of geodesics}\label{sec:classical}

As in Ref.~\cite{Fewster:2006uf}, we will not be able to rule out ANEC
violation on a single geodesic.  However, a single geodesic would
not lead to an exotic spacetime.  It would be necessary to have ANEC
violation along a finite congruence of geodesics in order to have a
physical effect.

So let us suppose that our spacetime contains a null geodesic $\gamma$
with tangent vector $\ell$ and that there is a ``tubular neighborhood''
$M'$ of $\gamma$ composed of a congruence of achronal null geodesics,
defined as follows.  Let $p$ be a point of $\gamma$, and let $M_p$ be
a normal neighborhood of $p$.  Let $v$ be a null vector at $p$,
linearly independent of $\ell$, and let $x$ and $y$ be spacelike
vectors perpendicular to $v$ and $\ell$.  Let $q$ be any point in
$M_p$ such that $p$ can be connected to $q$ by a geodesic whose
tangent vector is in the span of $\{v,x,y\}$.  Let $\gamma(q)$
be the geodesic through $q$ whose tangent vector is the vector $\bl$
parallel transported from $p$ to $q$.  If a neighborhood $M'$ of
$\gamma$ is composed of all geodesics $\gamma(q)$ for some choice of
$p$, $M_p$, $v$, $x$ and $y$, we will say that $M'$ is a tubular
neighborhood of $\gamma$.

\subsection{Coordinate system}\label{sec:coordinates}

Given the above construction, we can define Fermi-like coordinates described in Appendix \ref{Fermi} on $M'$ as follows.  Without loss of generality
we can take the vector $v$ to be normalized so that $v_a  \ell^a
=1$, and $x$ and $y$ to be unit vectors.  Then we have a
pseudo-orthonormal tetrad at $p$ given by $E_{(u)} = \ell$,
$E_{(v)} = v$, $E_{(x)} = x$, and $E_{(y)} = y$.  The
point $q = (u,v,x,y)$ in these coordinates is found as follows.  Let
$q^{(1)}$ be found by traveling unit affine parameter from $p$ along
the geodesic generated by $v E_{(v)} + x E_{(x)}+y E_{(y)}$.
Then $q$ is found by traveling unit affine parameter from $q^{(1)}$
along the geodesic generated by $u E_{(u)}$.  During this process
the tetrad is parallel transported.  All vectors and tensors will be
described using this transported tetrad unless otherwise specified.

The points with $u$ varying but other coordinates fixed form one of
the null geodesics of the previous section.

\subsection{Curvature}
\label{sec:curvature}
We suppose that the curvature inside $M'$ obeys the null convergence
condition, 
\be\label{eqn:nullconvergence}
R_{ab}V^a V^b\ge0
\ee
for any null vector $V^a$.  Equation~(\ref{eqn:nullconvergence})
holds whenever the curvature is generated by a ``classical
background'' whose stress tensor obeys the NEC of Eq.~(\ref{eqn:nec}).  
We will refer to this as
a ``classical background'', but the only way it need be classical is Eq.~(\ref{eqn:nec}).

We would not expect any energy conditions to hold when the curvature
is arbitrarily large, because then we would be in the regime of
quantum gravity, so we will require that the curvature be bounded.
In the coordinate system we described we require
\be \label{eqn:Rmax}
|R_{abcd}|< \Rmax 
\ee
and 
\be
|R_{abcd,\alpha}|<\Rmax' , \qquad |R_{abcd,\alpha \beta}|<\Rmax'', \qquad | R_{abcd,\alpha \beta \gamma}|< \Rmax'''
\ee
in $M'$, where the greek indices $\alpha, \beta, \gamma \dots=v,x,y$ and $\Rmax, \Rmax', \Rmax'', \Rmax'''$ are finite numbers but not necessarily small. Also we assume that the curvature is smooth.

\subsection{Causal structure}

We will also require that conditions outside $M'$ do not affect the
causal structure of the spacetime in $M'$
\cite{Fewster:2006uf}\footnote{This condition is equivalent to 
$J^-(p,M) \cap M'=J^-(p,M')$ for all $p \in M'$.}
\be\label{eqn:causal}
J^+(p,M) \cap M'=J^+(p,M')
\ee
for all $p \in M'$. 
Otherwise the curvature outside $M'$ may be
arbitrary.

\subsection{Quantum field theory} \label{sec:quantum}

We consider a quantum scalar field in $M$.  We will work entirely
inside $M'$, and there we require that the field be massless, free and minimally
coupled.  Outside $M'$, however, we
can allow different curvature coupling, interactions with other
fields, and even boundary surfaces with specified boundary conditions.

Because $M$ may not be globally hyperbolic, it is not completely
straightforward to specify what we mean by a quantum field theory on
$M$.  We will use the same strategy as Ref.~\cite{Fewster:2006uf}.
Our results will hold for any quantum field theory on $M$ that reduces
to the usual quantum field theory on each globally hyperbolic
subspacetime of $M$.  The states of interest will be those that
reduce to Hadamard states on each globally hyperbolic subspacetime,
and we will refer to any such state as ``Hadamard''.  See Sec.~II~B
of Ref.~\cite{Fewster:2006uf} for further details.

\section{The theorem} \label{sec:theorem}

\subsection{Stating the theorem}

\emph{Theorem 1.}  Let $(M,g)$ be a (time-oriented) spacetime and let
$\gamma$ be an achronal null geodesic on $(M,g)$, and suppose that $\gamma$ is
surrounded by a tubular neighborhood $M'$ in the sense of
Sec.~\ref{sec:classical}, obeying the null convergence condition,
Eq.~(\ref{eqn:nec}), and that we have constructed
coordinates by the procedure of Sec.~\ref{sec:coordinates}.  Suppose
that the curvature in this coordinate system is smooth and obeys the
bounds of Sec.~\ref{sec:curvature}, that the curvature in the system
is localized, i.e., in the distant past and future the spacetime is
flat, and that the causal structure of $M'$ is not affected by
conditions elsewhere in $M$, Eq.~(\ref{eqn:causal}).

Let $\omega$ be a state of the free minimally coupled quantum scalar
field on $M'$ obeying the conditions of Sec.~\ref{sec:quantum}, and
let $T$ be the renormalized expectation value of the
stress-energy tensor in state $\omega$.

\emph{Under these conditions, it is impossible for the ANEC integral,
\be
A=\int_{-\infty}^{\infty} d \lambda\, T_{ab}\ell^a \ell^b (\Gamma(\lambda)),
\ee
to converge uniformly to negative values on all geodesics
$\Gamma(\lambda)$ in $M'$.}

\subsection{The parallelogram}

We will use the $(u,v,x,y)$ coordinates of
Sec.~\ref{sec:coordinates}. Let $r$ be a positive number small enough such that
whenever $|v|,|x|,|y|<r$, the point $(0,v,x,y)$ is inside the
normal neighborhood $N_p$ defined in Sec.~\ref{sec:classical}.  Then
the point $(u,v,x,y) \in M'$ for any $u$.

\begin{figure}
\centering
\epsfysize=70mm
\epsfbox{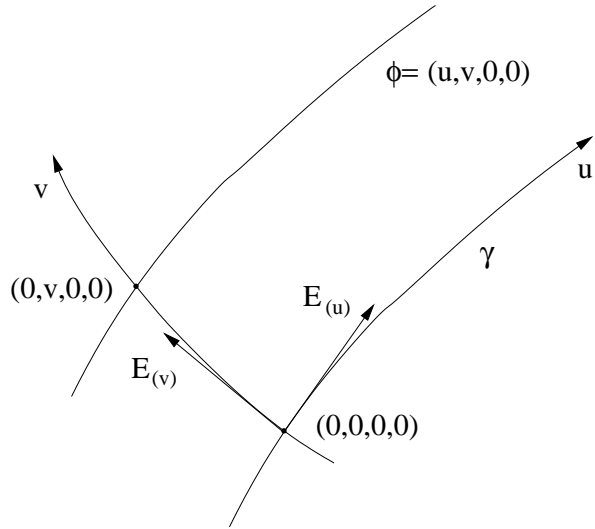}
\caption{Construction of the family of null geodesics $\Phi$ using Fermi normal coordinates}
\label{fig:fermi}
\end{figure}

Now consider the points
\be\label{eqn:Phi}
\Phi(u,v) = (u,v,0,0)\,.
\ee
With $v$ fixed and $u$ varying, these are null geodesics in $M'$.
(See Fig.~\ref{fig:fermi}.)
Write the ANEC integral
\be\label{eqn:Av}
A(v) = \int_{-\infty}^\infty du\, T_{uu}(\Phi(u,v))\,.
\ee
Suppose that, contrary to Theorem 1, Eq.~(\ref{eqn:Av}) converges uniformly to
negative values for all $|v| < r$.  We will prove that this leads to a
contradiction.

Since the convergence is uniform, $A(v)$ is continuous.  Then 
since $A(v)<0$ for all $|v| < r$, we can choose a positive number $v_0 < r$
and a negative number $-A$ larger than all $A(v)$ with $v \in
(-v_0, v_0)$.  Then it is possible to find some number $u_1$ large enough that
\be\label{eqn:uintegral}
\int_{u_-(v)}^{u_+(v)}du\,  T_{uu}(\Phi(u,v)) < -A/2
\ee
for any $v \in (-v_0, v_0)$ as long as \bml \label{eqn:uinequality}
\bea
u_+(v)&>&u_1 \\
u_-(v)&<&-u_1\,.
\elea

As in Ref.~\cite{Fewster:2006uf}, we will define a series of
parallelograms in the $(u, v)$ plane, and derive a contradiction by
integrating over each parallelogram in null and timelike directions.
Each parallelogram will have the form 
\bml\label{eqn:uvrange}
\bea
v &\in& (-v_0, v_0)\\
u &\in& (u_-(v),u_+(v))\,,
\elea
where $u_-(v),u_+(v)$ are
linear functions of $v$ obeying Eqs.~(\ref{eqn:uinequality}).  On each
parallelogram we will construct a weighted integral of
Eq.~(\ref{eqn:uintegral}) as follows.  Let $f(a)$ be a smooth function
supported only within the interval $(-1,1)$ and normalized
\be \label{eqn:normal}
\int_{-1}^1 da f(a)^2 = 1\,.
\ee
Then we can write
\be\label{eqn:uvintegral}
\int_{-v_0}^{v_0} dv\, f(v/v_0)^2 \int_{u_-(v)}^{u_+(v)} du\,T_{uu}(\Phi(u,v)) < -v_0A/2\,.
\ee

We can construct this same parallelogram as follows.  First choose
a velocity V.  Eventually we will take the limit $V \to 1$.  Define
the Doppler shift parameter
\be
\delta= \sqrt{\frac{1+V}{1-V}}\,.
\ee
Let $\alpha$ be some fixed number with $0 <\alpha <1/3$ and then let
\be\label{eqn:tau0}
t_0 =\delta^{-\alpha} r\,.
\ee
As $V \to 1$, $\delta\to \infty$ and $t_0 \to 0$.

\begin{figure}
\centering
\epsfysize=70mm
\epsfbox{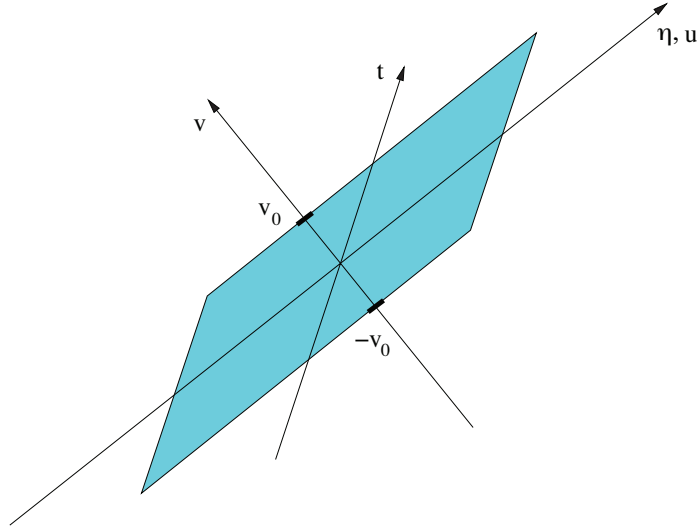}
\caption{The parallelogram $\Phi(u,v)$, $v \in (-v_0, v_0)$,
$u \in (u_-(v),u_+(v))$, or equivalently $\Phi_V(\eta,t)$,
  $t \in(-t_0,t_0)$, $\eta \in (-\eta_0,\eta_0)$}
\label{fig:parallelogram}
\end{figure}

Now define the set of points
\be\label{eqn:PhiV}
\Phi_V(\eta,t)=\Phi \left(\eta+\frac{\delta t}{\sqrt{2}},
\frac{t}{\sqrt{2} \delta}\right)\,.
\ee
We will be interested in the paths given by $\Phi_V(\eta,t)$ with
$\eta$ fixed and $t$ ranging from $-t_0$ to $t_0$.  In flat
space, such paths would be timelike geodesic segments, parameterized
by $t$ and moving at velocity $V$ with respect to the original
coordinate frame.  In our curved spacetime, this is nearly the case,
as we will show below.
Define 
\bml\label{eqn:etavu}\bea 
\eta_0&=&u_1+t_0 \delta/\sqrt{2}\label{eqn:eta0}\\
v_0&=&t_0/(\sqrt{2} \delta)\\
u_{\pm}(v)&=&\pm \eta_0+\delta^2 v
\elea
so that $u_{\pm}$ satisfies Eqs.~(\ref{eqn:uinequality}). Then the
range of points given by Eq.~(\ref{eqn:Phi}) with coordinate ranges
specified by Eqs.~(\ref{eqn:uvrange}) is the same as that given by
Eq.~(\ref{eqn:PhiV}) with coordinate ranges
\blea
-t_0&<&t<t_0\\
-\eta_0&<&\eta<\eta_0
\elea
The parallelogram is shown in Fig.~\ref{fig:parallelogram}.

The Jacobian
\be
\left|\frac {\partial (u,v)}{\partial (\eta,\tau)}\right| =
\frac{1}{\sqrt{2}\delta}
\ee
so Eq.~(\ref{eqn:uvintegral}) becomes
\be\label{eqn:etintegral}
\int_{-\eta_0}^{\eta_0} d\eta \int_{-t_0}^{t_0} d\tau\,
T_{uu}(\Phi_V(\eta,t)) f(t/t_0)^2 < -At_0/2\,.
\ee
We will show that this is impossible.

\subsection{Transformation of the Riemann tensor}
\label{sec:riemann}

Since we are taking $\delta \rightarrow \infty $, components of R with
more $u$'s than $v$'s diverge after the transformation. Components of R
with fewer $u$'s than $v$'s go to zero and components with equal
numbers of $u$'s and $v$'s remain the same.  We want the curvature to
be bounded by $\Rmax$ in the primed coordinate system, which will be
true if all components of the Riemann tensor with more $u$'s than
$v$'s are zero.  We will now show that this is the case in our system.

All points of interests are on achronal null geodesics, which thus
must be free of conjugate points.  Using
Eq.~(\ref{eqn:nullconvergence}) and proposition 4.4.5 of
Ref.~\cite{HawkingEllis}, each geodesic must violate the ``generic
condition''.  That is to say, we must have
\be\label{eqn:nongeneric}
\ell^c \ell^d \ell_{[a} R_{b]cd[e} \ell_{f]} = 0
\ee
everywhere in $M'$.

The only nonvanishing components of the metric in the tetrad basis
are $g_{uv} = g_{vu} = -1$ and $g_{xx} = g_{yy} = 1$.  The tangent
vector $\bl$ has only one nonvanishing component $\ell^u = 1$,
while the covector has only one nonvanishing component $\ell_v = -1$.
Thus Eq.~(\ref{eqn:nongeneric}) becomes
\be
\ell_{[a} R_{b]uu[e} \ell_{f]} = 0\,.
\ee
Let $j$, $k$, $l$, $m$ and $n$ denote indices chosen only from \footnote{This notation applies only in this subsection and not the rest of the thesis} $\{x,y\}$.
Choosing $a=m$, $e=n$, and $b=f=v$  we find
\be\label{eqn:Ruumn}
R_{muun} = 0
\ee
for all $m$ and $n$.  Thus
\be\label{eqn:Ruu}
R_{uu} = 0\,.
\ee
Equation~(\ref{eqn:Ruu}) also follows immediately from the fact that
since $R_{uu}$ cannot be negative, any positive $R_{uu}$ would lead to
conjugate points.

If we apply the null convergence condition,
Eq.~(\ref{eqn:nullconvergence}), to $V=E_{(u)}+
\epsilon E_{(m)}+ (\epsilon^2 /2) E_{(v)}$, where $\epsilon \ll 1$, we get
\be\label{eqn:NECepsilon}
R_{uu}+2R_{mu} \epsilon+O(\epsilon^2)\ge0\,.
\ee
Since $R_{uu} = 0$ from Eq.~(\ref{eqn:Ruu}), in order to have
Eq.~(\ref{eqn:NECepsilon}) hold for both signs of $\epsilon$, we must have
\be\label{eqn:Rmu}
R_{mu} = 0\,.
\ee
Since $R_{mu}=-R_{umvu}+g^{jk}R_{jmku}$,
\be\label{eqn:Ruuvm1}
R_{umvu} = g^{jk}R_{jmku}\,.
\ee

Now we use the Bianchi identity,
\be\label{eqn:Bianchi}
R_{luum;n}+R_{lunu;m}+R_{lumn;u}=0\,.
\ee
From Eq.~(\ref{eqn:Ruumn}), $R_{luum,n}=0$.  The correction to make
the derivatives covariant involves terms of the forms
$R_{auum}\nabla_n E^{(a)}_l$ and $R_{laum}\nabla_n E^{(a)}_u$.
Because of Eq.~(\ref{eqn:Ruumn}), the only contribution to the first
of these comes from $a=v$, which we can transform using
Eq.~(\ref{eqn:Ruuvm1}).  For the second, we observe that
$0=\nabla_n(E^{(v)}\cdot E^{(v)}) = 2
\nabla_n E^{(v)}\cdot E^{(v)} = 2\nabla_n E^{(v)}_u$, so $a=v$ does
not contribute.  Furthermore $R_{lumn;u} = R_{lumn,u}$, because the
$u$ direction is the single final direction in the coordinate
construction of Sec.~\ref{sec:coordinates}, and so in this direction
the tetrad vectors are just parallel transported.  Thus we find
\bea\label{eqn:Rulmndeq}
\frac{dR_{lumn}}{du} &=& g^{jk}[R_{jmku}\nabla_n E^{(v)}_l
  + R_{jlku}\nabla_n E^{(v)}_m
-R_{jnku}\nabla_m  E^{(v)}_l
-R_{jlku}\nabla_m E^{(v)}_n]\nonumber \\
&&+(R_{lkum}+ R_{lukm})\nabla_n E^{(k)}_u
+(R_{lknu}+ R_{lunk})\nabla_m E^{(k)}_u\,.
\eea
Eq.~(\ref{eqn:Rulmndeq}) is a first-order differential equation in
the pair of independent Riemann tensor components $R_{xuxy}$ and
$R_{yuxy}$.  By assumption, the curvature and its derivative vanish
in the distant past, and therefore the correct solution to these
equations is
\be\label{eqn:Rulmn}
R_{lumn}= 0\,.
\ee
Eqs.~(\ref{eqn:Ruuvm1}) and (\ref{eqn:Rulmn}) then give
\be\label{eqn:Ruuvm}
R_{umvu} = 0\,.
\ee

Combining Eqs.~(\ref{eqn:Ruumn}), (\ref{eqn:Rulmn}), and
(\ref{eqn:Ruuvm}) and their transformations under the usual Riemann
tensor symmetries, we conclude that all components of the Riemann
tensor with more $u$'s than $v$'s vanish as desired.

\subsection{Timelike paths} 
\label{sec:timelike}

The general quantum inequality of Ch.~\ref{ch:QI} we will use for this proof is applied on timelike paths. So we are going to show that the paths $\Phi(\eta+\delta t/\sqrt{2},
t/\sqrt{2} \delta)$ are indeed timelike. Differentiating Eq.~(\ref{eqn:PhiV}), we find the components of the
tangent vector $k=d\Phi_V/dt=(1/\stwo,1/\stwo)$ in the Fermi coordinate
basis. The squared length of $p$ in terms of these components is
$g_{ab}k^{a} k^{b}$.  We showed in Appendix \ref{Fermi}  that $g_{ab}=
\eta_{ab}+h_{ab}$, where $h_{ab}$ at some
point $X$ is a sum of a small number of terms (6 in the present case
of 2-step Fermi coordinates) each of which is a coefficient no greater
than 1 times an average of
\be\label{eqn:RXX}
R_{abcd}X^{d}X^{c}
\ee
over one of the geodesics used in the construction of the Fermi
coordinate system.  The summations over $d$ and $c$ in
Eq.~(\ref{eqn:RXX}) are only over restricted sets of indices depending
on the specific term under consideration.  From
Eqs.~(\ref{eqn:PhiV}) and (\ref{eqn:eta0}) the points under
consideration satisfy \bml\label{eqn:uvmax}
\bea
|u|&<&u_1/\delta+\sqrt{2}\tau_0\label{eqn:umax}\\
|v|&<&\tau_0/\sqrt{2}\\
x &=& y = 0\,.
\elea

From Eq.~(\ref{eqn:tau0}), the first term in Eq.~(\ref{eqn:umax})
decreases faster than the second, so we find that all components of
$X$ are $O(t_0)$.  Using the fact that the components of the Riemann tensor are bounded we find
\be
h_{ab} = O(\Rmax t_0^2)
\ee
so
\be \label{eqn:timelike}
g_{ab}k^{a}k^{b}=-1+O(\Rmax t_0^2)\,.
\ee
Thus for sufficiently large $\delta$, and thus small $t_0$, $k$
is timelike.

\subsection{Causal diamond}

The quantum inequality of Ch.~\ref{ch:QI}Kontou:2014tha is applied to timelike paths inside a globally hyperbolic region of the spacetime. So this region $N$ must
include the timelike path from $p = \Phi_V(\eta,-t_0)$ to
$q = \Phi_V(\eta,t_0)$, and to be globally hyperbolic it must
include all points in both the future of $p$ in the past of $q$, so we
can let $N$ be the ``double cone'' or ``causal diamond'',
\be
N=J^+(p) \cap J^-(q)\,.
\ee
We have shown that the curvature is small everywhere in the tube
$M'$, so we must show that $N \subset M'$.

From the previous section, we have that the metric can be written as
\be
g_{ab}=\eta_{ab}+h_{ab}\,,
\ee
where $h_{ab}$ consists of terms of the form
$R_{abcd}X^{d}X^{c}$.
The double cone in flat space obeys
\be\label{eqn:xyuvflat}
|x|,|y|,|v|<\tau_0\,,
\ee
so the same is true at zeroth order in the Riemann tensor $R$.
Thus at zeroth order,
\be
h_{ab} =O(\Rmax t_0^2)\,,
\ee
and so at first order in $R$,
\be
|x|,|y|,|v|<\tau_0(1+O(\Rmax t_0^2))\,.
\ee
Since $t_0 \ll r$ for large $\delta$, we have
\be
|x|,|y|,|v|<r\,.
\ee
and $N\subset M'$ as desired.

\section{The null-projected quantum inequality}

We can write  the general quantum inequality of Eq.~(\ref{qinequality}) for $w(t)=\Phi_V(\eta,t)$ for a specific value of $\eta$ and the stress energy tensor contracted with null vector field $\ell^a \equiv u$ as
\be\label{eqn:qinequality2}
\int_{-\infty}^\infty d\tau\,g(t)^2 \langle \Tren_{uu} \rangle (w(t))
\geq -B\,,
\ee 
where
\be\label{eqn:B}
B = \int_0^\infty\frac{d\xi}{\pi}\hat F(-\xi,\xi)
-\int_{-\infty}^\infty dt\,g^2(t) \left(2aR_{,uu}+b R_{,uu} \right)\,,
\ee
where we used that $g_{uu}=0$ so the term $Q$ doesn't contribute at all, $R_{uu}=0$ according to Sec.~\ref{sec:riemann} and 
\be\label{eqn:Ft}
F(t,t') = g(t)g(t') \Tsplit_{uu'} \tilde H_{(1)}(w(t),w(t'))\,,
\ee
$\hat F$ denotes the Fourier transform in both arguments according to
Eq.~(\ref{Fourier}).

\section{Calculation of $\Tsplit_{uu'} \tilde{H}_{(1)}$}
\label{sec:tildeH}

To simplify the calculation we will evaluate the $T\tilde{H}_{(1)}$ in the coordinate system $(t,x,y,z)$ where the timelike path $w(t)$ points only in the $t$ direction, $z$ direction is perpendicular to it and $x$ and $y$ are the previously defined ones. More specifically $t$ and $z$ are
\bea \label{eqn:tzuv}
t=\frac{\delta^{-1} u+\delta v}{\stwo}\,, \qquad z=\frac{\delta^{-1} u-\delta v}{\stwo} \,.
\eea
The new null coordinates $\tu$ and $\tv$ are defined by
\bea
\tu=\frac{t+z}{\stwo}\,, \qquad \tv=\frac{t-z}{\stwo} \,,
\eea
and are connected with $u$ and $v$,
\bea
\tu=\delta^{-1} u\,, \qquad \tv=\delta v \,.
\eea

The operator $\Tsplit_{uu'}$ can be written as
\be\label{eqn:Tsplit}
\Tsplit_{uu'}=\delta^{-2}\partial_{\tu} \partial_{\tu'} \,.
\ee
If we define $\zeta=z-z'$ and $\bar{u}$ as the $\tu$ coordinate of $\bar{x}$, the center point between $x$ and $x'$ we have
\be\label{eqn:tutz}
\Tsplit_{uu'}=\frac{1}{2}\delta^{-2} \left( \frac{1}{2} \partial_{\bar{u}}^2 -(\partial_\tau^2+2\partial_\zeta \partial_\tau+\partial_\zeta^2) \right)
\ee

\subsection{Derivatives on $\tilde{H}_{-1}$}
\label{sec:H-1}

For the derivatives of $\tilde{H}_{-1}$ it is simpler to use
Eq.~(\ref{eqn:Tsplit}).  We have
\bea
\partial_{\tu'} \partial_{\tu} \tilde{H}_{-1}&=&\frac{1}{4\pi^2} \left(\frac{\partial}{\partial x^{\tu}}\frac{\partial}{\partial x'^\tu}\right) \left(\frac{1}{\sigma_+}\right)\,.
\eea
In flat spacetime it is straightforward to apply the derivatives to
$\tilde{H}_{-1}$. However in curved spacetime, there will be
corrections first order in the Riemann tensor to both $\sigma$ and its
derivatives.  

We are considering a path $w$ whose tangent vector is
constant in the coordinate system described in
Sec.~\ref{sec:coordinates}.  The length of this path can be written
\be\label{eqn:s}
s(x,x')=\int_0^1 d\lambda \sqrt{g_{ab} (w(\lambda)) \frac{d x^a}{d\lambda} \frac{dx^b}{d\lambda}}
=\int_0^1 d\lambda \sqrt{g_{ab}(x'') \Delta x^a \Delta x^b} \,.
\ee
where $\Delta x=x-x'$ and $x''=x'+\lambda \Delta x$ since $dx^a/d\lambda$ is a constant.

Now $\sigma$ is the negative squared length of the geodesic connecting
$x'$ to $x$.  This geodesic might be slightly different from the path
$w$.  However, the deviation results from the Christoffel symbol
$\Gamma^a_{bc}$, which is $O(R)$.  Thus the distance between the two
paths is also $O(R)$, and the difference in the metric between the
two paths is thus $O(R^2)$.  Similarly, the difference in length in the
same metric due to the different path between the same two points is
$O(R^2)$.  All these effects can be neglected, and so we take
$\sigma=-s^2$.

Now using Eq.~(\ref{eqn:smetric}) of Appendix \ref{Fermi} we can write the first-order
correction to the metric,
\be
g_{ab}=\eta_{ab}+F_{ab}+F_{ba} \,,
\ee
where $F_{ab}$ is given by Eq.~(\ref{eqn:Flower2}) of the same Appendix because the first step for $x=y=0$ is in the $\tv$ direction and the
second in the $\tu$ direction. By the symmetries of the Riemann tensor
the only non-zero component is
\be
F_{\tv\tv}(x'')=\int_0^1 d\kappa (1-\kappa) R_{\tv\tu\tv\tu}(\kappa x''^\tu, x''^\tv) x''^\tu x''^\tu \,,
\ee
where we took into account the different sign conventions.
Putting this in Eq.~(\ref{eqn:s}) gives
\bea
s(x,x')&=&\int_0^1 d\lambda \sqrt{2 \Delta x^{\tu} \Delta x^{\tv}+2F_{\tv\tv} \Delta x^{\tv} \Delta x^{\tv}}\\
&=&\int_0^1 d\lambda \stwo\left( \sqrt{\Delta x^{\tu} \Delta x^{\tv}}+
\frac12 F_{\tv\tv} (\Delta x^{\tv})^{3/2} (\Delta x^{\tu})^{-1/2}\right)\,.\nonumber
\eea
So to first order in the curvature,
\be
\sigma(x,x')=-s(x,x')^2=-\tau^2+\zeta^2-2\int_0^1 d\lambda F_{\tv\tv}\Delta
x^{\tv} \Delta x^{\tv}\,.
\ee
We define the zeroth order $\sigma$,
\be
\sigma^{(0)}(x,x')=-\tau^2+\zeta^2 \,,
\ee
and the first order,
\bea
\sigma^{(1)}(x,x')&=&
-2 \int_0^1 d\lambda F_{\tv\tv} \Delta x^{\tv} \Delta x^{\tv} \\
&=&-2 \int_0^1 d\lambda  \int_0^1 d\kappa (1-\kappa)R_{\tv\tu\tv\tu}(\kappa x''^\tu, x''^\tv) x''^\tu x''^\tu \Delta x^{\tv} \Delta x^{\tv}\nonumber\\
&=&-2 \int_0^1 d\lambda \int_0^\ell dy (\ell-y)R_{\tv\tu\tv\tu}(y,x''^\tv) \Delta x^{\tv} \Delta x^{\tv}\nonumber \,,
\eea
where we defined $\ell \equiv x''^\tu$ and changed variables to $y=\kappa \ell$. Now to first order,
\be
\frac{1}{\sigma_+}=\frac{1}{\sigma^{(0)}}-\frac{\sigma^{(1)}}{(\sigma^{(0)})^2} \,,
\ee
and the derivatives,
\be
 \left(\frac{\partial}{\partial x^{\tu}}\frac{\partial}{\partial
   x'^\tu}\right)\left( \frac{1}{\sigma_+}\right)
 \bigg|_{\zeta=0}=\frac{4}{\tau_-^4}+\frac{12}{\tau_-^6}
 \sigma^{(1)}-\frac{2\stwo}{\tau_-^5}\left(\sigma^{(1)}_{,\tu}-\sigma^{(1)}_{,\tu'}\right)-\frac{1}{\tau_-^4}\sigma^{(1)}_{,\tu\tu'}\,.
\ee
Now we can take the derivatives of $\sigma$ ,
\bea
\sigma^{(1)}_{,\tu}&=&- 2 \int_0^1 d\lambda \lambda \frac{\partial}{\partial \ell} \int_0^\ell dy (\ell-y)R_{\tv\tu\tv\tu}(y,x''^\tv) \Delta x^{\tv} \Delta x^{\tv} \nonumber\\
&=&- 2 \int_0^1 d\lambda \lambda \int_0^\ell dy R_{\tu\tv\tu\tv}(y,x''^\tv)  \Delta x^{\tv} \Delta x^{\tv} \,.
\eea
Similarly,
\bea
\sigma^{(1)}_{,\tu'}&=&-2 \int_0^1 d\lambda (1-\lambda) \frac{\partial}{\partial \ell} \int_0^\ell dy (\ell-y)R_{\tv\tu\tv\tu}(y,x''^\tv) \Delta x^{\tv} \Delta x^{\tv} \nonumber\\
&=&-2\int_0^1 d\lambda (1-\lambda) \int_0^\ell dy R_{\tu\tv\tu\tv}(y,x''^\tv) \Delta x^{\tv} \Delta x^{\tv} \,.
\eea
For the two derivatives of $\sigma^{(1)}$,
\be
\sigma^{(1)}_{,\tu\tu'}=-2 \int_0^1 d\lambda (1-\lambda)\lambda R_{\tu\tv\tu\tv}(x''^\tu,x''^\tv) \Delta x^{\tv} \Delta x^{\tv}\,.
\ee
Now we can assume purely temporal separation, so $\Delta x^{\tu} =\Delta x^{\tv} =
\tau/\stwo$ and
\be\label{eqn:xpp}
x'' =\frac{1}{\stwo}(t''+\bar{z},t''-\bar{z})\,,
\ee
where $\bar{z}=(z+z')/2$ and $t''=t'+\lambda \tau$. 
Then the derivatives of $\tilde{H}_{-1}$ are
\bea \label{eqn:tsplity}
\Tsplit_{uu'} \tilde{H}_{-1}&=&\frac{\delta^{-2}}{4 \pi^2\tau_-^4}\bigg(4-12  \int_0^1 d\lambda  \int_0^\ell dy (\ell-y)R_{\tv\tu\tv\tu}(y, x''^\tv)  \\
&&-2 \stwo \int_0^1 d\lambda (1-2\lambda) \int_0^\ell dy R_{\tu\tv\tu\tv}(y,x''^\tv) \tau \nonumber\\
&&+\int_0^1 d\lambda (1-\lambda)\lambda R_{\tu\tv\tu\tv}(x''^\tu,x''^\tv) \tau^2\bigg)  \nonumber\,.
\eea
Let us define the locations $\bar{x}_\kappa=(\kappa \bar{x}^{\tu},  \bar{x}^{\tv})$ and
\be\label{eqn:xppk}
x''_\kappa=\frac{1}{\stwo}(\kappa(t''+\bar{z}),t''-\bar{z}) \,.
\ee
Then Eq.~(\ref{eqn:tsplity}) can be written
\bea \label{eqn:tsplittau}
\Tsplit_{uu'} \tilde{H}_{-1}&=&\frac{\delta^{-2}}{4 \pi^2\tau_-^4}\bigg(4-\int_0^1 d\lambda \bigg[12 \int_0^1 d\kappa  (1-\kappa)(x''^{\tu})^2 R_{\tv\tu\tv\tu}(x''_\kappa) \\
&&+2 \stwo (1-2\lambda) \int_0^1 d\kappa x''^{\tu} R_{\tu\tv\tu\tv}(x''_\kappa) \tau-(1-\lambda)\lambda R_{\tu\tv\tu\tv}(x'') \tau^2  \bigg] \bigg)\nonumber \,.
\eea

The derivatives of $\tilde{H}_{-1}$ can thus be written
\be \label{eqn:zero}
\Tsplit_{uu'} \tilde{H}_{-1}=\delta^{-2} \bigg[\frac{1}{\tau_-^4}\left(\frac{1}{\pi^2}+ y_1(\bar{t},\tau)\right)+\frac{1}{\tau_-^3}y_2(\bar{t},\tau)+\frac{1}{\tau_-^2}y_3(\bar{t},\tau)\bigg] \,,
\ee
where the $y_i$'s are smooth functions of the curvature,
\be\label{eqn:yi}
y_1(\bar{t},\tau)=\int_0^1 d\lambda Y_1 (t'') \qquad  y_2(\bar{t},\tau)=\int_0^1 d\lambda (1-2\lambda) Y_2 (t'') \qquad y_3(\bar{t},\tau)=\int_0^1 d\lambda  (1-\lambda) \lambda Y_3 (t'') \,,
\ee
with
\blea\label{eqn:Y1}
Y_1(t'')&=&-\frac{3}{2 \pi^2}  \int_0^1 d\kappa(1-\kappa) (t''+\bar z)^2 R_{\tv\tu\tv\tu}(x''_\kappa) \,,\\
\label{eqn:Y2}Y_2(t'')&=&\frac{1}{2\pi^2} \int_0^1 d\kappa (t''+\bar z) R_{\tv\tu\tv\tu}(x''_\kappa) \,,\\
\label{eqn:Y3}Y_3(t'')&=&-\frac{1}{4\pi^2} R_{\tu\tv\tu\tv}(x'') \,,
\elea
where $x''$ and $x''_\kappa$ are defined in terms of $t''$ by
Eqs.~(\ref{eqn:xpp}) and (\ref{eqn:xppk}).

\subsection{Derivatives with respect to $\tau$ and $\bar{u}$}

In Ch.~\ref{ch:curvature} we calculated $\tilde{H}_{(1)}$, but for points separated only in time.  Let us use coordinates
$(T,Z,X,Y)$ to denote a coordinate system where the coordinates of $x$
and $x'$ differ only in T.  Ref.~\cite{Kontou:2014tha} gives
\bea
\tilde{H}_{(1)}(T,T')&=&\tilde{H}_{-1}(T,T')+\tilde{H}_0(T,T')+\tilde{H}_1(T,T')+\frac{1}{2}iR_1(T,T')\,,\\
\tilde{H}_{(0)}(T,T')&=&\tilde{H}_{-1}(T,T')+\tilde{H}_0(T,T')+\frac{1}{2}iR_0(T,T') \,,
\eea
where
\blea\label{eqn:Ht-1}
\tilde{H}_{-1}(T,T')&=&-\frac{1}{4\pi^2 (T-T'-i\epsilon)^2} \,,\\
\label{eqn:Ht0}
\tilde{H}_0(T,T')&=&\frac{1}{48\pi^2} \left[ R_{TT}(\bar{x})-\frac{1}{2}R(\bar{x}) \ln{(-(T-T'-i\epsilon)^2)} \right]\,, \\
\label{eqn:Ht1}
\tilde{H}_1(T,T')&=&\frac{(T-T')^2}{640\pi^2} \bigg[ \frac{1}{3}R_{TT,TT}(\bar{x})-\frac{1}{2} \Box R(\bar{x})\\
&&-\frac{1}{3} \left( \Box R_{II}(\bar{x})+\frac{1}{2}
  R_{,TT}(\bar{x}) \right) \ln{(-(T-T'-i\epsilon)^2)} \bigg]
\nonumber\,.
\elea
The order-0 remainder term is
\bea
\label{eqn:R0}
R_0(T,T')=\frac{1}{32\pi^2}\int d\Omega \bigg\{&& \frac{1}{2}\left[G_{TT}^{(1)}(X'')-G_{RR}^{(1)}(X'')\right] \\
&&-\int_0^1 ds\,s^2 G_{TT}^{(1)}(X''_s) \bigg\} \sgn{(T-T')} \nonumber\,,
\eea
where $\int d\Omega$ means to integrate over solid angle with unit 3-vectors
$\hat\Omega$, the 4-vector $\Omega = (0, \hat\Omega)$, the subscript $R$
means the radial direction, and we define $X''=\bar{x}+(1/2)|T-T'|
\Omega$, $X''_s=\bar{x}+(s/2)|T-T'| \Omega$, and
\be \label{eqn:G1}
G_{AB}^{(1)}(X'')=G_{AB}(X'')-G_{AB}(\bar{x})=\int_0^{|T-T'|/2} dr \,G_{AB,I}(\bar{x}+r\Omega)\Omega^I \,.
\ee
The order-1 remainder term is
\bea\label{eqn:R1}
R_1(T,T')=\frac{1}{32\pi^2} \int d\Omega \bigg\{&& \frac{1}{2} \left[G_{TT}^{(3)}(X'')-G_{RR}^{(3)}(X'') \right] \nonumber\\
&&-\int_0^1 ds \, s^2 G_{TT}^{(3)}(X''_s)  \bigg\} \sgn{(T-T')} \,,
\eea
where $G_{AB}^{(3)}$ is the remainder after subtracting the
second-order Taylor series.  We can write
\be\label{eqn:G3}
G_{AB}^{(3)}(X'')=\frac{1}{2}\int_0^{|T-T'|/2} dr G_{AB,IJK}(\bar{x}+r\Omega) \left( \frac{T-T'}{2}-r\right)^2 \Omega^I \Omega^J \Omega^K \,.
\ee

When we apply the $\tau$ and $\bar{u}$ derivatives from
Eq.~(\ref{eqn:tutz}), we can take $(T,Z,X,Y) = (t,z,x,y)$ and
calculate $\partial_{\bar{u}}^2 \tilde{H}_{0}$, $\partial_\tau^2
\tilde{H}_0$, $\partial_\tau^2 \tilde{H}_1$, $\partial_{\bar{u}}^2
R_0$, and $\partial_\tau^2 R_1$.  Applying $\bar{u}$ derivatives to
$\tilde{H}_0$ gives
\be\label{eqn:one}
\partial_{\bar{u}}^2 \tilde{H}_0=\frac{1}{48\pi^2}\left[ R_{tt,\tu\tu}(\bar{x})-\frac{1}{2}R_{,\tu\tu}\ln{(-\tau_-^2)}\right] \,.
\ee
For the derivatives with respect to $\tau$ we have
\be\label{eqn:two}
\partial_\tau^2 \tilde{H}_0=\frac{1}{48 \pi^2\tau_-^2} R(\bar{x}) \,,
\ee
and
\be\label{eqn:three}
 \partial_\tau^2 \tilde{H}_1=\frac{1}{320 \pi^2} \left[ \frac{1}{3} R_{tt,tt}(\bar{x})-\frac{1}{2} \Box R(\bar{x})-\frac{1}{3} \left( \Box R_{ii}(\bar{x})+\frac{1}{2} R_{,tt}(\bar{x}) \right) (3+\ln{(-\tau_-^2)})\right] \,.
\ee
in the $\tau\to0$ limit.

Applying $\bar{u}$ derivatives to $R_0$ gives
\bea\label{eqn:four}
 \partial_{\bar{u}}^2 R_0=\frac{1}{32 \pi^2} \int d\Omega\, \int_0^{|\tau|/2} dr \, \partial_{\bar{u}}^2 \bigg\{&& \frac{1}{2} \left[ G_{tt,i}(x''')-G_{rr,i}(x''') \right] \nonumber\\
&&-\int_0^1 ds \, s^2 G_{tt,i}(x'''_s) \bigg\}\Omega^i  \sgn{\tau} \,,
\eea
where $x'''=\bar{x}+r\Omega$ and $x'''_s=\bar{x}+sr\Omega$. 

Now we have to take the second derivative of $R_1$ with respect to
$\tau$, which is $T-T'$ in this case.  This appears in three places:
the argument of $\sgn$ in Eq.~(\ref{eqn:R1}), the limit of integration
in Eq.~(\ref{eqn:G3}), and the term in parentheses in
Eq.~(\ref{eqn:G3}).  When we differentiate the $\sgn$, we get
$\delta(\tau)$ and $\delta'(\tau)$. but since $G_{AB}^{(3)} \sim
\tau^3$, there are enough powers of $\tau$ to cancel the $\delta$ or
$\delta'$, so this gives no contribution.  When we differentiate the
limit of integration, the term in parentheses in Eq.~(\ref{eqn:G3})
vanishes immediately. The one remaining possibility gives
\bea\label{eqn:five}
 \partial_\tau^2 R_1=\frac{1}{128 \pi^2} \int d\Omega \int_0^{|\tau|/2} dr
\bigg\{&& \frac{1}{2} \left[ G_{tt,ijk}(x''')-G_{rr,ijk}(x''') \right] \nonumber\\
&&-\int_0^1 ds \,  s^2 G_{tt,ijk}(x'''_s) \bigg\} \Omega^i \Omega^j \Omega^k \sgn{\tau} \,.
\eea

\subsection{Derivatives with respect to $\zeta$}

To differentiate with respect to $\zeta$, we must consider the
possibility that $x$ and $x'$ are not purely temporally separated.  We
will suppose that the separation is only in the $t$ and $z$ directions
and construct new coordinates $(T, Z)$ using a Lorentz transformation
that leaves $\bar x$ unchanged and maps the interval
$(T-T',0)$ in the new coordinates to $(\tau,\zeta)$ in the old
coordinates.  Then
\be\label{eqn:TTtz}
T-T'=\sgn\tau\sqrt{\tau^2-\zeta^2}\,,
\ee
and the transformation from $(T, Z)$ to $(t, z)$ is given by
\be 
\Lambda=\frac{1}{\sgn\tau\sqrt{\tau^2-\zeta^2}}
\left(
\begin{array}{cc}
\tau & \zeta\\
\zeta & \tau
\end{array}
\right) \,.
\ee
with the $x$ and $y$ coordinates unchanged.  Then
\be
\left(
\begin{array}{c}
\tau\\
\zeta 
\end{array}
\right)=\Lambda \left(
\begin{array}{c}
T-T'\\
0
\end{array}
\right) \,.
\ee 

Now let $M$ be some tensor appearing in $\tilde{H}_{(1)}$.  The components
in the new coordinate system are given in terms of those in the old by
\be
M_{ABC\dots}= \Lambda^a_A \Lambda^b_B \Lambda^c_C \dots M_{abc\dots}
\ee
We would like to differentiate such an object with respect to $\zeta$
and then set $\zeta = 0$.  The only place $\zeta$ can appear is in
the Lorentz transformation matrix, where we see
\be
\partial_\zeta \Lambda^a_A \bigg|_{\zeta=0} = \tau^{-1}
\left(\begin{array}{cc}
0 & 1\\
1 & 0
\end{array} \right) 
\ee
and similarly,
\be
\partial_\zeta^2 \Lambda^a_A \bigg|_{\zeta=0} = \tau^{-2}
\left(\begin{array}{cc}
1 & 0\\
0 & 1
\end{array} \right) \,.
\ee
To simplify notation, we will define $P$ and $Q$ to be the matrices on
the right hand sides.  Reinstating $x$ and $y$,
\bea
P &=& \left(\begin{array}{cccc}
0 & 1 & 0 & 0\\
1 & 0 & 0 & 0\\
0 & 0 & 0 & 0\\
0 & 0 & 0 & 0
\end{array} \right)\\
Q &=& \left(\begin{array}{cccc}
1 & 0 & 0 & 0\\
0 & 1 & 0 & 0\\
0 & 0 & 0 & 0\\
0 & 0 & 0 & 0
\end{array} \right)\,.
\eea

Now we can write the derivative of $M_{ABC\dots}$ as
\bea\label{eqn:oneder}
\partial_\zeta M_{ABC\dots}\bigg|_{\zeta=0}&=&\partial_\zeta (\Lambda^a_A \Lambda^b_B \Lambda^c_C \dots)M_{abc \dots} \bigg|_{\zeta=0}\\
&=& \bigg[ (\partial_\zeta \Lambda^a_A) \delta^b_B \delta^c_C \dots+\delta^a_A (\partial_\zeta \Lambda^b_B) \delta^c_C \dots+\dots  \bigg] M_{abc \dots}\bigg|_{\zeta=0} \nonumber \\
&=&\frac{1}{\tau_-} (\underbrace{P^a_A \delta^b_B \delta^c_C \dots+\delta^a_A P^b_B \delta^c_C \dots+\dots}_n )M_{abc\dots}=\frac{1}{\tau_-}p^{abc\dots}_{ABC\dots}  M_{abc\dots} \nonumber
\eea
where $p^{abc\dots}_{ABC\dots}$ is a rank-$n$ matrix of 0's and 1's.
With two derivatives, we have
\bea\label{eqn:twoder}
\partial_\zeta^2 M_{ABC \dots}\bigg|_{\zeta=0}&=&\partial_\zeta^2 (\Lambda^a_A \Lambda^b_B \Lambda^c_C \dots)M_{abc \dots} \bigg|_{\zeta=0}\\
&=& \bigg[ (\partial_\zeta^2 \Lambda^a_A) \delta^b_B \delta^c_C \dots+\delta^a_A (\partial_\zeta^2 \Lambda^b_B)\delta^c_C \dots+\dots \nonumber\\
&&+2(\partial_\zeta \Lambda^a_A)(\partial_\zeta \Lambda^b_B)\delta^c_C \dots+2(\partial_\zeta \Lambda^a_A) \delta^b_B (\partial_\zeta \Lambda^c_C) \dots+\dots \bigg] M_{abc \dots}\bigg|_{\zeta=0}\nonumber \\
&=&\frac{1}{\tau_-^2} (\underbrace{Q^a_A \delta^b_B \delta^c_C\dots+\delta^a_A Q^b_B \delta^c_C \dots+\dots}_n \nonumber\\
&&+\underbrace{P^a_A P^b_B\delta^c_C\dots+P^a_A \delta^b_B P^c_C \dots+\dots}_{(n-1)n} )M_{abc\dots} \nonumber \\
&=&  \frac{1}{\tau_-^2} q^{abc\dots}_{ABC\dots} M_{abc\dots} \nonumber
\eea
where $q^{abc\dots}_{ABC\dots}$ is a rank-$n$ matrix of nonnegative
integers.

There are also places where $T-T'$ appears explicitly in $\tilde{H}_1$.  We
can differentiate it using Eq.~(\ref{eqn:TTtz}),
\blea\label{eqn:dTTdz}
\partial_\zeta(T-T')\bigg|_{\zeta=0}&=&0\,,\\
\label{eqn:dTTdz2}
\partial_\zeta^2(T-T')\bigg|_{\zeta=0}&=&-\tau^{-1}\,.
\elea

Now we apply the operators $\partial_\zeta^2$ and
$\partial_\tau \partial_\zeta$ to $\tilde{H}_0$, $\tilde{H}_1$, and
$R_1$.  First we apply one $\zeta$ derivative\footnote{The Lorentz
  transformation technique we use here is not quite sufficient to
  determine the singularity structure of the distribution
  $\partial_\zeta\tilde{H}_0$ at coincidence.  Instead we can use
  Eq.~(\ref{deltaexp}) to compute the
  non-logarithmic term in $\tilde{H}_0$ for arbitrary $x$ and $x'$,
  which is then $-R_{ab}(\bar{x})(x-x')^a(x-x')^b/(48\pi^2\sigma_+)$.
  Differentiating this term gives Eq.~(\ref{eqn:six}) and explains the
  presence of $\tau_-$ instead of $\tau$ in the denominator.  The
  first term of Eq.~(\ref{eqn:seven}) arises similarly.} to
Eq.~(\ref{eqn:Ht0}) using Eq.~(\ref{eqn:oneder}),
\be\label{eqn:six}
\partial_\tau \left( \partial_\zeta \tilde{H}_0\bigg|_{\zeta=0}\right)=-\frac{1}{48\pi^2\tau_-^2} p^{ab}_{tt}R_{ab}(\bar{x}) \,,
\ee
and two $\zeta$ derivatives using Eqs.~(\ref{eqn:twoder}) and (\ref{eqn:dTTdz}),
\be\label{eqn:seven}
\partial_\zeta^2 \tilde{H}_0\bigg|_{\zeta=0}=\frac{1}{48 \pi^2\tau_-^2} \left[ q^{ab}_{tt}R_{ab}(\bar{x})+R(\bar{x}) \right] \,.
\ee
Then we apply one $\zeta$ derivative to $\tilde{H}_1$,
\be\label{eqn:eight}
\partial_\tau \left( \partial_\zeta \tilde{H}_1\bigg|_{\zeta=0}\right)=\frac{1}{1920 \pi^2} \left[p^{abcd}_{tttt} R_{ab,cd}(\bar{x})-\left( p^{ab}_{ii} \Box R_{ab}(\bar{x})+ \frac{1}{2}p^{ab}_{tt} R_{,ab}(\bar{x}) \right)(\ln{(-\tau_-^2)}+2) \right] \,, 
\ee
and two $\zeta$ derivatives to $\tilde{H}_1$,
\bea\label{eqn:nine}
\partial_\zeta^2 \tilde{H}_1\bigg|_{z=0}=\frac{1}{640 \pi^2} \bigg[&& \frac{1}{3} q^{abcd}_{tttt} R_{ab,cd}(\bar{x})-\frac{2}{3}R_{tt,tt}(\bar{x})+\Box R(\bar{x})-\frac{1}{3}\bigg( q^{ab}_{ii} \Box R_{ab}(\bar{x}) \nonumber \\
&&+ \frac{1}{2} q^{ab}_{tt} R_{,ab}(\bar{x}) \bigg)\ln{(-\tau_-^2)}+\frac{2}{3}\left( \Box R_{ii}(\bar{x})+\frac{1}{2} R_{,tt}(\bar{x})\right)(1+\ln{(-\tau_-^2)}) \bigg]  \,. \nonumber\\
&&
\eea

Finally we have to apply the derivatives to the remainder $R_1$. We
can apply the $\zeta$ derivatives in two places, the Lorentz
transformations and $G_{AB}^{(3)}$. Since the three terms are very
similar we will apply the derivatives to one of them
\be \label{eqn:zetaR1}
\partial_\zeta \int d\Omega G^{(3)}_{TT}(\bar{x}+Y) \bigg|_{\zeta=0}=\int d\Omega \left( \frac{1}{\tau}p^{ab}_{tt} G^{(3)}_{ab}(x'')+ \frac{\partial}{\partial Y^a}  G^{(3)}_{tt}(\bar{x}+Y)\partial_\zeta Y^a \bigg|_{\zeta=0}\right) \,,
\ee
where we defined $Y^a \equiv (1/2) |T-T'| \Lambda^a_I\Omega^I$. 
Then using Eqs.~(\ref{eqn:oneder}) and (\ref{eqn:dTTdz}), we find that
that $\partial_\zeta Y^a |_{\zeta=0}=(1/2) p^a_i \Omega^i \sgn\tau$ and taking into account the properties of Taylor expansions,
\be
 \frac{\partial}{\partial Y^a} G^{(3)}_{tt}(\bar{x}+Y)\bigg|_{\zeta=0}=G^{(2)}_{tt,a}(x'')\,,
\ee
where $G^{(2)}_{ab,c}$ is the remainder of the Taylor expansion of
$G_{ab,c}$ after subtracting the first-order Taylor series.

Thus Eq.~(\ref{eqn:zetaR1}) becomes
\be \label{eqn:zeta}
\partial_\zeta \int d\Omega G^{(3)}_{TT}(\bar{x}+Y) \bigg|_{\zeta=0}=\int d\Omega \left( \frac{1}{\tau}p^{ab}_{tt} G^{(3)}_{ab}(x'')+ \frac{1}{2} G^{(2)}_{tt,a}(x'') p^a_i \Omega^i \sgn\tau\right)\,.
\ee
Using $G^{(3)}$ from Eq.~(\ref{eqn:G3}) and
\be\label{eqn:G2}
G_{ab,c}^{(2)}(x'')=\int_0^{|\tau|/2} dr G_{ab,ijc}(\bar{x}+r\Omega) \left( \frac{|\tau|}{2}-r\right) \Omega^i \Omega^j \,,
\ee
Eq.~(\ref{eqn:zeta}) becomes
\bea
\partial_\zeta \int d\Omega G^{(3)}_{TT}(\bar{X}+Y) |_{\zeta=0}=\int
d\Omega \int_0^{|\tau|/2} dr \left( \frac{|\tau|}{2}-r\right)&& \bigg[p^{ab}_{tt}G_{ab,ijk}(x''')\frac{1}{\tau} \left(\frac{|\tau|}{2}-r \right)\\
&&+\frac{1}{2}p^a_i G_{tt,ija}(x''')\sgn\tau\bigg] \Omega^i \Omega^j
\Omega^k\,.\nonumber
\eea
We could simplify further by using the explicit values of the $p$
matrices, but our strategy here is to show that all terms are bounded
by some constants without computing the constants explicitly, since
the actual constant values will not matter to the proof.

Applying the $\tau$ derivative gives
\bea
\partial_\tau\left(\partial_\zeta \int d\Omega G^{(3)}_{TT}(\bar{x}+Y)
\bigg|_{\zeta=0}\right)=\int d\Omega \int_0^{|\tau|/2} dr \bigg[& & \left(\frac{1}{4}-\frac{r^2}{\tau^2} \right) p^{ab}_{tt}G_{ab,ijk}(x''')\\
&&+\frac{1}{4}p^a_i G_{tt,ija}(x''')\bigg]  \Omega^i \Omega^j \Omega^k \,. \nonumber
\eea
We do not have to differentiate $\sgn\tau$ here, because the rest of
the term is $O(\tau^2)$ and so a term involving $\delta(\tau)$ would
not contribute.

The same procedure can be applied to all three terms.  Terms involving
$X''_s$ will get an extra power of $s$ each time $G$ is
differentiated.  The final result is
\bea \label{eqn:ten}
\lefteqn{\partial \tau \left(\partial_\zeta R_1(T,T') \bigg|_{\zeta=0}\right)}\\
&=&\frac{1}{32\pi^2} \int d\Omega \int_0^{|\tau|/2} dr \bigg\{ \left(\frac{1}{4}-\frac{r^2}{\tau^2} \right) \bigg[ \frac{1}{2}(p^{ab}_{tt}-p^{ab}_{rr}) G_{ab,ijk}(x''')-\int_0^1 ds s^2  p^{ab}_{tt}G_{ab,ijk}(x_s''') \bigg]\nonumber\\
&&\qquad\qquad+\frac{1}{4}\bigg[ \frac{p^a_i}{2} (G_{tt,ija}(x''')
-G_{rr,ija}(x'''))-\int_0^1 ds s^3  p^a_i G_{tt,ija}(x_s''') \bigg] \bigg\} \Omega^i \Omega^j \Omega^k \sgn{\tau} \,.\nonumber
\eea

For two $\zeta$ derivatives we can apply both on the Lorentz transforms,
both on the Einstein tensor or one on each,
\bea \label{eqn:zetazetaR1}
\partial_\zeta^2 \int d\Omega G^{(3)}_{TT}(\bar{x}+Y) \bigg|_{\zeta=0}&=&\int d\Omega \bigg( \frac{q^{ab}_{tt} }{\tau^2}G^{(3)}_{ab}(x'')+ \frac{\partial^2}{\partial Y^a \partial Y^b} G^{(3)}_{tt}(\bar{x}+Y) \partial_\zeta Y^a \partial_\zeta Y^b \bigg|_{\zeta=0}\nonumber \\
&&+  \frac{\partial}{\partial Y^a} G^{(3)}_{tt}(\bar{x}+Y)  \partial_\zeta^2 Y^a \bigg|_{\zeta=0} \\
&&+2 \frac{p^{ab}_{tt}}{\tau}  \frac{\partial}{\partial Y^c} G^{(3)}_{ab}(\bar{x}+Y) \partial_\zeta Y^c \bigg|_{\zeta=0}\bigg)\,.\nonumber
\eea
Using Eqs.~(\ref{eqn:dTTdz2}) and (\ref{eqn:twoder}),
\be
\partial_\zeta^2 Y^j \bigg|_{\zeta=0}=\frac{q^j_i \Omega^i}{2\tau}-\frac{\Omega^j}{2\tau}=\frac{1}{2\tau}h^j_i \Omega^i \,,
\ee
with $h^j_i \equiv q^j_i-\delta^j_i$, while
\be
\partial_\zeta^2 Y^t \bigg|_{\zeta=0}=0\,,
\ee
since $q^t_i = 0$ and $\Omega^t = 0$.
Using properties of the Taylor series as before, we can write
\be
 \frac{\partial^2}{\partial Y^a \partial Y^b}G^{(3)}_{tt}(\bar{x}+Y)=G^{(1)}_{tt,ab}(x'')\,,
\ee
so Eq.~(\ref{eqn:zetazetaR1}) becomes
\bea
\partial_\zeta^2 \int d\Omega G^{(3)}_{TT}(\bar{x}+Y) \bigg|_{\zeta=0}&=&\int d\Omega \bigg[ \frac{q^{ab}_{tt} }{\tau^2}G^{(3)}_{ab}(x'')+\frac{1}{4}p^a_i p^b_j G^{(1)}_{tt,ab}(x'')  \Omega^i \Omega^j\nonumber \\
&& \qquad +\frac{1}{2|\tau|} \bigg(2 p^{ab}_{tt} p^c_i G^{(2)}_{ab,c}(x'')+G_{tt,j}^{(2)}(x'') h^j_i\bigg) \Omega^i \bigg]\,.
\eea
Using $G^{(1)}$ as in Eq.~(\ref{eqn:G1}) and $G^{(2)}$ and $G^{(3)}$
from Eqs.~(\ref{eqn:G2}) and (\ref{eqn:G3}) this becomes
\bea
\partial_\zeta^2 \int d\Omega G^{(3)}_{TT}(\bar{x}+Y) \bigg|_{\zeta=0}&=&\int d\Omega \int_0^{|\tau|/2} dr \bigg[ \bigg(\frac{1}{2}-\frac{r}{|\tau|}\bigg)^2 q^{ab}_{tt}G_{ab,ijk}(x''') \nonumber \\
&&+\frac{1}{4}p^a_i p^b_j G_{tt,kab}(x''')+ \left(\frac{1}{4}-\frac{r}{2|\tau|}\right) \bigg(2p^{ab}_{tt}p^c_i G_{ab,jkc}(x''')  \nonumber\\
&&+h^l_i G_{tt,ljk}(x''') \bigg) \bigg] \Omega^i \Omega^j \Omega^k \,.
\eea
For all three terms
\bea \label{eqn:eleven}
\lefteqn{\partial_\zeta^2 R_1(T,T') \bigg|_{\zeta=0}}\\
&=&\frac{1}{32\pi^2} \int d\Omega \int_0^{|\tau|/2} dr \bigg\{\bigg(\frac{1}{2}-\frac{r}{|\tau|}\bigg)^2 \bigg[ \frac{1}{2}(q^{ab}_{tt}-q^{ab}_{rr}) G_{ab,ijk}(x''')-\int_0^1 ds s^2 q^{ab}_{tt}G_{ab,ijk}(x'''_s) \bigg]\nonumber\\
&&\qquad\qquad\qquad+\frac{1}{4} p^a_i p^b_j \bigg[ \frac{1}{2}(G_{tt,kab}(x''')-G_{rr,kab}(x'''))-\int_0^s ds s^4 G_{tt,kab}(x'''_s) \bigg] \nonumber\\
&&\qquad\qquad\qquad+ \left(\frac{1}{4}-\frac{r}{2|\tau|}\right) \bigg[ p^c_i (p^{ab}_{tt}-p^{ab}_{rr}) G_{ab,jkc}(x''')+\frac{1}{2}h^l_i (G_{tt,ljk}(x''')-G_{rr,ljk}(x''')) \nonumber\\
&&\qquad\qquad\qquad\qquad\qquad-\int_0^1 ds s^3 (2 p^c_i p^{ab}_{tt} G_{ab,jkc}(x'''_s)+h^l_i  G_{tt,ljk}(x_s''')) \bigg] \bigg\} \Omega^i \Omega^j \Omega^k\sgn\tau\,. \nonumber
\eea

\section{The Fourier transform}

Eqs.~(\ref{eqn:zero}), (\ref{eqn:one}), (\ref{eqn:two}),
(\ref{eqn:three}), (\ref{eqn:four}), (\ref{eqn:five}),
(\ref{eqn:six}), (\ref{eqn:seven}), (\ref{eqn:eight}),
(\ref{eqn:nine}), (\ref{eqn:ten}) and (\ref{eqn:eleven}) include all
the $\Tsplit_{uu'} \tilde{H}_{(1)}$ terms. To perform the Fourier
transform we expand
$\Tsplit_{uu'} \tilde{H}_{(1)}$ according to Eqs.~(\ref{eqn:Tsplit})
and (\ref{eqn:tutz}) and separate the terms by their $\tau$
dependence,
\bea\label{eqn:TsplitH}
\Tsplit_{uu'} \tilde{H}_{(1)}&=&\delta^{-2}\bigg[
\partial_{\tu} \partial_{\tu'}\tilde{H}_{-1}
+\frac{1}{4}\bigg(\partial_{\bar{u}}^2\tilde{H}_{0}+ \frac{1}{2} iR_0\bigg)\nonumber \\ 
&&\qquad
  -\frac{1}{2}(\partial_\tau^2+\partial_\zeta^2+2\partial_\tau
  \partial_\zeta)\left( \tilde{H}_{0}+\tilde{H}_1+\frac{1}{2}i R_1\right)\bigg]
\nonumber\\ &=&\delta^{-2} \bigg[ \frac{1}{\tau_-^4}
\left(\frac{1}{\pi^2}+ y_1(\bar{t},\tau)\right)+\frac{1}{\tau_-^3}y_2(\bar{t},\tau)+\frac{1}{\tau_-^2}
  (c_1(\bar{t})+y_3(\bar{t},\tau))
  \nonumber\\ &&\qquad+\ln{(-\tau_-^2)}
  c_2(\bar{t})+c_3(\bar{t})+c_4(\bar{t},\tau) \bigg] \,,
\eea
where $c_1$, $c_2$, and $c_3$ are smooth and have no $\tau$ dependence
and $c_4$ is odd, $C_1$ and bounded. As mentioned in
Sec.~\ref{sec:H-1}, the functions $y_i$ depend on $\tau$ but are
smooth. Explicit expressions for the $c_i$ are 
\blea
c_1&=&\frac{1}{48\pi^2}\bigg(-R(\bar{x})+\left(p_{tt}^{ab}-\frac{q_{tt}^{ab}}{2}\right)R_{ab}(\bar{x}) \bigg) \\
c_2&=& \frac{1}{1920\pi^2}\bigg(-5 R_{,\tu\tu}(\bar{x})+\left(p^{ab}_{ii}+\frac{q^{ab}_{ii}}{2} \right) \Box R_{ab}(\bar{x})+\frac{1}{2}\left(p^{ab}_{tt}+\frac{q^{ab}_{tt}}{2} \right) R_{,ab}(\bar{x}) \bigg)\\
c_3&=&\frac{1}{960\pi^2}\bigg(5R_{tt,\tu\tu}(\bar{x})-\frac{1}{2}\left(p^{abcd}_{tttt}+\frac{q^{abcd}_{tttt}}{2}\right)R_{ab,cd}(\bar{x})+\Box R_{ii}(\bar{x})+\frac{1}{2} R_{,tt}(\bar{x}) \nonumber \\
&&\qquad \qquad +p^{ab}_{ii} \Box R_{ab}(\bar{x})+\frac{1}{2} p^{ab}_{tt}R_{,ab}(\bar{x}) \bigg)\\
c_4&=&\frac{1}{256\pi^2}\int_0^{|\tau|/2} dr \int d\Omega \bigg( \partial_{\bar{u}}^2 \bigg\{ \frac{1}{2} \left[ G_{tt,i}(x''')-G_{rr,i}(x''') \right]-\int_0^1 ds \, s^2 G_{tt,i}(x_s''')\bigg\}\nonumber \\
&&-\bigg\{ \frac{1}{4} \left[ G_{tt,ijk}(x''')-G_{rr,ijk}(x''') \right]-\frac{1}{2}\int_0^1 ds \,  s^2 G_{tt,ijk}(x_s''') \nonumber\\
&&+\left(1-\frac{4 r^2}{\tau^2} \right) \bigg[ \frac{1}{2}(p^{ab}_{tt}-p^{ab}_{rr}) G_{ab,ijk}(x''')- \int_0^1 ds s^2  p^{ab}_{tt}G_{ab,ijk}(x_s''') \bigg] \nonumber\\
&&+ \frac{p^a_i}{2} (G_{tt,jka}(x''')-G_{rr,jka}(x'''))-\int_0^1 ds s^3  p^a_i G_{tt,jka}(x_s''') 
 \nonumber\\
&&+2 \bigg(\frac{1}{2}-\frac{r}{|\tau|}\bigg)^2 \bigg[\frac{1}{2}(q^{ab}_{tt}-q^{ab}_{rr}) G_{ab,ijk}(x''')- \int_0^1 ds s^2 q^{ab}_{tt}G_{ab,ijk}(x'''_s) \bigg] \nonumber\\
&&+\frac{1}{2} p^a_i p^b_j \bigg[ \frac{1}{2}(G_{tt,kab}(x''')-G_{rr,kab}(x'''))-\int_0^s ds s^4 G_{tt,kab}(x'''_s) \bigg] \nonumber \\
&&+ \left(\frac{1}{2}-\frac{r}{|\tau|}\right) \bigg[ p^c_i(p^{ab}_{tt}-p^{ab}_{rr}) G_{ab,jkc}(x''')+\frac{1}{2}h^l_i (G_{tt,ljk}(x''')-G_{rr,ljk}(x''')) \nonumber \\
&&-\int_0^1 ds s^3 (2 p^c_i p^{ab}_{tt} G_{ab,jkc}(x'''_s)+ h^l_i G_{tt,ljk}(x_s'''))   \bigg\} \Omega^j \Omega^k \bigg) \Omega^i \sgn{\tau} \,. 
\elea

We now put the terms of Eq.~(\ref{eqn:TsplitH}) into
Eq.~(\ref{eqn:B}), and Fourier transform them, following
the procedure of Sec.~\ref{sec:Fourier} , to obtain the
bound $B$ in the form
\be\label{eqn:BI}
B=\delta^{-2}\sum_{i=0}^{6} B_i \,.
\ee
The first term in Eq.~(\ref{eqn:TsplitH}) is $1/(\pi^2\tau_-^4)$, and we
proceed exactly as Sec.~\ref{sec:Fourier}, except for the
different numerical coefficient, to obtain
\be
\label{eqn:B0}
B_0=\frac{1}{24\pi^2}\int_{-\infty}^\infty d\bar{t} \, g''(\bar{t})^2\,.
\ee
Putting only Eq.~(\ref{eqn:B0}) into Eq.~(\ref{eqn:BI}) gives the result
for flat space. Fewster and Eveson \cite{Fewster:1998pu} found a
result of the same form, but they considered $T_{tt}$ instead of
$T_{uu}$, so the multiplying constant is different. Fewster and
Roman \cite{Fewster:2002ne} found the result for null projection.
Where we have $1/24$, they had $(v\cdot\ell)^2/12$, where $v$ is the
unit tangent vector to the path of integration.  Here $v\cdot\ell =
\ell^t = 1/(\delta\sqrt{2})$, from Eq.~(\ref{eqn:tzuv}), so the
results agree.

The remaining $\tau_-^{-4}$ term requires more attention, because of
the $\tau$ dependence in $y_1$.  We write
\be
B_1= \int_0^\infty \frac{d \xi}{\pi} \int_{-\infty}^{\infty} d\tau
G_1(\tau) \frac{1}{\tau_-^4} e^{-i\xi \tau} \,,
\ee
with
\be
G_1(\tau)=\int_{-\infty}^\infty d\bar{t}(y_1(\bar{t},\tau))g\left(\bar{t}-\frac{\tau}{2}\right)g\left(\bar{t}+\frac{\tau}{2}\right) \,.
\ee
Then 
\be
B_1=\frac{1}{24}G_1''''(0)\,.
\ee
Applying the $\tau$ derivatives to $G_1$ gives
\bea
G''''_1(\tau)\bigg|_{\tau=0}=\int_{-\infty}^\infty d\bar{t} \bigg[&& \frac{d^4}{d\tau^4} y_1(\bar{t},\tau)\bigg|_{\tau=0} g(\bar{t})^2+3 \frac{d^2}{d\tau^2} y_1(\bar{t},\tau)\bigg|_{\tau=0} (g''(\bar{t})g(\bar{t})-g'(\bar{t})^2) \nonumber\\
&&+ \frac{1}{8}y_1(\bar{t})(g''''(\bar{t})g(\bar{t})-4g'''(\bar{t})g(\bar{t})+3g''(\bar{t})^2)\bigg] \,,
\eea
where the terms with an odd number of derivatives of the product of the sampling functions vanish after taking $\tau=0$.

Now $y_1$ depends on $\tau$ and $\bar t$ only through
  $t''=\bar t+ (\lambda-1/2)\tau$, so using Eq.~(\ref{eqn:yi}), we can write
\be
\frac{d}{d\tau} y_1(\bar{t},\tau)=\frac{d}{d\tau} \int_0^1 d\lambda
Y_1(t'')=\frac{d}{d\bar{t}}\int_0^1 d\lambda (\lambda-1/2) Y_1(t'') \,.
\ee
Then we integrate by parts and put all the derivatives on the sampling functions $g$,
\bea
B_1=\frac{1}{24} \int_{-\infty}^\infty d\bar{t} \bigg[&&2 \int_0^1 d\lambda \left(\lambda-\frac{1}{2}\right)^4 Y_1(\bar{t}) (3 g''(\bar{t})^2+4g'(\bar{t})g'''(\bar{t})+g(\bar{t})g''''(\bar{t})) \nonumber\\
&&+ 3 \int_0^1 d\lambda \left(\lambda-\frac{1}{2}\right)^2 Y_1(\bar{t})(g''''(\bar{t}) g(\bar{t})-g''(\bar{t})^2) \nonumber\\
&&+\frac{1}{8}y_1(\bar{t})(g''''(\bar{t})g(\bar{t})-4g'''(\bar{t})g'(\bar{t})+3g''(\bar{t})^2)\bigg] \,.
\eea
Since we set $\tau=0$, $Y_1$ has no $\lambda$ dependence and we can perform the integral. The result is
\be\label{eqn:B1}
B_1=\frac{1}{120} \int_{-\infty}^\infty  d\bar{t} \, Y_1(\bar{t}) (g''(\bar{t})^2-2g'''(\bar{t})g'(\bar{t})+2g''''(\bar{t})g(\bar{t})) \,.
\ee

For the term proportional to $\tau_-^{-3}$, we have
\be
B_2=\int_0^\infty \frac{d \xi}{\pi} \int_{-\infty}^{\infty} d\tau G_2(\tau) \frac{1}{\tau_-^3} e^{-i\xi \tau} \,.
\ee
where
\be
G_2(\tau)=\int_{-\infty}^\infty d\bar{t} y_2(\bar{t},\tau) g\left(\bar{t}-\frac{\tau}{2}\right)g\left(\bar{t}+\frac{\tau}{2}\right) \,.
\ee
We calculate this Fourier transform in Appendix \ref{sec:fourierodd} and the result is
\be
B_2=\frac{1}{6} G'''_2(0) \,.
\ee
Applying the derivatives to $G_2$ gives
\be
G'''_2(\tau)\bigg|_{\tau=0}=\int_{-\infty}^\infty d\bar{t} \bigg[ \frac{d^3}{d\tau^3} y_2(\bar{t},\tau)\bigg|_{\tau=0}g(\bar{t})^2+\frac{3}{2}\frac{d}{d\tau} y_2(\bar{t},\tau)\bigg|_{\tau=0}(g''(\bar{t})g(\bar{t})-g'(\bar{t})^2) \bigg] \,.
\ee
Again the only dependence of $y_2$ on $\tau$ is in the form of $t''$ so we can integrate by parts
\bea
B_2=-\frac{1}{3} \int_{-\infty}^\infty d\bar{t} \int_0^1 d\lambda \bigg[&& 2 \left(\lambda-\frac{1}{2}\right)^4  Y_2(\bar{t}) (3g'(\bar{t}) g''(\bar{t})+g(\bar{t})g'''(\bar{t})) \nonumber\\
&&+\frac{3}{2} \left(\lambda-\frac{1}{2}\right)^2 Y_2(\bar{t})(g'''(\bar{t})g(\bar{t})-g''(\bar{t})g'(\bar{t})) \bigg] \,,
\eea
and perform the $\lambda$ integrals
\be\label{eqn:B2}
B_2=\frac{1}{60} \int_{-\infty}^\infty d\bar{t} \, Y_2(\bar{t})( g'(\bar{t})g''(\bar{t})-3 g'''(\bar{t})g(\bar{t}))\,.
\ee

For the term proportional to $\tau_-^{-2}$, we have
\be
B_3=\int_0^\infty \frac{d \xi}{\pi} \int_{-\infty}^{\infty} d\tau G_3(\tau) \frac{1}{\tau_-^2} e^{-i\xi \tau} \,.
\ee
where
\be
G_3(\tau)=\int_{-\infty}^\infty d\bar{t} (c_1(\bar{t})+y_3(\bar{t},\tau)) g\left(\bar{t}-\frac{\tau}{2}\right)g\left(\bar{t}+\frac{\tau}{2}\right) \,.
\ee
The result from Sec.~\ref{sec:Fourier} is
\be
B_3=\frac{1}{2} G''_3(0) \,.
\ee
Applying the derivatives to $G_3$ gives
\be
G''_3(\tau)\bigg|_{\tau=0}=\int_{-\infty}^\infty d\bar{t} \bigg[ \frac{d^2}{d\tau^2} y_3(\bar{t},\tau)\bigg|_{\tau=0} g(\bar{t})^2+\frac{1}{2}(c_1(\bar{t})+y_3(\bar{t}))(g''(\bar{t})g(\bar{t})-g'(\bar{t})^2) \bigg] \,.
\ee
As before, we integrate by parts
\bea
B_3=\frac{1}{2} \int_{-\infty}^\infty d\bar{t} \int_0^1 d\lambda
\bigg[& &2  \left(\lambda-\frac{1}{2}\right)^2(1-\lambda)\lambda Y_3(\bar{t}) (g'(\bar{t})^2+g(\bar{t})g''(\bar{t}))\nonumber\\
&&+\frac{1}{2} (c_1(\bar{t})+(1-\lambda)\lambda Y_3(\bar{t})) (g''(\bar{t})g(\bar{t})-g'(\bar{t})^2) \bigg]\,.
\eea
Integrating in $\lambda$ gives
\be\label{eqn:B3}
B_3=\frac{1}{4} \int_{-\infty}^\infty d\bar{t} \left[c_1(\bar{t}) (g''(\bar{t})g(\bar{t})-g'(\bar{t})^2)+\frac{1}{15} Y_3(\bar{t})(3g''(\bar{t})g(\bar{t})-2g'(\bar{t})^2) \right]\,.
\ee

The three remaining terms have Fourier transforms given in Sec.~\ref{sec:Fourier}
\bml\label{eqn:B46}\bea\label{eqn:B4}
B_4 &=& - \int_{-\infty}^{\infty} d\bar{t} \int_{-\infty}^\infty d\tau g'\left(\bar{t}+\frac{\tau}{2}\right)g\left(\bar{t}-\frac{\tau}{2}\right)\ln{|\tau|} c_2(\bar{t})\sgn{\tau} \\
B_5 &=& \int_{-\infty}^\infty d\bar{t} g(\bar{t})^2 (c_3(\bar{t})+c_5(\bar{t})) \\
B_6 &=&\frac{1}{\pi} \int_{-\infty}^\infty d\bar{t} \int_{-\infty}^\infty d\tau \frac{1}{\tau} g\left(\bar{t}+\frac{\tau}{2}\right)g\left(\bar{t}-\frac{\tau}{2}\right) c_4(\bar{t},\tau) \,,
\elea
where we added
\bea
c_5(\bar{t})= -(2a+b)R_{,\tu\tu}(\bar{t}) \,,
\eea
which is the local curvature term from Eq.~(\ref{eqn:B}).

The bound is now given by Eqs.~(\ref{eqn:BI}), (\ref{eqn:B0}),
(\ref{eqn:B1}), (\ref{eqn:B2}), (\ref{eqn:B3}), (\ref{eqn:B46}).

\section{The inequality}
\label{sec:theinequality}

We would like to bound the correction terms $B_1$ through $B_6$ using
bounds on the curvature and its derivatives.  Using
Eq.~(\ref{eqn:Rmax}) in Eq.~(\ref{eqn:Y1}), we find
\be\label{eqn:Y1bound}
|Y_1(\bar t)|<\frac{3}{2\pi^2}|\bar{x}^{\tu}|^2\Rmax\,.
\ee
We can use Eq.~(\ref{eqn:Y1bound}) in Eq.~(\ref{eqn:B1}) to get a
bound on $|B_1|$.  But will not be interested in specific numerical
factors, only the form of the quantities that appear in our bounds.
So we will write
\be
|B_1| \leq J_1^{(3)}[g] |\bar{x}^{\tu}|^2 \Rmax  \,,
\ee
where $J_1^{(3)}[g]$ is an integral of some combination of the
sampling function and its derivatives appearing in
Eq.~(\ref{eqn:B1}).  We will need many similar functionals
$J_n^{(k)}[g]$, which are listed at the end of the section.
The number in the parenthesis shows the
dimension of the integral,
\be
J_n^{(k)}[g] \sim \frac{1}{[L]^{k}} \,.
\ee

Similar analyses apply to $B_2$ and $B_3$ and the results are
\blea
|B_2| &\leq& J_2^{(2)}[g] |\bar{x}^{\tu}| \Rmax  \\
|B_3| &\leq& J_3^{(1)}[g] \Rmax \,.
\elea

Among the rest of the terms in $B$ there are some components of the form $R_{abcd,\tu}$ which diverge after boosting to the null geodesic, as shown in Ref.~\cite{Kontou:2012ve}. However we can show that these derivatives are not a problem since we can integrate them by parts. Suppose we have a term of the form
\be
B_n=\int_{\infty}^\infty d\bar{t} \int_{-\infty}^\infty d\tau L_n(\tau,\bar{t})  R_{abcd,\tu}(\bar{x}) \,,
\ee
where $L_n(\tau, \bar{t})$ is a function that contains the sampling function $g$ and its derivatives. The $\tu$ derivative on the Riemann tensor can be written
\be
R_{abcd,\tu}=R_{abcd,t}-R_{abcd,\tv} \,.
\ee
The term can be reorganized the following way by grouping the terms with $t$ and $\tv,x,y$ derivatives
\be
B_n=\int_{\infty}^\infty d\bar{t} \int_{-\infty}^\infty d\tau L_n(\tau,\bar{t}) (A^{abcd}_n R_{abcd,t}(\bar{x})+A^{abcd\alpha}_n R_{abcd,\alpha}(\bar{x}) ) \,,
\ee
where $A^{abcd \dots}_n$ are arrays with constant components and the subscript $n$ denotes the term they come from. Here the greek indices $\alpha,\beta,\dots=\tv,x,y$. The term with one derivative on $\alpha$ can be bounded while the term with one derivative on $t$ can be integrated by parts,
\be
B_n=-\int_{\infty}^\infty d\bar{t} \int_{-\infty}^\infty d\tau  (L_n'(\bar{t},\tau) A^{abcd}_n R_{abcd}(\bar{x})+L_n(\bar{t},\tau) A^{abcd\alpha}_n R_{abcd,\alpha}(\bar{x})) \,.
\ee
where the primes denote derivatives with respect to $\bar{t}$. The sampling function is $C_0^\infty$ so  $L'(\tau,\bar{t})$ is still smooth and the boundary terms vanish. Now it is possible to bound this term,
\be
|B_n| \leq \int_{\infty}^\infty d\bar{t} \int_{-\infty}^\infty d\tau(|L_n'(\bar{t},\tau)| a_n^{(0)} \Rmax +|L_n(\bar{t},\tau)| a_n^{(1)} \Rmax') \,,
\ee
where we defined
\be
a_n^{(m)} = \sum_{abcd \underbrace{\scriptstyle\alpha \beta \dots}_{m}} \left| A_n^{abcd \overbrace{\scriptstyle \alpha \beta \dots}^{m}} \right| \,.
\ee
The same method can be applied with more than one $\tu$ derivative. 

Now we apply this method to the integrals $B_4$, $B_5$ and $B_6$ of Eq.~(\ref{eqn:B46}). We start with $B_4$, which has the form
\bea
B_4&=&\int_{-\infty}^\infty d\tau \ln{|\tau|} \sgn{\tau} \int_{-\infty}^\infty d\bar{t} L_4(\bar{t},\tau) \bigg( A_4^{abcd} R_{abcd,tt}(\bar{x})+A_4^{abcd\alpha} R_{abcd,\alpha t}(\bar{x}) \nonumber\\
&& \qquad \qquad \qquad \qquad \qquad \qquad \qquad \qquad +A_4^{abcd\alpha \beta} R_{abcd,\alpha \beta}(\bar{x}) \bigg) \,,
\eea
where
\be
L_4(\bar{t},\tau)=g(\bar{t}+\tau/2)g'(\bar{t}-\tau/2) \,.
\ee
After integration by parts
\bea
B_4&=&\int_{-\infty}^\infty d\tau \ln{|\tau|} \sgn{\tau} \int_{-\infty}^\infty d\bar{t} \bigg( L_4''(\bar{t},\tau)  A_4^{abcd} R_{abcd}(\bar{x})-L_4'(\bar{t},\tau) A_4^{abcd\alpha} R_{abcd,\alpha}(\bar{x}) \nonumber\\
&&\qquad \qquad \qquad \qquad \qquad \qquad+L_4(\bar{t},\tau)A_8^{abcd\alpha \beta} R_{abcd,\alpha \beta}(\bar{x}) \bigg) \,.
\eea
Taking the bound gives
\be
|B_4| \leq \sum_{m=0}^2 J_4^{(1-m)}[g] \Rmax^{(m)} \,.
\ee

Reorganizing $B_5$ based on the number of $t$ derivatives gives
\bea
B_5&=&\int_{-\infty}^\infty d\bar{t} L_5(\bar{t}) \bigg( A_5^{abcd} R_{abcd,tt}(\bar{x})+A_5^{abcd\alpha} R_{abcd,\alpha t}(\bar{x})+A_5^{abcd\alpha \beta} R_{abcd,\alpha \beta}(\bar{x})\bigg) \\
&=&  \int_{-\infty}^\infty d\bar{t} \bigg( L_5''(\bar{t})  A_5^{abcd} R_{abcd}(\bar{x})-L_5'(\bar{t}) A_5^{abcd\alpha} R_{abcd,\alpha}(\bar{x})+L_5(\bar{t})A_5^{abcd\alpha \beta} R_{abcd,\alpha \beta}(\bar{x})\bigg)\nonumber \,,
\eea
where
\be
L_5(\bar{t})=g(\bar{t})^2 \,,
\ee
and the bound is
\be
|B_5| \leq \sum_{m=0}^2 J_5^{(1-m)}[g]  \Rmax^{(m)} \,.
\ee

Finally the remainder term is
\bea
B_6&=& \int_{-\infty}^\infty  d\bar{t} \int_{-\infty}^\infty d\tau L_6(\bar{t},\tau)  \int d\Omega\, \int_0^{1} d\lambda  \bigg\{  A^{abcd}_6(\lambda,\Omega) R_{abcd,ttt} (\lambda\Omega) \nonumber\\
&&+A^{abcd\alpha}_6(\lambda,\Omega) R_{abcd,\alpha tt}(\lambda \Omega)+A^{abcd\alpha \beta}_6 (\lambda,\Omega)R_{abcd,\alpha \beta t}(\lambda\Omega) \nonumber\\
&&+A_6^{abcd\alpha \beta \gamma}(\lambda,\Omega)R_{abcd,\alpha \beta \gamma}(\lambda \Omega)  \bigg\} \sgn{\tau} \,, 
\eea
where we changed variables to $\lambda=r/\tau$ and now arrays $A^{abcd \dots}_6$ have components that depend on $\lambda$ and $\Omega$, and 
\be
L_6(\bar{t},\tau)=g(\bar{t}-\tau/2)g(\bar{t}+\tau/2) \,.
\ee
After integration by parts
\bea
B_6&=& \int_{-\infty}^\infty  d\tau  \int_{-\infty}^\infty d\bar{t}  \int d\Omega  \int_0^1 d\lambda \bigg\{  L_6'''(\tau,\bar{t})A^{abcd}_6(\lambda,\Omega) R_{abcd} (\lambda\Omega) \nonumber\\
&&+L_6(\tau,\bar{t})'' A^{abcd\alpha}_6(\lambda,\Omega) R_{abcd,\alpha}(\lambda \Omega)+L_6(\tau,\bar{t})' A^{abcd\alpha \beta}_6(\lambda,\Omega)R_{ab,\alpha \beta}(\lambda\Omega)\nonumber\\
&&+L_6(\tau,\bar{t}) A_6^{abcd\alpha \beta \gamma}(\lambda,\Omega)R_{ab,\alpha \beta \gamma}(\lambda \Omega) \bigg\} \sgn{\tau} \,.
\eea
We define constants $a^{(m)}_6$ 
\be
a_6^{(m)}=\sum_{abcd\underbrace{\scriptstyle\alpha\beta\dots}_{m}} \left| \int d\Omega \int_0^1 d\lambda A_6^{abcd\overbrace{\scriptstyle\alpha\beta\dots}^{m}}(\lambda,\Omega) \right| \,,
\ee
and now we can take the bound
\bea
|B_6| \leq \sum_{m=0}^3 J_6^{(1-m)}[g]  \Rmax^{(m)} \,.
\eea

Putting everything together gives
\be\label{eqn:inequality}
B \leq \delta^{-2} \bigg(B_0+\sum_{n=1}^3 J^{(4-n)}_n [g] |\bar{x}^{\tu}|^{3-n} \Rmax+\sum_{n=4}^{6} \sum_{m=0}^3 J_n^{(1-m)}[g]  \Rmax^{(m)}\bigg) \,.
\ee
The functionals $J_n^{(k)}[g]$ are
\bml\label{eqn:Jmn}\bea
J_1^{(3)}[g]&=&\int_{-\infty}^\infty dt (a_{11}|g''''(t)|g(t)+a_{12}|g'''(t)g'(t)|+a_{13} g''(t)^2) \\
J_2^{(2)}[g]&=&\int_{-\infty}^\infty dt (a_{21}|g'''(t)|g(t)+a_{22}|g''(t)g'(t)|) \\
J_3^{(1)}[g]&=&\int_{-\infty}^\infty dt (a_{31}|g''(t)|g(t)+a_{32}g'(t)^2) \\
J_4^{(1)}[g]&=&\int_{-\infty}^\infty dt \int_{-\infty}^\infty dt' \left|\ln{|t-t'|}\right| \left(a_{41}|g'''(t')|g(t)+a_{42}|g''(t)g'(t')| \right) \\
J_4^{(0)}[g]&=&\int_{-\infty}^\infty dt \int_{-\infty}^\infty dt' \left|\ln{|t-t'|}\right| \left( a_{43}|g''(t')|g(t)+a_{44}|g'(t)g'(t')| \right)\\
J_4^{(-1)}[g]&=&\int_{-\infty}^\infty dt \int_{-\infty}^\infty dt' \left|\ln{|t-t'|}\right| a_{45} |g'(t')|g(t) \\
J_5^{(1)}[g]&=& \int_{-\infty}^\infty dt (a_{51}|g''(t)|g(t)+a_{52}g'(t)^2)\\
J_5^{(0)}[g]&=& \int_{-\infty}^\infty dt \, a_{53}|g'(t)|g(t)\\
J_5^{(-1)}[g]&=&\int_{-\infty}^\infty dt \, a_{54}g(t)^2 \\
J_6^{(1)}[g]&=&\int_{-\infty}^\infty dt \int_{-\infty}^\infty dt' (a_{61}|g'''(t)|g(t')+a_{62}|g''(t)g'(t')|) \, \\
J_6^{(0)}[g]&=&\int_{-\infty}^\infty dt \int_{-\infty}^\infty dt' (a_{63}|g''(t)|g(t')+a_{64} |g'(t)g'(t')|)\\
J_6^{(-1)}[g]&=&\int_{-\infty}^\infty dt \int_{-\infty}^\infty dt'  \, a_{65}|g'(t)|g(t')\\
J_6^{(-2)}[g]&=&\int_{-\infty}^\infty dt \int_{-\infty}^\infty dt'  \, a_{66} g(t)g(t') \,,
\elea
where $a_{nk}$ are positive constants that may depend on $a_n^{(m)}$.

We can change the argument of the sampling function, writing $g(t)=f(t/t_0)$, where $f$ is defined in Sec.~\ref{sec:theorem} and normalized according to Eq.~(\ref{eqn:normal}), so Eq.~(\ref{eqn:inequality}) becomes
\bea\label{eqn:QI}
\int dt T_{uu}(w(t)) g(t)^2 &\geq& -\frac{\delta^{-2}}{t_0^3} \bigg\{ \frac{1}{24\pi^2 t_0} \int_{-t_0}^{t_0} dt f''(t/t_0)^2+\sum_{n=1}^3 J^{(4-n)}_n [f] |\bar{x}^{\tu}|^{3-n} \Rmax t_0^{n-1} \nonumber\\
&&\quad+\sum_{n=4}^{6} \sum_{m=0}^3 J_n^{(1-m)}[f]  \Rmax^{(m)} t_0^{m+2} \bigg\} \,,
\eea
where we used $J^{(k)}_n[g]=t_0^{-k} J^{(k)}_n [f]$. We can simplify the inequality by defining
\be
F=\int f''(\alpha)^2 d\alpha=\frac{1}{t_0} \int f''(t/ t_0)^2 dt \,,
\ee
\be
F^{(m)}=\sum_{n=4}^{6} J_n^{(1-m)}[f]  \,,
\ee
and
\be
\bar{F}^{(n)}=J^{(4-n)}_n [f]  \,.
\ee
Then Eq.~(\ref{eqn:QI}) becomes
\bea \label{eqn:QIF}
&&\int dt T_{uu}(w(t)) g(t)^2 \geq  \\
&&\qquad \qquad -\frac{\delta^{-2}}{t_0^3}  \left\{ \frac{1}{24\pi^2} F+\sum_{m=0}^3 F^{(m)} \Rmax^{(m)} t_0^{m+2}+\sum_{n=1}^3 |\bar{x}^{\tu} |^{3-n} \bar{F}^{(n)} \Rmax t_0^{n-1}  \right\}\,. \nonumber
\eea

We will use this result to prove the achronal ANEC.

\section{The proof of the theorem}
\label{sec:proof}

We use Eq.~(\ref{eqn:QIF}) with $w(t)=\Phi_V(\eta,t)$ and
integrate in $\eta$ to get
\bea\label{eqn:lowerbound}
&&\int_{-\eta_0}^{\eta_0} d\eta \int_{-t_0}^{t_0} T_{uu} (\Phi_V(\eta,t)) f(t/t_0)^2 \geq \\
&& \qquad \qquad-\frac{\eta_0}{\delta^2 t_0^3} \left\{ \frac{1}{24
  \pi^2} F+\sum_{m=0}^3 F^{(m)} \Rmax^{(m)} t_0^{m+2}+\sum_{n=1}^3
|\bar{x}^{\tu} |^{3-n} \bar{F}^{(n)} \Rmax t_0^{n-1} \right\}\,. \nonumber
\eea
As $\delta\to\infty$, $t_0 \to 0$ but $F^{(m)}$, $\bar{F}^{(n)}$,
$\Rmax$, and $\Rmax^{(m)}$ are constant.  Now
$\bar{x}^{\tu}=\bar{x}^u/\delta$, and using Eqs.~(\ref{eqn:uvrange}),
(\ref{eqn:etavu}), $|\bar{x}^u|<u_1+\stwo\delta t_0$.  Thus
as $\delta\to\infty$, $\bar{x}^{\tu}\to 0$.  Therefore only the first
term in braces in Eq.~(\ref{eqn:lowerbound}) survives, so the bound
goes to zero as
\be\label{eqn:lbound}
\frac{\eta_0}{\delta^2 t_0^3} \sim \delta^{2\alpha-1}\,.
\ee
Equation~(\ref{eqn:lowerbound}) is a lower bound.  It says that its
left-hand side can be no more negative than the bound, which declines
as $\delta^{2\alpha-1}$.  But Eq.~(\ref{eqn:etintegral}) gives an upper
bound on the same quantity, saying that it must be more negative than
$-At_0/2$, which goes to zero as $t_0 \sim \delta^{-\alpha}$. Since
$\alpha<1/3$, the lower bound goes to zero more rapidly, and therefore
for sufficiently large $\delta$, the lower bound will be closer to
zero than the upper bound, and the two inequalities cannot be
satisfied at the same time. This contradiction proves Theorem 1.

The ambiguous local curvature terms do not contribute in the limit
$\eta_0 \to \infty$ because they are total derivatives proportional to
\be
\int_{-\eta_0}^{\eta_0}  d\eta R_{,uu}(\bar{x}))=0\,.
\ee

\chapter{Conclusions}
\label{ch:conclusions}

In this thesis we presented the derivation of quantum inequalities and a proof of the averaged null energy condition in curved spacetimes.  Using a general quantum inequality derived by Fewster and Smith \cite{Fewster:2007rh}, we first derived a bound for a quantum inequality in flat spacetime with a background potential, a case with similarities to the curved spacetime case,
\bea
\int_{\mathbb{R}} dt \,g(t)^2\langle T^{ren}_{tt}\rangle_{\omega}
(t,0) \geq- \frac{1}{16\pi^2} \bigg\{& &I_1+\frac12\Vmax J_2
+\Vmax''\left[\frac{1}{2}J_3+\left(\frac{11}{24}+48\pi^2|C|\right) J_4\right]
\nonumber\\
&&+\Vmax''' \left[\frac{11\pi+1}{16\pi}J_5
+\frac{2\pi+1}{64\pi}(4J_6+J_7)\right] \bigg\}\,,
\eea
where the $J_i$ integrals are given in Eq.~(\ref{J17V}). We then calculated the bound for a timelike projected quantum inequality in curved spacetime, 
\bea
\int_{\mathbb{R}} dt \,g(t)^2\langle T^{ren}_{tt}\rangle_{\omega}
(t,0) \geq- \frac{1}{16\pi^2} \bigg\{& &I_1+\frac{5}{6}\Rmax J_2\\
&+&\Rmax''\left[\frac{23}{60}J_3+\left(\frac{43}{40}+16\pi^2(24|a|+11|b|)\right) J_4\right]
\nonumber\\
&+&\Rmax''' \left[\frac{163\pi+14}{96\pi}J_5
+\frac{7(2\pi+1)}{192\pi}(4J_6+J_7)\right] \bigg\}\,,\nonumber
\eea
where the $J_i$'s in these case are given in Eq.~(\ref{J17}). The importance and application of these results, for example in the special case of vacuum spacetimes, is discussed in Chapters \ref{ch:potential} and \ref{ch:curvature}.
Next we presented the derivation of a null projected quantum inequality in curved spacetime 
\bea
\int_{-\infty}^\infty dt \,g(t)^2 \langle \Tren_{uu} \rangle (w(t)) &\geq& \delta^{-2} \bigg(B_0+\sum_{n=1}^3 J^{(4-n)}_n [g] |\bar{x}^{\tu}|^{3-n} \Rmax+\sum_{n=4}^{6} \sum_{m=0}^3 J_n^{(1-m)}[g]  \Rmax^{(m)}\bigg)\nonumber\\ \,,
\eea
where $J_n^{(m)}$ integrals are presented in Eq.~(\ref{eqn:Jmn}). Finally we used this result to prove achronal ANEC in spacetimes with curvature,
\be
\int_\gamma T_{ab} \ell^a \ell^b d\lambda \geq 0 \,.
\ee

As discussed in the introduction, Ref.~\cite{Graham:2007va} showed that to have an exotic spacetime
there would have to be violation of ANEC on achronal geodesics,
generated by a state of quantum fields in that same spacetime.  The result discussed above concerns integrals of the stress-energy
tensor of a quantum field in a background spacetime; we have so far
not been concerned about the back-reaction of the stress-energy tensor
on the spacetime curvature. Thus we have shown that no spacetime
that obeys NEC can be perturbed by a minimally-coupled quantum scalar
field into one which violates achronal ANEC.  This analysis is correct in the case
where the quantum field under consideration produces only a small
perturbation of the spacetime. Thus no such
perturbation of a classical spacetime would allow wormholes,
superluminal travel, or construction of time machines.

What possibilities remain for the generation of such exotic phenomena?

Could it be that a single effect both violates ANEC and produces the curvature that allows ANEC
to be violated?  The following heuristic argument casts doubt on this
possibility. Suppose ANEC violation and NEC violation have the same source.  We
will say that they are produced by an exotic stress-energy tensor
$\Texotic$.  This $\Texotic$ gives rise to an exotic
Einstein curvature tensor,
\be\label{eqn:G}
\Gexotic = 8\pi\lpl^2 \Texotic\,.
\ee
It is $\Gexotic$ that permits $\Texotic$
to arise from the quantum field.  Without $\Gexotic$, the spacetime
would obey the null convergence condition, and so, since $\Texotic$
violates ANEC, it would have to vanish.  A reasonable conjecture is
that as $\Gexotic \to 0$, $\Texotic \to 0$ at least
linearly.\footnote{Not, for example, changing discontinuously for
  infinitesimal but nonzero $\Gexotic$ or going as $\Gexotic^{1/2}$.}
We can then write schematically
\be\label{eqn:Texoticbound}
|\Texotic| \lesssim  l^{-2} |\Gexotic|\,,
\ee
where $l$ is a constant length obeying $l\gg \lpl$.  The parameter
$l$, needed on dimensional grounds, might be the wavelength of some
excited modes of the quantum field.  Equation~(\ref{eqn:Texoticbound})
is schematic because we have not said anything about the places
at which these tensors should be compared, or in what coordinate
system they should be measured.
Combining Eqs.~(\ref{eqn:G}) and (\ref{eqn:Texoticbound}), we find
\be
|\Texotic| \lesssim (\lpl/l)^2 |\Texotic|\,,
\ee
which is impossible since $l\gg \lpl$.

Given the assumptions of this work, it appears that the only
remaining possibility for self-consistent achronal ANEC violation
using minimally coupled free fields is to have first a quantum field
that violates NEC but obeys ANEC, and then a second quantum field (or
a second, weaker effect produced by the same field) that violates ANEC
when propagating in the spacetime generated by the first field.  The
stress-energy tensor of the second field would be a small correction
to that of the first, but this correction might lead to ANEC
violation on geodesics that were achronal (and thus obeyed ANEC only
marginally), taking into account only the first field.

There is also the possibility of different fields.  We have not
studied higher-spin fields, but these typically obey the same energy
conditions as minimally-coupled scalars. 

If one considers quantum scalar fields with non-minimal curvature
coupling, the situation is rather different.  Even classical
non-minimally coupled scalar fields can violate ANEC
\cite{Barcelo:1999hq,Barcelo:2000zf}, with large enough (Planck-scale)
field values.  However, as the field values increase toward such
levels, the effective Newton's constant first diverges and then becomes
negative. Recently an even stronger result has been proven; the effective Newton's constant has to change sign between the two asymptotic regions on
different ends of the wormhole \cite{Butcher:2015sea}. Such situations may not be physically realizable.   If one
excludes such field values, some restrictions are known, but there are
no quantum inequalities of the usual sort
\cite{Fewster:2006ti,Fewster:2007ec}, and there are general
\cite{Visser:1994jb} and specific \cite{Urban:2009yt,Urban:2010vr}
cases where conformally coupled quantum scalar fields violate ANEC in
curved space.  It may be possible to control such situations by
considering only cases where a spacetime is produced self-consistently
by fields propagating in that spacetime, but the status of this
``self-consistent achronal ANEC'' for non-minimally coupled scalar
fields outside the large-field region is not known.

\appendix

\chapter{Multi-step Fermi coordinates}
\label{Fermi}

In this Appendix, we generalize the Fermi coordinate construction to allow an
arbitrary number of arbitrary subspaces (and an arbitrary number of
dimensions $d$), rather than just a timelike geodesic and the
perpendicular space.  First, we construct the
generalized coordinate system, then we compute
the connection, and finally we compute the metric in
the generalized Fermi coordinates. Results shown in this Appendix are used throughout the main body of the thesis.
Here Greek indices $\alpha,\beta \dots $ refer to tetrad components while latin $a,b, \dots$ to coordinate basis. This notation is not used in the rest of the thesis.

\section{Multi-step Fermi coordinates}\label{sec:Fermi}

Consider a $d$-dimensional Riemannian or Lorentzian manifold $(\calM,
g)$.  We will start our construction by choosing a base point
$p\in\calM$.  We decompose the tangent space $T_p$ into $n$ subspaces,
$T_p = A^{(1)}_p \times A^{(2)}_p \times A^{(3)}_p \ldots \times
A_p^{(n)}$ so that any $V\in T_p$ can be uniquely written as $V=
V_{(1)} + V_{(2)} + V_{(3)} + \ldots + V_{(n)}$.  We choose,
as a basis for $T_p$, $d$ linearly independent vectors
$\{E_{(\alpha)}\}$ adapted to the decomposition of $T_p$ so that for
each $m=1\ldots n$, $\{E_{(\alpha)}|\alpha\in c_m\}$ is a basis for
$A_p^{(m)}$, where ${c_1,c_2, \ldots c_n}$ is an ordered partition of
$\{1\ldots d\}$. Thus each $V_{(m)}=\sum_{\alpha \in c_m}
x^\alpha E_{(\alpha)}$. The vectors $\{E_{(\alpha)}\}$ need not be
normalized or orthogonal.

\begin{figure}
\centering
\epsfysize=31mm
\epsfbox{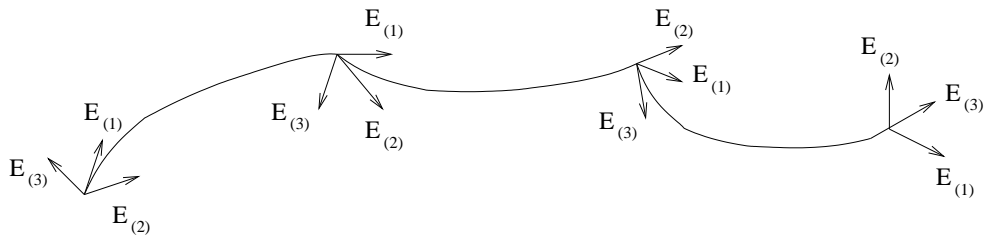}
\caption{Construction of 3-step Fermi coordinates in 3 dimensions.  We
  travel first in the direction of $E_{(1)}$, then $E_{(2)}$, then
  $E_{(3)}$, parallel transporting the triad as we go.}
\label{fig:3step}
\end{figure}

The point corresponding to coordinates $x^a$ is then found by starting
from $p$ and going along the geodesic whose whose tangent vector is
$V_{(1)}$, parallel transporting the rest of the vectors, then along
the geodesic whose tangent vector is $V_{(2)}$, and so on.  An
example is shown in Fig.~\ref{fig:3step}.
With that general construction of multi-step Fermi coordinates we can
define a general Fermi mapping $q=\Fermi_p(V)$ given by
\bea
q_{(0)}&=&p \nonumber\\
q_{(1)}&=&\exp_p(V_{(1)}) \nonumber\\
q_{(2)}&=&\exp_{q_{(1)}}(V_{(2)}) \\
\dots \nonumber\\
q=q_{(n)}&=&\exp_{q_{(n-1)}}(V_{(n)})\nonumber
\eea
From that general construction we can return to the original Fermi
case by choosing $c_1=\{t\}$ and $c_2=\{x,y,z\}$.  In the Lorentzian
case, we could also choose a pseudo-orthonormal tetrad
${E_u, E_v, E_x, E_y}$, with $E_u$ and $E_v$ null, $E_u
\cdot E_v = -1$, and other inner products vanishing, and $c_0=\{u\}$
and $c_1=\{v,x,y\}$.

For later use we will define
\bea
V_{(\leq m)}&=&\sum_{\alpha\in c_1 \cup c_2 \cup \dots \cup c_m} x^\alpha
E_{(\alpha)} = \sum_{l=1}^m V_{(l)}\\
V_{(<m)}&=&V_{(\le (m-1))} = V_{(\le m))} - V_{(m)}
\eea
Then we can write $q_{(m)} = \Fermi_p(V_{(\leq m)})$.

\section{Connection}\label{sec:connection}

We will parallel transport our orthonormal basis vectors
$E_{(\alpha)}$ along the geodesics that generate the coordinates,
and use them as a basis for vectors and tensors throughout the region
of $\calM$ covered by our coordinates.  Components in this basis will
be denoted by Greek indices.  We will use Latin letters from the
beginning of the alphabet to denote indices in the Fermi coordinate
basis.  Of course at $p$, there is no difference between these bases.

Latin letters from the middle of the alphabet will denote the
subspaces of $T_p$ or equivalently the steps of the Fermi mapping
process.

We would like to calculate the covariant derivatives of the basis
vectors, $\nabla_\beta E_{(\alpha)}$, which are connected with connection
one-forms, see for example Ref.~\cite{WaldGRbook} by
\be \label{eqn:oneforms}
\omega_{\beta \alpha \delta}=\eta_{\gamma \delta} \nabla_{\beta} E_{(\alpha)}^{\gamma}\,,
\ee
because we are using a orthonormal tetrad basis.

We can then calculate the covariant derivative of any vector field
$V = V^\beta E_{(\beta)}$ along a curve $f(\lambda)$ as
\be \label{eqn:covariant}
\frac{DV^\beta}{d\lambda} =\frac{dV^\beta}{d\lambda}  
+ V^\gamma \left(\frac{\partial}{\partial\lambda}\right)^\alpha
 \nabla_\alpha E_{(\gamma)}^\beta
\ee

To evaluate $\nabla_\beta E_{(\alpha)}$ at some point
$q_1=\exp_p(X)$, consider an infinitesimally separated point
$q_2=\exp_p(X+E_{(\eta)} dx)$.  The covariant derivative of
$E_{(\alpha)}$ at $q_1$ is the difference between
$E_{(\alpha)}(q_2)$ parallel transported to $q_1$ and the actual
$E_{(\alpha)}(q_1)$, divided by $dx$.  That difference is the same as
the change in $E_{(\alpha)}$ by parallel transport around a loop
following the geodesics from $q_1$ backward to $p$, the
infinitesimally different geodesics forward from $p$ to $q_2$, and the
infinitesimal distance back to $q_1$.  We can write this loop parallel
transport as an integral over the Riemann tensor.

\begin{figure}
\centering
\epsfysize=60mm
\epsfbox{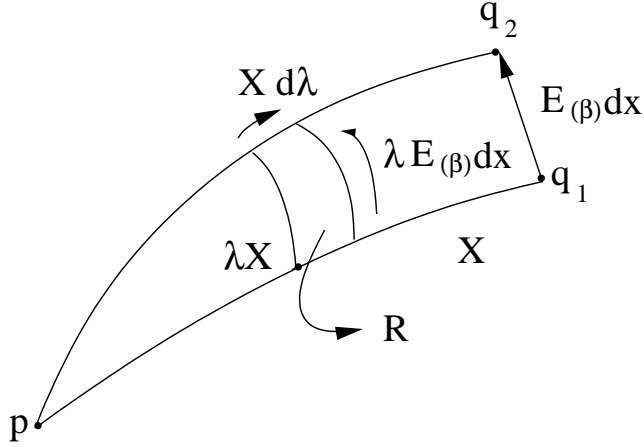}
\caption{The covariant derivative $\nabla_\beta E_{(\alpha)}$ is the
change in $E_{(\alpha)}$ under parallel transport along the path
$q_1\to p \to q_2 \to q_1$, divided by $dx$.  The parallel transport can
be decomposed into a series of transports clockwise around trapezoidal
regions with sides $\lambda E_{(\beta)} dx$ and $X d\lambda$.}
\label{fig:Riemann}
\end{figure}

Let us first consider the Riemannian case, as shown in
Fig.~\ref{fig:Riemann}.  The total parallel transport can be
written as the sum of parallel transport around a succession of small
trapezoidal regions whose sides are $\lambda E_{(\beta)}$ and $X
d\lambda$.   By using the definition of the Riemann tensor we have
\be\label{eqn:Riemanncovariant}
\nabla_\beta E_{(\alpha)}^\gamma = \int_0^1
d\lambda\, {R^\gamma}_{\alpha \delta \beta}(\lambda X) \lambda X^\delta\,.
\ee
Here $R$ is evaluated at the point $\exp_p(\lambda X)$, which we
have denoted merely $\lambda X$ for compactness.

Equation~(\ref{eqn:Riemanncovariant}) reproduces Eq.~(13) of
Ref.~\cite{Nesterov:1999ix}.  Note, however, that
Eq.~(\ref{eqn:Riemanncovariant}) is exact and does not require $R$ to
be smooth, whereas that of Ref.~\cite{Nesterov:1999ix} was given as
first order in $R$ and was derived by means of a Taylor series.

We see immediately that the covariant derivative of any $E_{(\alpha)}$
at $X$ in the direction of $X$ vanishes.  This
happens simply because changes with $dX$ in the direction of
$X$ correspond to additional parallel transport of
$E_{(\alpha)}$.

\begin{figure}
\centering
\epsfysize=40mm
\epsfbox{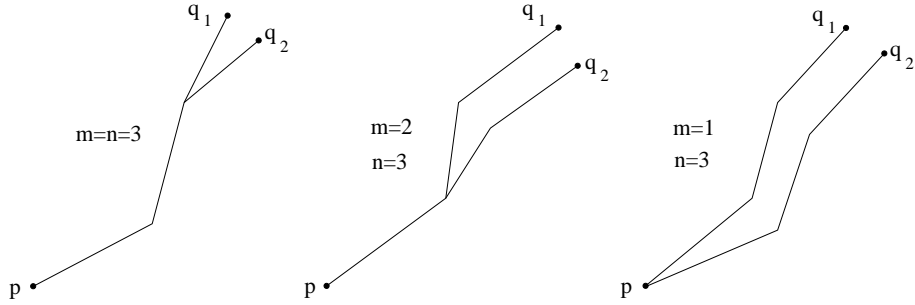}
\caption{Original and displaced geodesics for Fermi coordinates with
  $n=3$ and $m=3$, 2, and 1.}
\label{fig:fermipaths}
\end{figure}

Let us now consider the general case where there are $n$ steps, and
compute $\nabla_\beta E_{(\alpha)}$.  Since the coordinates are
adapted to our construction, the index $\beta$ must be in some
specific set $c_m$, which is to say that the direction of the
covariant derivative, $E_{(\beta)}$, is part of step $m$ in the
Fermi coordinate process.  We will write the function that gives that
$m$ as $m(\beta)$.  Some particular cases are shown in
Fig.~\ref{fig:fermipaths}.

If $m=n$ (leftmost in Fig.~\ref{fig:fermipaths}), only the last step
is modified.  The integration is exactly as shown in
Fig.~\ref{fig:Riemann}, except that it covers only the final geodesic
from $X_{(<n)}$ to $X_{(n)}$,
\be \label{eqn:covariantlast}
\nabla_\beta E_{(\alpha)}^\gamma= \int_0^1 d\lambda\, 
 {R^\gamma}_{ \alpha \delta \beta}(X_{(<n)} + \lambda X_{(n)})
 \lambda X_{(n)}^\delta.
\ee

\begin{figure}
\centering
\epsfysize=60mm
\epsfbox{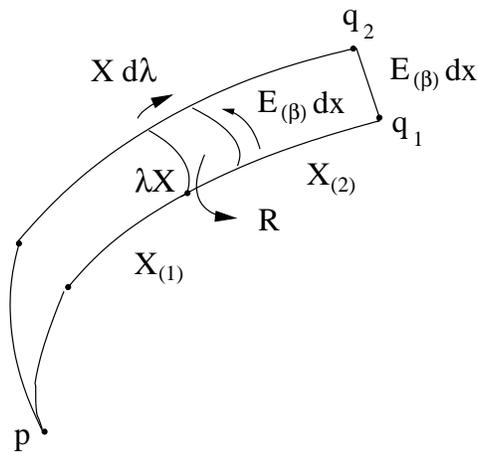}
\caption{Part of the calculation of $\nabla_\beta
  E_{(\alpha)}$ in the case $n=2$, $m(\beta)=1$.  The geodesic of the
  first step has been modified, causing the geodesic in the second
  step to be displaced.  The parallel transport integrates the Riemann
  tensor over a series of rectangular regions between the 2
  second-step geodesics.}
\label{fig:Fermi}
\end{figure}

If $m<n$, then we are modifying some intermediate step, and the path
followed at later steps is displaced parallel to itself.  In that case
we get an integral over rectangular rather than trapezoidal regions,
as shown in  Fig.~\ref{fig:Fermi}.
For general $m$ there is a contribution for each step $j \ge m$.  The
$j=m$ contribution integrates over trapezoids that grow with
$\lambda$, while the $j>m$ contributions integrate over rectangles
with fixed width $dx$.  We can write the complete result
\be \label{eqn:completecovariant}
\nabla_\eta E^\gamma_{(\alpha)}=\sum_{j=m}^n 
\int_0^1 d\lambda\, a_{jm}(\lambda)
{R^\gamma}_{ \alpha \delta \beta}(X_{(< j)} + \lambda X_{(j)}) X_{(j)}^\delta
\ee
where $m = m(\beta)$ and
\be
a_{jm}(\lambda)=\begin{cases}
1 & j \neq m\\
\lambda & j=m\,.
\end{cases}
\ee
Equation (\ref{eqn:completecovariant}) is exact and
includes Eqs.~(\ref{eqn:Riemanncovariant},\ref{eqn:covariantlast}) as
special cases.

Consider the case where $c_1$ consists only of one index.  If $m>1$,
there is no $j=1$ term in Eq.~(\ref{eqn:completecovariant}).  If
$m=1$, then $\beta$ is the single index in $c_1$, and
$X_{(1)}^\delta= 0$ unless $\delta =\beta$, so the $j=1$ term in
vanishes because $R_{\alpha\gamma\delta\beta}$ is antisymmetric
under $\delta \leftrightarrow\beta$.  Thus there is never a $j=1$
contribution to Eq.~(\ref{eqn:completecovariant}) when there is only
one index in $c_1$.

Now suppose $X$ lies on the first generating geodesic, so $X_{(j)} =
0$ for $j>1$.  Then all $j>1$ terms vanish in
Eq.~(\ref{eqn:completecovariant}).  So if $c_1$ consists only of one
index, all Christoffel symbols vanish at $X$.  This is well known in
the case of the usual Fermi coordinates.

\section{Metric}\label{sec:metric}

Now we would like to compute the metric $g$ at some point $X$.
Specifically, we would like to compute the metric component $g_{ab}$
in our generalized Fermi coordinates.

We will start by considering the vectors $Z_{(a)} =
\partial/\partial x^a$.  These are the basis vectors of the Fermi
coordinate basis for the tangent space, so the metric is given by
$g_{ab} =Z_{(a)} \cdot Z_{(b)}$.  Thus if we compute the
orthonormal basis components $Z_{(a)}^\alpha$ we can write
$g_{ab}=\eta_{\alpha \beta} Z_{(a)}^\alpha Z_{(b)}^\beta$.

Again we will start with the case of Riemann normal coordinates.  Let
$W(t,s)$ be the point $\exp_p s(X+tE_{(a)})$.  Define $Y =
\partial W/\partial t$ and $V = \partial W/\partial s$.  Then 
$Y(X) = Z_{(a)}$ and $V^\beta = X^\beta + t\delta^\beta_a$.
The components of $Z_{(a)}$ at $X$ can be calculated by integration,
\be\label{eqn:Yintegral}
Z_{(a)}^\beta(X)=Y^\beta(X)=\int_0^1 ds
\frac{\partial Y^\beta(s X)}{\partial s} .
\ee
Because the orthonormal basis is parallel transported we can write
\be\label{eqn:dYDV}
\frac{d}{ds} Y^\beta=\frac{D Y^\beta}{ds}.
\ee
By construction, the Lie
derivative $L_VY = 0$ and thus \cite[Ch.~4]{HawkingEllis}
\be\label{eqn:DYDV}
\frac{DY}{ds} = \frac{DV}{dt}
\ee
From Eq.~(\ref{eqn:covariant}) we have
\be\label{eqn:Vcovariant}
\frac{DV^\beta}{dt}=\frac{dV^\beta}{dt}
+V^\gamma Y^\alpha \nabla_\alpha E_{(\gamma)}^\beta 
= \delta_a^\beta+s \delta_a^\alpha V^\gamma \nabla_\alpha
E_{(\gamma)}^\beta + \order(R^2).
\ee
where we have retained $\delta_a^\alpha$ instead of writing $\nabla_a
E_{(\gamma)}^\beta$ to make it clear
that the covariant derivative is with respect to the orthonormal
basis.

From Eq.~(\ref{eqn:Riemanncovariant}) we have
\be\label{eqn:covariantE}
\nabla_\alpha E_{(\gamma)}^\beta( s X)=
\int_0^1 d\lambda\,\lambda {R^\beta}_{\gamma \delta \alpha}(\lambda s X)  sX^\delta
=\frac{1}{s}\int_0^s d \lambda\, \lambda {R^\beta}_{\gamma \delta \alpha}(\lambda  X)X^\delta
\ee
Taking $t=0$, $V$ is just $X$. Combining Eqs.~(\ref{eqn:Yintegral}-\ref{eqn:covariantE}), we
find
\bea\label{eqn:ZRiemann}
Z_{(a)}^\beta(X)&=& \int_0^1 ds \left[\delta_a^\beta
+ \delta_a^\alpha \int_0^s d \lambda\, \lambda {R^\beta}_{\gamma \delta
  \alpha}(\lambda  X)X^\delta X^\gamma\right]+\order(R^2)\nonumber\\
 &=&\delta_a^\beta+\delta_a^\alpha\int_0^1  d\lambda\,
\lambda (1-\lambda) {R^\beta}_{\gamma \delta \alpha}(\lambda X)X^\delta X^\gamma+\order(R^2).
\eea

From Eq.~(\ref{eqn:ZRiemann}), the metric is given by
\be\label{eqn:Riemannmetric}
g_{ab}=\eta_{ab}+2 \delta_a^\alpha \delta_b^Eta \int_0^1  d\lambda\, \lambda (1-\lambda) R_{\alpha \gamma \delta \beta}(\lambda X)X^\delta X^\gamma+\order(R^2).
\ee
Equation~(\ref{eqn:Riemannmetric}) reproduces Eq.~(14) of
Ref.~\cite{Nesterov:1999ix}\footnote{Ref.~\cite{Nesterov:1999ix} uses
  the same sign convention for ${R^\alpha}_{\beta\gamma\delta}$ as the
  present thesis, but the opposite convention for $g_{ab}$ and
  consequently also for $R_{\alpha\beta\gamma\delta}$.}.

Next let us consider the case where there are $n$ steps in our
procedure.  We will define a set of functions $W_j$ as
\be
W_j(s)=\Fermi_p(X_{(<j)}+sX_{(j)}).
\ee
The path $W_j(s), j=1\ldots n, s=0\ldots 1$ traces the geodesics
generating the Fermi coordinates for the point $X$.
Now consider $Z_{(a)} = \partial/\partial x^a$.  Let $m=m(a)$,
so $Z_{(a)}(p)\in A_p^{(m)}$.  Then let
\be
W_j(s,t)=\Fermi_p \begin{cases}
s X_{(j)} & j<m\\
X_{(<j)}+s(X_{(j)}+tE_{(a)}) & j=m\\
X_{(<j)}+tE_{(a)}+sX_{(j)} & j>m 
\end{cases}
\ee
Let $Y = \partial W/\partial t$ and $V = \partial W/\partial
s$ as before.
To find $Z_{(a)}$ we now must integrate over a multi-step path from $p$,
\be\label{eqn:YintegralFermi}
Z_{(a)}^\beta(X)=\sum_{j=1}^n \int_0^1 ds \frac{\partial
Y^\beta(W_j(s))}{\partial s}.
\ee
The generalized version of Eq.~(\ref{eqn:Vcovariant}) is
\be\label{eqn:VcovariantFermi}
\frac{DV^\beta (W_j(s,t))}{dt}=\frac{dV^\beta}{dt}
+V^\gamma Y^\alpha \nabla_\alpha E_{(\gamma)}^\beta 
=\begin{cases}
0 & j<m\\
\delta_a^\beta+s \delta_a^\alpha V^\gamma \nabla_\alpha
E_{(\gamma)}^\beta + \order(R^2) & j=m\\
\delta_a^\alpha V^\gamma \nabla_\alpha
E_{(\gamma)}^\beta + \order(R^2) & j>m.
\end{cases}
\ee 
Now
\be\label{eqn:covariantEfermi}
\nabla_\alpha E_{(\gamma)}^\beta(W_j(s))=
\sum_{k=m}^{j} \frac{1}{s_{kj}(s)}  \int_0^{s_{kj}(s)} d\lambda\, a_{km}(\lambda) {R^\beta}_{\gamma \delta \alpha}(X_{(<k)}
+\lambda X_{(k)}) X_{(k)}^\delta
\ee
where
\be
s_{kj}(s)=\begin{cases}
1 & k \neq j \\
s & k=j.
\end{cases}
\ee
The $k = j$ term is analogous to Eq.~(\ref{eqn:covariantE}), while
the others have no dependence on $s$.

Combining
Eqs.~(\ref{eqn:dYDV},\ref{eqn:DYDV},\ref{eqn:YintegralFermi}--\ref{eqn:covariantEfermi})
we get
\be
Z_{(a)}^\beta(X)= \delta_a^\beta+ F_a^\beta+ \order(R^2)
\ee
where
\bea
F_\alpha^\beta&=&\sum_{j=m}^n \sum_{k=m}^{j} 
\int_0^1 ds \int_0^{s_{kj}(s)} d\lambda\, a_{km}(\lambda) 
{R^\beta}_{\gamma \delta \alpha}(X_{(<k)} + \lambda X_{(k)}) X_{(k)}^\delta X_{(j)}^\gamma\nonumber\\
&&= \sum_{j=m}^n \sum_{k=m}^{j} \int_0^1 d\lambda\,a_{km}(\lambda)b_{kj}(\lambda) 
{R^\beta}_{\gamma \delta \alpha}(X_{(<k)} + \lambda X_{(k)}) X_{(k)}^\delta X_{(j)}^\gamma
\eea
where $m = m(\alpha)$ and
\be
b_{kj}(\lambda)= \begin{cases}
1 & k \neq j\\
1-\lambda & k=j
\end{cases}.
\ee
Thus the metric is
\be\label{eqn:smetric}
g_{ab}=\eta_{\alpha \beta} Z_{(a)}^\alpha Z_{(b)}^\beta
= \eta_{ab}+F_{ab}+F_{ba}+\order(R^2)
\ee
where
\be\label{eqn:Flower}
F_{\alpha\beta}= \sum_{j=m}^n \sum_{k=m}^{j} \int_0^1 d\lambda\,
a_{km}(\lambda)b_{kj}(\lambda) R_{\alpha\gamma\delta\beta}(X_{(<k)} + \lambda X_{(k)}) X_{(k)}^\delta
X_{(j)}^\gamma
\ee
where $m = m(\beta)$. 

Florides and Synge \cite{Florides} construct coordinates by taking
geodesics perpendicular to an embedded submanifold.  Our construction
for $n = 2$ is of this kind in the case where all basis vectors with
indices in $c_1$ lie tangent to the surface generated by all
first-step geodesics.  This will be so if
$R_{\alpha\gamma\delta\beta} = 0$ everywhere on this surface whenever
$\gamma,\delta,\beta\in c_1$ and $\alpha\in c_2$.  In that case,
Eq.~(\ref{eqn:Flower}) agrees with Theorem I of Ref.~\cite{Florides}.

Now, consider again the case where $c_1$ contains only one index.  As
discussed with respect to Eq.~(\ref{eqn:completecovariant}), if
$\beta\in c_1$, there is no nonvanishing $k=1$ term in
Eq.~(\ref{eqn:Flower}).  Thus $g_{ab} = \eta_{ab}$ everywhere on the
first generating geodesic.  This is also well known in the usual Fermi
case.

Now suppose $c_1$ consists only of one index and furthermore $n = 2$.
The only possible term in Eq.~(\ref{eqn:Flower}) is then
$j=k=2$, so
\be\label{eqn:Flower2}
F_{\alpha\beta}= \int_0^1 d\lambda\,
a_{2m}(\lambda)(1-\lambda) R_{ \alpha\gamma\delta\beta}(X_{(1)} + \lambda X_{(2)}) X_{(2)}^\delta
X_{(2)}^\gamma.
\ee
where $m = m(\beta)$.
Equation Eq.~(\ref{eqn:Flower2}) is equivalent to Eq.~(28) in
Ref.~\cite{Nesterov:1999ix} in the case where the generating curve of
the Fermi coordinates is a geodesic.

\section{Regularity of the coordinates}\label{sec:regularity}

Riemann normal coordinates cannot in general be defined over the
entirety of a manifold, because there might be points conjugate to the
base point $p$.  At such a point, some infinitesimal change to the
coordinates would yield no change to the resulting point, and the
metric would be singular.

Similar considerations apply to Fermi normal coordinates
\cite{Manasse:1963zz}.  No trouble can occur in the first step,
because that consists merely of traveling down a geodesic.  Along the
geodesic, and thus by continuity in some neighborhood surrounding the
geodesic, Fermi normal coordinates are regular.  If we attempt to
extend beyond this neighborhood, we may find points that are conjugate
to the generating geodesic.  In such places the metric will become
singular.

The situation here is more complicated.  The metric will be singular
whenever an infinitesimal change in coordinates fails to yield a
change in the location of the resulting point.  But when there are
more than two steps, the result can no longer be described in terms of
conjugate points.  Nevertheless, it is easy to see that one if one
chooses a sufficiently small neighborhood around $p$, all multi-step
Fermi coordinates will be well defined, since any such coordinates
approach Riemann normal coordinates when all coordinate values are
sufficiently small.

In case $c_1$ contains only one index, multi-step Fermi coordinates
will be well defined in a neighborhood of the initial geodesic,
because when all coordinate values except for the first are small, the
multi-step Fermi coordinates approach the regular Fermi coordinates.
There is no particular advantage to having a single index in any later
$c_m$.

One can get a simple condition sufficient for the existence of
multi-step Fermi coordinates in a small region by looking at
Eqs.~(\ref{eqn:smetric}) and (\ref{eqn:Flower}).  As long as
$|F_{ab}|\ll 1$, the metric $g_{ab}$ cannot degenerate.  Thus the
coordinates will be well defined if \cite{Nesterov:1999ix}
\be\label{eqn:Rlimit}
|R_{ \alpha\gamma\beta\delta}| (X^\epsilon)^2 \ll 1
\ee
throughout the region of interest, for all
$\alpha,\gamma,\beta,\delta,\epsilon$.  In the case where there is
only one index in $c_1$, there is no contribution to $F_{ab}$ from
$X_{(1)}$.  Then it is sufficient for Eq.~(\ref{eqn:Rlimit}) to hold
for $\epsilon>1$.  Thus if the first step is one-dimensional, it can
be arbitrarily long \cite{Manasse:1963zz}, as discussed above.

\chapter{Fourier transforms of some distributions}
\label{app:fouriergen}

\section{Fourier transforms of some distributions involving logarithms}
\label{app:fourier}

In this appendix will compute the Fourier transforms of the
distributions given by
\bea
u(\tau) &=& \ln|\tau|\\
v(\tau) &=& \ln(-\tau_-^2)\,.
\eea
We write $u$ as a distributional limit,
\be
u =  \lim_{\epsilon\to0^+} u_\epsilon\,,
\ee
where
\be
u_\epsilon(\tau) = \ln|\tau|e^{-\epsilon|\tau|}\,,
\ee
so its Fourier transform is
\be
\hat u_\epsilon(k) =\int_{-\infty}^\infty d\tau \ln|\tau|e^{-\epsilon|\tau|}e^{i k \tau}
= 2 \Re\int_0^\infty d\tau \ln \tau \,e^{(ik-\epsilon)\tau}
= -2 \Re\frac{\gamma + \ln(\epsilon-ik)}{\epsilon-ik}\,.
\ee
Thus the action of $\hat u$ on a test function $f$ is
\be
\hat u[f] = -2\lim_{\epsilon\to0^+} \Re\int_{-\infty}^\infty dk
\frac{\gamma + \ln(\epsilon-ik)}{\epsilon-ik} f(k)\,.
\ee
The term involving $\gamma$ is
\be
- 2\gamma \lim_{\epsilon\to0^+} \int_{-\infty}^\infty dk
\frac{\epsilon}{k^2+\epsilon^2}f(k)
= - 2\pi\gamma f(0)\,.
\ee
In the other term we integrate by parts,
\bea
- 2 \lim_{\epsilon\to0^+}\Re \int_{-\infty}^\infty dk
\frac{\ln(\epsilon-ik)}{\epsilon-ik} f(k)
&=& -\lim_{\epsilon\to0^+}\Im\int_{-\infty}^\infty dk\,f'(k) [\ln(\epsilon-ik)]^2\nonumber\\
&=& -\Im\int_{-\infty}^\infty dk\,f'(k) [\ln|k|-i(\pi/2)\sgn k]^2\nonumber\\
&=& \pi \int_{-\infty}^\infty dk\,f'(k) \ln|k|\sgn k\,,
\eea
and thus
\be\label{uhat}
\hat u[f] = \pi\int_{-\infty}^\infty dk\,f'(k) \ln|k|\sgn k -2\pi \gamma f(0)\,.
\ee
Since the Fourier transform of the constant $\gamma$ is just
$2\pi\gamma\delta(k)$, the transform of
\be\label{distw}
w(\tau) = \ln|\tau| + \gamma
\ee
is just
\be\label{distwhat}
\hat w[f] = \pi\int_{-\infty}^\infty dk\,f'(k) \ln|k|\sgn k\,.
\ee

Now
\be\label{vdist1}
v(\tau) = \lim_{\epsilon\to0}\ln(-(\tau-i\epsilon)^2) = 2\ln|\tau|+\pi
i \sgn \tau\,.
\ee
The Fourier transform of $\sgn$ acts on $f$ as \cite{Gelfand:functions}
\be\label{sgnhat}
2iP \int_{-\infty}^\infty dk\,\frac{f(k)}{k}
= -2i\int_{-\infty}^\infty dk\,f'(k) \ln |k|\,,
\ee
Putting Eqs.~(\ref{uhat},\ref{sgnhat}) in Eq.~(\ref{vdist1}) gives
\be
\hat v[f] = 4\pi\int_0^\infty dk\,f'(k) \ln|k| -4\pi \gamma f(0)\,.
\ee

\section{Fourier transform of distribution $\tau_-^{-3}$} 
\label{sec:fourierodd}

We follow the procedure of Sec.~\ref{sec:Fourier} to calculate 
\be
B_2=\int_0^\infty \frac{d\xi}{\pi} \int_{-\infty}^\infty d\tau G_2(\tau) s_2(\tau) e^{-i\xi\tau} \,.
\ee
where
\be
G_2(\tau)=\int_{-\infty}^\infty d\bar{t} y_2(\bar{t},\tau)g\left(\bar{t}-\frac{\tau}{2}\right)g\left(\bar{t}+\frac{\tau}{2}\right) \,.
\ee
and
\be
s_2(\tau)=\frac{1}{\tau_-^3} \,.
\ee
This is the Fourier transform of a product so we can write it as a
convolution. The function $s_2$ is real and odd, so its Fourier transform is imaginary, but $G_2$ is also real and odd, thus the Fourier transform of their product is real. We have
\be
B_2=\frac{1}{2\pi^2} \int_0^\infty d\xi \int_{-\infty}^\infty d\zeta  \hat{G}_2(-\xi-\zeta)\hat{s}_2(\zeta) \,.
\ee
We can change the order of integrals and change variables to $\eta=-\xi-\zeta$ which gives
\bea \label{eqn:convolution}
B_2&=&-\frac{1}{2\pi^2} \int_{-\infty}^\infty d\zeta \int_\zeta^{\infty} d\eta\, \hat{G}_2(\eta)\hat{s}_2(\zeta)\nonumber\\
&=& -\frac{1}{2\pi^2} \int_{-\infty}^\infty d\eta\, \hat{G}_2(\eta)\int_{-\infty}^\eta d\zeta \hat{s}_2(\zeta) \,.
\eea
The Fourier transform of $s_2$ is \cite{Gelfand:functions}
\be
\hat{s}_2(\zeta)=-i \pi \zeta^2 \Theta(\zeta) \,,
\ee
and
\be
\int_0^\eta d\zeta (-i \pi \zeta^2)=-\frac{i\pi}{3} \eta^3 \Theta(\eta) \,.
\ee
From Eq.~(\ref{eqn:convolution}) we have
\be
B_2=-\frac{i}{6\pi}\int_0^\infty d\eta\, \hat{G}_2(\eta) \eta^3 \,.
\ee
Using $\widehat{f'} (\xi)=-i\xi \hat{f}(\xi)$, we get
\be
B_2=\frac{1}{6\pi}\int_0^\infty d\eta\, \widehat{G'''_2}(\eta) \,.
\ee
The function $G_2$ is odd but with three derivatives it becomes even,
so we can extend the intergal
\be
B_2=\frac{1}{12\pi}\int_{-\infty}^\infty d\eta\, \widehat{G'''_2}(\eta)=\frac{1}{6}G'''_2(0) \,.
\ee

\bibliographystyle{plain}
{
\newpage
\addcontentsline{toc}{chapter}{Bibliography}
\bibliography{refs/thesisbib}

\begin{thebibliography}{10}

\bibitem{Barcelo:1999hq}
Carlos Barcelo and Matt Visser.
\newblock {Traversable wormholes from massless conformally coupled scalar
  fields}.
\newblock {\em Phys.Lett.}, B466:127--134, 1999.

\bibitem{Barcelo:2000zf}
Carlos Barcelo and Matt Visser.
\newblock {Scalar fields, energy conditions, and traversable wormholes}.
\newblock {\em Class.Quant.Grav.}, 17:3843--3864, 2000.

\bibitem{Birrellbook}
N.~D. Birrell and P.~C.~W. Davies.
\newblock {\em {Quantum Fields in Curved Space}}.
\newblock Cambridge University Press, 1982.

\bibitem{arXiv:1304.7682}
Henning Bostelmann, Daniela Cadamuro, and Christopher~J. Fewster.
\newblock {Quantum Energy Inequality for the Massive Ising Model}.
\newblock {\em Phys.Rev.}, D88(2):025019, 2013.

\bibitem{Butcher:2015sea}
Luke~M. Butcher.
\newblock {Traversable Wormholes and Classical Scalar Fields}.
\newblock 2015.

\bibitem{Decanini:2005gt}
Yves Decanini and Antoine Folacci.
\newblock {Off-diagonal coefficients of the Dewitt-Schwinger and Hadamard
  representations of the Feynman propagator}.
\newblock {\em Phys.Rev.}, D73:044027, 2006.

\bibitem{Epstein:1965zza}
H.~Epstein, V.~Glaser, and A.~Jaffe.
\newblock {Nonpositivity of energy density in Quantized field theories}.
\newblock {\em Nuovo Cim.}, 36:1016, 1965.

\bibitem{Fewster:1999gj}
Christopher~J. Fewster.
\newblock {A General worldline quantum inequality}.
\newblock {\em Class.Quant.Grav.}, 17:1897--1911, 2000.

\bibitem{Fewster:1998pu}
Christopher~J. Fewster and S.P. Eveson.
\newblock {Bounds on negative energy densities in flat space-time}.
\newblock {\em Phys.Rev.}, D58:084010, 1998.

\bibitem{math-ph/0412028}
Christopher~J. Fewster and Stefan Hollands.
\newblock {Quantum energy inequalities in two-dimensional conformal field
  theory}.
\newblock {\em Rev.Math.Phys.}, 17:577, 2005.

\bibitem{Fewster:2006uf}
Christopher~J. Fewster, Ken~D. Olum, and Michael~J. Pfenning.
\newblock {Averaged null energy condition in spacetimes with boundaries}.
\newblock {\em Phys.Rev.}, D75:025007, 2007.

\bibitem{Fewster:2006ti}
Christopher~J. Fewster and Lutz~W. Osterbrink.
\newblock {Averaged energy inequalities for the non-minimally coupled classical
  scalar field}.
\newblock {\em Phys.Rev.}, D74:044021, 2006.

\bibitem{Fewster:2007ec}
Christopher~J. Fewster and Lutz~W. Osterbrink.
\newblock {Quantum Energy Inequalities for the Non-Minimally Coupled Scalar
  Field}.
\newblock {\em J.Phys.}, A41:025402, 2008.

\bibitem{Fewster:2002ne}
Christopher~J. Fewster and Thomas~A. Roman.
\newblock {Null energy conditions in quantum field theory}.
\newblock {\em Phys.Rev.}, D67:044003, 2003.

\bibitem{Fewster:2007rh}
Christopher~J. Fewster and Calvin~J. Smith.
\newblock {Absolute quantum energy inequalities in curved spacetime}.
\newblock {\em Annales Henri Poincare}, 9:425--455, 2008.

\bibitem{Florides}
P.~S. Florides and J.~L. Synge.
\newblock Coordinate conditions in a riemannian space for coordinates based on
  a subspace.
\newblock {\em Proc. R. Soc. Lond. A}, 323:1--10, 1971.

\bibitem{Ford:1978qya}
L.H. Ford.
\newblock {Quantum Coherence Effects and the Second Law of Thermodynamics}.
\newblock {\em Proc.Roy.Soc.Lond.}, A364:227--236, 1978.

\bibitem{Ford:1994bj}
L.H. Ford and Thomas~A. Roman.
\newblock {Averaged energy conditions and quantum inequalities}.
\newblock {\em Phys.Rev.}, D51:4277--4286, 1995.

\bibitem{Ford:1995wg}
L.H. Ford and Thomas~A. Roman.
\newblock {Quantum field theory constrains traversable wormhole geometries}.
\newblock {\em Phys.Rev.}, D53:5496--5507, 1996.

\bibitem{Gelfand:functions}
I.~M. Gel'fand and G.~E. Shilov.
\newblock {\em Generalized functions}.
\newblock Academic press, New York and London, 1964.

\bibitem{Graham:2007va}
Noah Graham and Ken~D. Olum.
\newblock {Achronal averaged null energy condition}.
\newblock {\em Phys.Rev.}, D76:064001, 2007.

\bibitem{HawkingEllis}
Stephen~W. Hawking and G.~F.~R. Ellis.
\newblock {\em The Large Scale Structure of Space-time}.
\newblock Cambridge University Press, London, 1973.

\bibitem{Kontou:2012ve}
Eleni-Alexandra Kontou and Ken~D. Olum.
\newblock {Averaged null energy condition in a classical curved background}.
\newblock {\em Phys.Rev.}, D87(6):064009, 2013.

\bibitem{Kontou:2012kx}
Eleni-Alexandra Kontou and Ken~D. Olum.
\newblock {Multi-step Fermi normal coordinates}.
\newblock {\em Class.Quant.Grav.}, 30:175018, 2013.

\bibitem{Kontou:2014eka}
Eleni-Alexandra Kontou and Ken~D. Olum.
\newblock {Quantum inequality for a scalar field with a background potential}.
\newblock {\em Phys.Rev.}, D90(2):024031, 2014.

\bibitem{Kontou:2014tha}
Eleni-Alexandra Kontou and Ken~D. Olum.
\newblock {Quantum inequality in spacetimes with small curvature}.
\newblock 2014.

\bibitem{Manasse:1963zz}
F.K. Manasse and C.W. Misner.
\newblock {Fermi Normal Coordinates and Some Basic Concepts in Differential
  Geometry}.
\newblock {\em J.Math.Phys.}, 4:735--745, 1963.

\bibitem{MTW}
Charles~W. Misner, K.S. Thorne, and J.A. Wheeler.
\newblock {\em Gravitation}.
\newblock W. H. Freeman, San Francisco, 1973.

\bibitem{Nesterov:1999ix}
Alexander~I. Nesterov.
\newblock {Riemann normal coordinates, Fermi reference system and the geodesic
  deviation equation}.
\newblock {\em Class.Quant.Grav.}, 16:465--477, 1999.

\bibitem{Poisson:2011nh}
Eric Poisson, Adam Pound, and Ian Vega.
\newblock {The Motion of point particles in curved spacetime}.
\newblock {\em Living Rev.Rel.}, 14:7, 2011.

\bibitem{Urban:2009yt}
Douglas Urban and Ken~D. Olum.
\newblock {Averaged null energy condition violation in a conformally flat
  spacetime}.
\newblock {\em Phys. Rev.}, D81:024039, 2010.

\bibitem{Urban:2010vr}
Douglas Urban and Ken~D. Olum.
\newblock {Spacetime Averaged Null Energy Condition}.
\newblock {\em Phys. Rev.}, D81:124004, 2010.

\bibitem{Visser:1992pz}
Matt Visser.
\newblock {van Vleck determinants: Geodesic focusing and defocusing in
  Lorentzian space-times}.
\newblock {\em Phys.Rev.}, D47:2395--2402, 1993.

\bibitem{Visser:1994jb}
Matt Visser.
\newblock {Scale anomalies imply violation of the averaged null energy
  condition}.
\newblock {\em Phys. Lett.}, B349:443--447, 1995.

\bibitem{Visser:1996iw}
Matt Visser.
\newblock {Gravitational vacuum polarization. 1: Energy conditions in the
  Hartle-Hawking vacuum}.
\newblock {\em Phys.Rev.}, D54:5103--5115, 1996.

\bibitem{Visser:1996iv}
Matt Visser.
\newblock {Gravitational vacuum polarization. 2: Energy conditions in the
  Boulware vacuum}.
\newblock {\em Phys.Rev.}, D54:5116--5122, 1996.

\bibitem{Visser:1997sd}
Matt Visser.
\newblock {Gravitational vacuum polarization. 4: Energy conditions in the Unruh
  vacuum}.
\newblock {\em Phys.Rev.}, D56:936--952, 1997.

\bibitem{Wald:1978pj}
Robert~M. Wald.
\newblock {Trace Anomaly of a Conformally Invariant Quantum Field in Curved
  Space-Time}.
\newblock {\em Phys.Rev.}, D17:1477--1484, 1978.

\bibitem{WaldGRbook}
Robert~M. Wald.
\newblock {\em General Relativity}.
\newblock Chicago University Press, 1984.

\bibitem{Wald:qft}
Robert~M. Wald.
\newblock {\em Quantum Field Theory in Curved Spacetime and Black Hole
  Thermodynamics}.
\newblock Chicago University Press, 1994.

\end{thebibliography}
}

\end{document}